\documentclass[fleqn]{elsart}
\usepackage{amsmath}
\usepackage{bm}
\usepackage{url}
\usepackage{graphicx}

\renewcommand{\vec}[1]{\hm{#1}}
\def\B{\text{B}}

\newcommand{\lvec}[1]{\line(-1,0){#1}\put(-#1,0)%
{\put(-0.8,0){\line(4,1){4}}\put(-0.8,0){\line(4,-1){4}}}}
\newcommand{\rvec}[1]{\line(1,0){#1}\put(0,0)%
{\put(0.8,0){\line(-4,1){4}}\put(0.8,0){\line(-4,-1){4}}}}

\def\drawgrid{%
\put(2,1){$0$}%
\put(0,5){\vector(1,0){130}}%
\put(124,1){$\pi$}%
\put(128,7){$k$}%
\put(5,0){\vector(0,1){45}}%
\put(1,42){$A$}%
\put(5,17){\line(-1,0){1}}%
\put(2,13){$1$}%
\put(5,29){\line(-1,0){1}}%
\put(2,25){$2$}%
\put(5,41){\line(-1,0){1}}%
\put(2,37){$3$}%
\put( 15,5){\line(0,-1){1}}%
\put( 25,5){\line(0,-1){1}}%
\put( 35,5){\line(0,-1){1}}%
\put( 45,5){\line(0,-1){1}}%
\put( 55,5){\line(0,-1){1}}%
\put( 65,5){\line(0,-1){1}}%
\put( 75,5){\line(0,-1){1}}%
\put( 85,5){\line(0,-1){1}}%
\put( 95,5){\line(0,-1){1}}%
\put(105,5){\line(0,-1){1}}%
\put(115,5){\line(0,-1){1}}%
\put(125,5){\line(0,-1){1}}%
\put( 15,5){\dashbox{1}(0,36){}}% 1
\put( 25,5){\dashbox{1}(0,36){}}% 2
\put( 35,5){\dashbox{1}(0,36){}}% 3
\put( 45,5){\dashbox{1}(0,36){}}% 4
\put( 55,5){\dashbox{1}(0,36){}}% 5
\put( 65,5){\dashbox{1}(0,36){}}% 6
\put( 75,5){\dashbox{1}(0,36){}}% 7
\put( 85,5){\dashbox{1}(0,36){}}% 8
\put( 95,5){\dashbox{1}(0,36){}}% 9
\put(105,5){\dashbox{1}(0,36){}}% 10
\put(115,5){\dashbox{1}(0,36){}}% 11
\put(125,5){\dashbox{1}(0,36){}}% 12
}

\def\drawgridA{%
\put(2,1){$0$}%
\put(0,5){\vector(1,0){130}}%
\put(124,1){$\pi$}%
\put(128,7){$k$}%
\put(5,0){\vector(0,1){45}}%
\put(1,42){$A$}%
\put(5,41){\line(-1,0){1}}%
\put(2,37){$1$}%
\put( 15,5){\line(0,-1){1}}%
\put( 25,5){\line(0,-1){1}}%
\put( 35,5){\line(0,-1){1}}%
\put( 45,5){\line(0,-1){1}}%
\put( 55,5){\line(0,-1){1}}%
\put( 65,5){\line(0,-1){1}}%
\put( 75,5){\line(0,-1){1}}%
\put( 85,5){\line(0,-1){1}}%
\put( 95,5){\line(0,-1){1}}%
\put(105,5){\line(0,-1){1}}%
\put(115,5){\line(0,-1){1}}%
\put(125,5){\line(0,-1){1}}%
\put( 15,5){\dashbox{1}(0,36){}}% 1
\put( 25,5){\dashbox{1}(0,36){}}% 2
\put( 35,5){\dashbox{1}(0,36){}}% 3
\put( 45,5){\dashbox{1}(0,36){}}% 4
\put( 55,5){\dashbox{1}(0,36){}}% 5
\put( 65,5){\dashbox{1}(0,36){}}% 6
\put( 75,5){\dashbox{1}(0,36){}}% 7
\put( 85,5){\dashbox{1}(0,36){}}% 8
\put( 95,5){\dashbox{1}(0,36){}}% 9
\put(105,5){\dashbox{1}(0,36){}}% 10
\put(115,5){\dashbox{1}(0,36){}}% 11
\put(125,5){\dashbox{1}(0,36){}}% 12
}

\def\drawgridB{%
\put(2,1){$0$}%
\put(0,5){\vector(1,0){130}}%
\put(124,1){$\pi$}%
\put(128,7){$k$}%
\put(5,0){\vector(0,1){45}}%
\put(1,42){$A$}%
\put(5,23){\line(-1,0){1}}%
\put(2,19){$1$}%
\put(5,41){\line(-1,0){1}}%
\put(2,37){$2$}%
\put( 15,5){\line(0,-1){1}}%
\put( 25,5){\line(0,-1){1}}%
\put( 35,5){\line(0,-1){1}}%
\put( 45,5){\line(0,-1){1}}%
\put( 55,5){\line(0,-1){1}}%
\put( 65,5){\line(0,-1){1}}%
\put( 75,5){\line(0,-1){1}}%
\put( 85,5){\line(0,-1){1}}%
\put( 95,5){\line(0,-1){1}}%
\put(105,5){\line(0,-1){1}}%
\put(115,5){\line(0,-1){1}}%
\put(125,5){\line(0,-1){1}}%
\put( 15,5){\dashbox{1}(0,36){}}% 1
\put( 25,5){\dashbox{1}(0,36){}}% 2
\put( 35,5){\dashbox{1}(0,36){}}% 3
\put( 45,5){\dashbox{1}(0,36){}}% 4
\put( 55,5){\dashbox{1}(0,36){}}% 5
\put( 65,5){\dashbox{1}(0,36){}}% 6
\put( 75,5){\dashbox{1}(0,36){}}% 7
\put( 85,5){\dashbox{1}(0,36){}}% 8
\put( 95,5){\dashbox{1}(0,36){}}% 9
\put(105,5){\dashbox{1}(0,36){}}% 10
\put(115,5){\dashbox{1}(0,36){}}% 11
\put(125,5){\dashbox{1}(0,36){}}% 12
}

\begin{document}
\begin{frontmatter}
\journal{Physica D}
\title{Stability of low-dimensional bushes of vibrational modes in the
Fermi-Pasta-Ulam chains}
\author{G.M.~Chechin},
\ead{chechin@phys.rsu.ru}
\author{D.S.~Ryabov},
\author{K.G.~Zhukov}
\address{Department of Physics, Rostov State University, Zorge 5,
Rostov-on-Don, 344090 Russia}
\begin{abstract}
Bushes of normal modes represent the exact excitations in the nonlinear
physical systems with discrete symmetries~[Physica~D \textbf{117} (1998)
43]. The present paper is the continuation of our previous paper~[Physica~D
\textbf{166} (2002) 208], where these dynamical objects of new type were
discussed for the monoatomic nonlinear chains. Here, we develop a simple
crystallographic method for finding bushes in nonlinear chains and
investigate stability of one-dimensional and two-dimensional vibrational
bushes for both FPU-$\alpha$ and FPU-$\beta$ models, in particular, of
those revealed recently in~[Physica~D \textbf{175} (2003) 31].
\end{abstract}
\end{frontmatter}

\maketitle

\section{Introduction}

This paper expands upon our previous work~\cite{FPU1} where an outline of
the general theory of the bushes of modes~\cite{DAN1,DAN2,PhysD} was
presented in connection with the FPU-chain dynamics. Nevertheless, it seems
reasonable to recapitulate here some basic notions and ideas.

\subsection{The concept of bushes of modes}

Bushes of modes can be considered as a new type of \textit{exact}
excitations in \textit{nonlinear} systems with discrete symmetries, such as
molecules and crystals. The simplest way for introducing this concept can
be described as follows.

Let us have an $N$-particle Hamiltonian system characterized by a symmetry
group $G_0$ in its equilibrium state. Let us also assume that this system
permits the harmonic approximation and, therefore, we can introduce a
complete set of normal modes. Note that every normal mode possesses its own
symmetry group $G_j$ which is a subgroup of the group $G_0$. Let us now
excite only one normal mode at the initial instant by imposing the
appropriate initial conditions. We call this mode by the term ``root
mode''. Normal modes are independent of each other in the harmonic
approximation. However, if we take into account some anharmonic terms of
the Hamiltonian, the excitation will transfer from the root mode to a
number of other normal modes with zero amplitudes at the initial instant,
so-called ``secondary modes''. Because of certain \textit{selection rules}
for the excitation transfer between modes of different symmetry, the number
of the secondary modes can often be rather \textit{small}.

\textit{Definition}. The complete collection of the root mode and all
secondary modes, corresponding to it, forms a \textit{bush of normal
modes}. The number of modes in this collection is the \textit{dimension} of
the given bush.

To avoid any misunderstanding at this point, let us note that one can
consider normal modes, determined for a linear system, as a basis for a
decomposition of different dynamical regimes in the nonlinear system (for
more details, see below).

The energy of the initial excitation turns out to be trapped in the bush
simply because of the above definition. The number of modes in the bush
does not change in time while the amplitudes of these modes do change.

For many systems, we can find one-dimensional, two-dimensional,
three-dimensional and so on bushes of modes. Note that one-dimensional
bushes can be treated as the \textit{similar nonlinear normal modes}
introduced by Rosenberg forty years ago~\cite{Rosenberg}.

Every bush possesses its own symmetry group $G$ which is a
\textit{subgroup} of the symmetry group of the Hamiltonian of the
considered system. In the above discussed case, when the bush is excited by
perturbing the root mode, the symmetry group $G$ of the bush coincides with
that of this root mode. The symmetries of all other modes of the bush are
\textit{greater} than or \textit{equal} to its full symmetry $G$. The
development of the above ideas leads to the following important statement
of the general theory~\cite{FPU1,DAN1,PhysD}:

\textit{Proposition}. Different nonlinear dynamical regimes of a physical
system can be classified by subgroups of its symmetry group $G_0$
(so-called ``parent'' group)\footnote{We prefer to use this term because
the considered theory is valid not only for the Hamiltonian systems, but
also for the dissipative systems.}.

It is important to emphasize that the considered bushes of modes are
\textit{symmetry determined} objects: the sets of their modes \textit{do
not depend} on the interparticle interactions in the physical system.
Taking into account the specific character of these interactions can only
\textit{reduce} the dimension of the bush.

\subsection{Some properties of bushes of modes}

In general case, we must speak about bushes of \textit{symmetry-adapted}
modes rather than normal modes. Indeed, we consider the symmetry determined
bushes whose finding does not need any information about the interactions
in the system. The symmetry-adapted coordinates are the \textit{basis
vectors} of the \textit{irreducible representations} (irreps) of the parent
symmetry group $G_0$ and they can be found by group-theoretical methods
only, without any information about the interactions. In contrast, the
normal coordinates, used for construction of normal modes, must be
obtained, generally, by diagonalization of the Hamiltonian matrix and,
therefore, the interparticle interactions must be taken into account.

In the geometrical sense, a bush of modes is an \textit{invariant manifold}
singled out by its symmetry and decomposed into the basis vectors of the
irreducible representations of the appropriate parent symmetry group.

This decomposition plays a very important role in our approach. Indeed,
different irreps describe the transformation properties of the dynamical
variables corresponding to \textit{different physical characteristics} of
the considered system. For example, some irreps are active in the infrared
or Raman experiments, while others are nonactive in optics, but play the
essential role in neutronographics, etc.

In this paper, we discuss bushes of \textit{vibrational} modes describing
the time-dependent displacements of the particles of nonlinear chains from
their equilibrium positions. Speaking about the symmetry group of a single
mode or of the bush as integral object, we mean the specific symmetry of
the \textit{patterns} of the instantaneous displacements of all particles
which correspond to the mode or to the bush. Considering the symmetry group
of the displacement pattern and the collection of the normal modes
contained in a given bush, we deal with the \textit{geometrical} aspect of
the bush. This aspect is independent of the interparticle interactions in
the chain.

On the other hand, we deal with the \textit{dynamical} aspect of the bush
when consider the differential equations describing the time-dependence of
the amplitudes of the bush modes. In this sense, the given bush represents
a \textit{reduced} dynamical system whose dimension is, in general, less
(often, considerably less!) than the dimension of the original physical
system. Obviously, the dynamical aspect of the bush, in contrast to the
geometrical aspect, depends essentially on the interparticle interactions.

The modes of the bush are coupled by the so-called ``force interactions'',
while their coupling to all other modes is brought about by ``parametric
interactions''~\cite{PhysD}. The latter interactions can lead to a loss of
stability of the bush, if the amplitudes of its modes become sufficiently
large~\cite{PhysD,FPU1,Okta}, because of the phenomenon similar to the
well-known parametric resonance. In this case, the spontaneous breaking of
the symmetry of the system's vibrational state takes place and, as a
consequence, the given bush transforms into a bush of larger dimension.
Precisely this mechanism of stability loss by the bushes of the FPU chains
is discussed in this paper. The other mechanisms of the loss of bush
stability will be considered elsewhere.

\subsection{Some history}

The general concept of bushes of modes for physical systems with discrete
symmetry was introduced in~\cite{DAN1}. Actually, this concept appeared as
a generalization of the notion of the complete condensate of primary and
secondary order parameters, whose theory was developed in our papers
devoted to phase transitions in crystals~\cite{PhysStatSol,Acta}. The
discussion of properties of bushes of modes, the group-theoretical methods
for their construction, some theorems about the bush structure can be found
in~\cite{DAN1,DAN2,PhysD}. The bushes of modes for different classes of
mechanical systems with point and space symmetry were investigated
in~\cite{DAN2,PhysD,Okta,NNM,ENOC3,C60}. In particular, let us mention the
investigation of

-- all possible ``irreducible bushes'' and symmetry determined similar
nonlinear normal modes for the mechanical systems with any of the 230 space
groups~\cite{NNM};

-- low-dimensional bushes for all point groups of crystallographic
symmetry~\cite{ENOC3};

-- bushes of vibrational modes for the fullerene $C_{60}$ (``buckyball''
structure)~\cite{C60};

-- bushes of modes and their stability for the octahedral structure with
Lennard-Jones potential~\cite{Okta}.

A review of the group-theoretical methods and the appropriate computer
algorithms for treating the condensates of order parameters and bushes of
modes can be found in~\cite{Comput}. A remarkable computer program
\texttt{ISOTROPY} by Stokes and Hatch, capable to deal with many
group-theoretical methods of crystal physics and, in particular, with
bushes of modes, is now available on the Internet as free
software~\cite{Isotropy}.

In~\cite{PoggiRuffo}, Poggi and Ruffo revealed some exact solutions for the
FPU-$\beta$ chain dynamics, by analyzing the appropriate dynamical
equations, and investigated their stability. These solutions correspond to
certain subsets of normal modes (so-called ``subsets of I type'') ``where
energy remains trapped for suitable initial conditions''.

In~\cite{FPU1}, we showed that these subsets are the simplest cases of the
bushes of modes for the monoatomic nonlinear chains and that they can be
found without taking into account the information about the specific
interactions in the FPU-$\beta$ model. We discussed there the
symmetry-determined bushes of modes with respect to the translational group
$T$, in detail, and the bushes with respect to the dihedral group $D$,
partially. Also, in the paper~\cite{FPU1}, the stability of the former
bushes for the FPU-$\alpha$ chain was studied.

In a recent paper~\cite{BRink}, Bob Rink developed a method for finding all
possible symmetric invariant manifolds for the monoatomic chains. In spite
of the fact that this paper was carried out independently of our approach,
based on the concept of bushes of modes, and that it is written in a more
rigorous mathematical style, the main idea and the group-theoretical method
are very similar to those in our previous papers. Actually, Bob Rink found
all symmetry-determined bushes of modes for the monoatomic chains with
respect to the dihedral symmetry group. Moreover, he revealed some bushes
for the FPU-$\beta$ chain which are brought about by the additional
symmetry of this mechanical system connected with the evenness of its
potential (hereafter we call them ``additional bushes''). However, let us
note that the general idea of the classification of the symmetric manifolds
(bushes of modes) by the \textit{subgroups} of the symmetry group of the
Hamiltonian was not emphasized explicitly in this paper.

Finally, we would like to comment on two papers by
Shinohara~\cite{Shinohara}, devoted to finding ``type I subsets of modes''
(bushes, in our terminology) in nonlinear chains with fixed endpoints. The
author does not use any symmetry-related method and prefers to analyze the
specific structure of the dynamical equations (practically, in the manner
similar to that of Poggi and Ruffo in~\cite{PoggiRuffo}). Moreover, he
writes that the ``approach, which relies on the direct analysis of the
mode-coupling coefficients, is much simpler'' than the one based on the
group-theoretical methods. In contrast to this opinion, we think that the
situation is quite the opposite! In particular, we have tried to
demonstrate in the present paper that the symmetry-related methods are much
simpler and much more transparent than those based on investigating the
structure of the dynamical equations. This simplicity becomes especially
apparent when one deals with physical objects more complex than monoatomic
chains, such as molecules and crystals characterized by nontrivial symmetry
groups.

Note that comparing our results and those from~\cite{Shinohara}, one must
take into account that we use periodic boundary conditions, while Shinohara
discusses the nonlinear chains with fixed endpoints (the former boundary
conditions are more general, since the latter conditions can be considered
as their special case). We would also like to note that the stability
problem for the bushes of modes (invariant manifolds or ``modes subsets of
I type'') is not considered in~\cite{Shinohara}. On the other hand, this
problem seems to be important for treating the induction phenomenon
discussed in those papers.

\subsection{The structure of the present paper}

In Section~2, we consider a simple crystallographic method for obtaining
bushes of modes in the space of atomic displacements which differ from that
in our previous paper~\cite{FPU1} and in the paper~\cite{BRink} by Bob
Rink. A major advantage of this method is its remarkable clarity. Actually,
we obtain the invariant manifolds corresponding to the subgroups of the
dihedral group and then decompose them into the normal modes of the
monoatomic chain. Here we restrict ourselves to the discussion of one- and
two-dimensional bushes only (about other bushes see~\cite{FPU2a}).

The dynamical description of the bushes of normal modes is presented
briefly in Section~3.

The general approach to analyzing the parametric stability of the bushes of
normal modes is discussed in Section~4.

Sec.~5 is devoted to investigation of the stability of the one-dimensional
(1D) bushes in the FPU-$\alpha$ and FPU-$\beta$ chains. Only one 1D bush,
B$[\hat{a}^2,\hat{i}]$, can be obtained for the monoatomic chain with an
even number of particles\footnote{It was denoted by the symbol B$[2a]$ in
the paper~\cite{FPU1}.}, if the translational group $T$ is considered as
the appropriate parent group of this mechanical system. The stability of
the bush B$[\hat{a}^2,\hat{i}]$ was studied for the FPU-$\alpha$ and
FPU-$\beta$ chains in the papers~\cite{FPU1} and~\cite{PoggiRuffo},
respectively.\footnote{The stability of this bush, consisting of only one
so-called $\pi$-mode, was also discussed in~\cite{Page}.} However, we
obtain some additional one-dimensional bushes considering the dihedral
group $D$ as the parent symmetry group~\cite{FPU1,FPU2a}. Moreover, as was
already mentioned, there exist still more 1D bushes for the FPU-$\beta$
model because of the evenness of its potential~\cite{BRink}. Therefore, it
is interesting to compare the regions of stability of all these bushes in
both FPU-$\alpha$ and FPU-$\beta$ models. We perform such a comparison in
this section. An exceptional case is represented by the bush
B$[\hat{a}^4,\hat{a}\hat{i}]$ for the FPU-$\alpha$ chain for which the
threshold of stability loss is exactly equal to zero (this result can be
proved analytically).

We discuss here the stability of 1D bushes not only for the FPU chains with
the finite number of particles~($N$), but for the continuum limit
($N\rightarrow\infty$), as well.

In Section~6 we discuss the shape of the stability regions of the
two-dimensional (2D) bushes for the FPU-$\alpha$ and FPU-$\beta$ chains.
Indeed, thresholds of the loss of stability of the 1D bushes depend only on
their energy, while the similar thresholds for multi-dimensional bushes
depend on the complete set of the initial conditions for the dynamical
equations of the bush. We demonstrate, with the aid of the appropriate
plots, how nontrivial the stability regions of the 2D bushes for the FPU
chains can be. The dependence of these regions on the total number of
particles in the chains is partially discussed.

In Conclusion (Sec.~7), the obtained results are summarized and some new
directions for the bush theory are outlined.

\section{Simple crystallographic method for finding bushes of vibrational
modes in the monoatomic chains}

We consider the longitudinal vibrations of $N$-particle monoatomic chains
with periodic boundary conditions. Let the $N$-dimensional vector
\[\vec{X}(t)=\{x_1(t),x_2(t),\dotsc,x_N(t)\}\]
describe all the displacements, $x_i(t)$, of the individual particles
(atoms) from their equilibrium states at the instant $t$. In accordance
with the above-mentioned periodic boundary conditions, we demand for
arbitrary time $t$:
\begin{equation}
x_{N+1}(t)\equiv x_1(t).
\label{eq1}
\end{equation}

In the equilibrium state, a given chain is invariant under the action of
the operator $\hat{a}$ which shifts the chain by the lattice spacing $a$.
This operator generates the translational group
\begin{equation}
T=\{\hat{E},\hat{a},\hat{a}^2,\dotsc,\hat{a}^{N-1}\},\quad\hat{a}^N=\hat{E},
\label{eq2}
\end{equation}
where $\hat{E}$ is the identity element and $N$ is the order of the cyclic
group~$T$. The operator $\hat{a}$ induces the cyclic permutation of all
particles of the chain and, therefore, it acts on the ``configuration
vector'' $\vec{X}(t)$ as follows:
\[\hat{a}\vec{X}(t)\equiv
\hat{a}\{x_1(t),x_2(t),\dotsc,x_{N-1}(t),x_N(t)\}=
\{x_N(t),x_1(t),x_2(t),\dotsc,x_{N-1}(t)\}.\]

The full symmetry group of the monoatomic chain contains also the inversion
$\hat{i}$, with respect to the center of the chain, which acts on the
vector $\vec{X}(t)$ in the following manner:
\[\hat{i}\vec{X}(t)\equiv
\hat{i}\{x_1(t),x_2(t),\dotsc,x_{N-1}(t),x_N(t)\}=
\{-x_N(t),-x_{N-1}(t),\dotsc,-x_2(t),-x_1(t)\}.\]

The complete set of all products $\hat{a}^k\hat{i}$ of the pure
translations $\hat{a}^k$ ($k=1,2,\dotsc,N-1$) with the inversion $\hat{i}$
forms the so-called dihedral group $D$ which can be written as a direct sum
of two cosets $T$ and $T\cdot\hat{i}$:
\begin{equation}
D=T\oplus T\cdot\hat{i}.
\label{eq500}
\end{equation}
The dihedral group is non-Abelian group induced by two generators
($\hat{a}$ and $\hat{i}$) with the following generating relations
\begin{equation}
\hat{a}^N=\hat{E},\quad\hat{i}^2=\hat{E},\quad\hat{i}\hat{a}=\hat{a}^{-1}\hat{i}.
\label{eq170a}
\end{equation}

Let us consider the geometrical interpretation of the symmetry elements of
the dihedral group $D$. For simplification, we will use the reflection
$\hat{\sigma}$ in the mirror plane orthogonal to the chain instead of the
inversion $\hat{i}$. Indeed, the actions of $\hat{\sigma}$ and $\hat{i}$ on
the vectors of the one-dimensional space are equivalent to each other
($\hat{\sigma}x=\hat{i}x=-x$). Then we can use the well-known theorem of
crystallography: ``Any mirror plane and the translation $\vec{A}$
orthogonal to it generate a new mirror plane, parallel to the former plane
and displaced away from it by $\vec{A}/2$''.

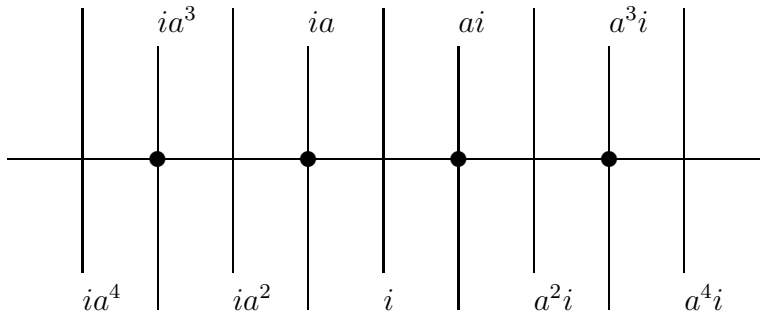
\begin{figure}
\centering
\setlength{\unitlength}{1mm}
\begin{picture}(100,40)(0,0)
\put(0,20){\line(1,0){100}}
\put(10,5){\line(0,1){35}}
\put(10,0){$ia^4$}
\put(20,0){\line(0,1){35}}
\put(20,20){\circle*{2}}
\put(20,37){$ia^3$}
\put(30,5){\line(0,1){35}}
\put(30,0){$ia^2$}
\put(40,0){\line(0,1){35}}
\put(40,20){\circle*{2}}
\put(40,37){$ia$}
\put(50,5){\line(0,1){35}}
\put(50,0){$i$}
\put(60,0){\line(0,1){35}}
\put(60,20){\circle*{2}}
\put(60,37){$ai$}
\put(70,5){\line(0,1){35}}
\put(70,0){$a^2i$}
\put(80,0){\line(0,1){35}}
\put(80,20){\circle*{2}}
\put(80,37){$a^3i$}
\put(90,5){\line(0,1){35}}
\put(90,0){$a^4i$}
\end{picture}
\caption{\label{fig1} Symmetry elements of the dihedral group $D$ near the
center of the monoatomic chain.}
\end{figure}

Let the number of particles in the chain ($N$) be an \textit{even} integer.
Then it is easy to check that the products
$\hat{a}^k\hat{\sigma}\equiv\hat{a}^k\hat{i}$ ($k=\pm 1,\pm 2,\pm
3,\dotsc$) generate the family of parallel planes depicted by vertical
lines\footnote{For simplicity, the hats above the operators are dropped in
all our pictures.} in Fig.~\ref{fig1}. For odd $k$, these planes pass
\textit{through} the atoms, while for the even $k$ they pass
\textit{between} atoms, exactly at the middle of the neighboring particles.
Note that, for odd $N$, everything will be vice versa: for even $k$, the
planes pass through the atoms and for the add $k$ -- between them. In
Fig.~\ref{fig1}, we show a part of the family of the above mentioned planes
near the center of the chain. It can be verified that the symmetry elements
$\hat{a}^k\hat{\sigma}\equiv\hat{a}^k\hat{i}$ ($k=1,2,3,\dotsc$) are
located to the right of the center of the chain, while the elements
$\hat{\sigma}\hat{a}^k\equiv\hat{i}\hat{a}^k$ are located to the left from
it\footnote{Note that
$\hat{a}^k\hat{i}=\hat{i}\hat{a}^{-k}=\hat{i}\hat{a}^{N-k}$.}.

Let us now consider the subgroups $G_j$ of the dihedral group $D$, keeping
in mind that every subgroup $G_j\subset D$ induces the certain bush of
modes.

Each group $G_j$ contains its own translational subgroup $T_j\subset T$,
where $T$ is the above discussed full translational group~(\ref{eq2}). If
$N$ is divisible by 4 (for example, we consider below the case $N=12$)
there exists the subgroup $T_4=[\hat{a}^4]$ of the group $T=[\hat{a}]$.
Note that in square brackets we write down only generators of the
considered group, while the complete set of group elements is written in
curly brackets (see, for example, Eqs.~(\ref{eq2})).

If a vibrational state of the chain possesses the symmetry group
$T_4=[\hat{a}^4]\equiv\{\hat{E},\hat{a}^4,\hat{a}^8,\dotsc,\hat{a}^{N-4}\}$,
displacements of the atoms, which are apart by the distance $4a$ from each
other in the equilibrium state, turn out to be equal. Therefore, for the
case $N=12$, we can write the following atomic displacement pattern:
\begin{equation}
\vec{X}(t)=\{x_1(t),x_2(t),x_3(t),x_4(t)~|~x_1(t),x_2(t),x_3(t),x_4(t)~|~x_1(t),%
x_2(t),x_3(t),x_4(t)\},
\label{eq11}
\end{equation}
where $x_i(t)$ ($i=1,2,3,4$) are arbitrary functions of time. The
pattern~(\ref{eq11}) determines the bush corresponding to the symmetry
group $T_4=[\hat{a}^4]$. Thus, the complete set of the atomic displacements
can be divided into $N/4$ (in our case, $N/4=3$) identical subsets, and
each of them determines the so-called ``Extended Primitive Cell''
(EPC)\footnote{This term is used in the theory of phase transitions in
crystals.}. For the bush~(\ref{eq11}), the EPC contains four atoms, and the
vibrational state of the whole chain is described by three such EPC. In
other words, the EPC for the vibrational state with the symmetry group
$T_4=[\hat{a}^4]$ (its size is equal to $4a$) is four times larger than the
primitive cell ($a$) of the chain in the equilibrium state.

It is essential that some symmetry elements of the dihedral group $D$
disappear as a result of the symmetry reduction
$D=[\hat{a},\hat{i}]\rightarrow T_4=[\hat{a}^4]$. In our case, there are no
nontrivial symmetry elements of the group $D$ inside the EPC and, as a
consequence, there are no restrictions on the displacements belonging to
one and the same extended cell (this fact is obvious from
Eq.~(\ref{eq11})).

There are four other subgroups of the dihedral group $D$ to which the same
translational subgroup $T_4=[\hat{a}^4]$ correspond:
\begin{equation}
[\hat{a}^4,\hat{i}],\quad [\hat{a}^4,\hat{a}\hat{i}],\quad
[\hat{a}^4,\hat{a}^2\hat{i}],\quad [\hat{a}^4,\hat{a}^3\hat{i}].
\label{eq50}
\end{equation}
These subgroups possess \textit{two generators}, namely, the common
translational generator $\hat{a}^4$ and different inversion elements
$\hat{a}^k\hat{i}$ ($k=0,1,2,3$). These inversion elements differ from each
other by the \textit{position} of the center of inversion (see
Fig.~\ref{fig1} where they are depicted by the vertical lines representing
the mirror planes).

Note that subgroups $[\hat{a}^4,\hat{a}^k\hat{i}]$ with $k>3$ are
equivalent to those from the list~(\ref{eq50}), because the second
generator $\hat{a}^k\hat{i}$ can be multiplied by $\hat{a}^{-4}$,
representing the inverse element with respect to the first generator
$\hat{a}^4$. Thus, there exist only five subgroups of the dihedral group
(with $N\mod 4=0$) constructed on the basis of the translational group
$T_4=[\hat{a}^4]$, namely, this group and the four groups from the
list~(\ref{eq50}).

Now, let us consider the bushes corresponding to the
subgroups~(\ref{eq50}).

1. The subgroup $[\hat{a}^4,\hat{i}]$ consists of the following six
elements:
\begin{equation}
\hat{E},\hat{a}^4,\hat{i},\hat{a}^4\hat{i},
\hat{a}^8\hat{i}\equiv\hat{i}\hat{a}^4.
\label{eq51a}
\end{equation}

\begin{figure}
\centering
\setlength{\unitlength}{1mm}
\begin{picture}(100,40)(0,0)
\put(0,20){\line(1,0){100}}
\put(10,5){\line(0,1){35}}
\put(10,0){$ia^4$}
\put(20,20){\circle*{2}}
\put(20,22){$x_1$}
\put(40,20){\circle*{2}}
\put(40,22){$x_2$}
\put(50,5){\line(0,1){30}}
\put(50,0){$i$}
\put(60,20){\circle*{2}}
\put(60,22){$x_3$}
\put(80,20){\circle*{2}}
\put(80,22){$x_4$}
\put(90,5){\line(0,1){35}}
\put(90,0){$a^4i$}
\put(40,37){\vector(-1,0){30}}
\put(60,37){\vector(1,0){30}}
\put(43,36){EPC ($4a$)}
\linethickness{0.5mm}
%\put(20,20){\vector(1,0){10}}
\put(20,20){\rvec{10}}
%\put(40,20){\vector(1,0){7}}
\put(40,20){\rvec{7}}
%\put(60,20){\vector(-1,0){7}}
\put(60,20){\lvec{7}}
%\put(80,20){\vector(-1,0){10}}
\put(80,20){\lvec{10}}
\end{picture}
\caption{\label{fig2} Vibrational bush corresponding to the subgroup
$G=[\hat{a}^4,\hat{i}]$.}
\end{figure}
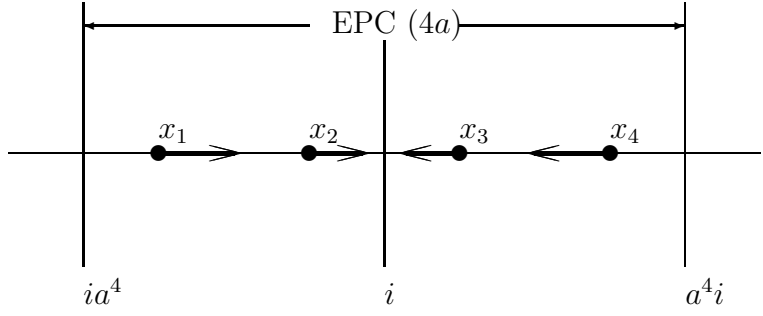

The diagram of this subgroup, similar to that of the full dihedral group
$D$ in Fig.~\ref{fig1}, is depicted in Fig.~\ref{fig2}.

Two symmetry elements ($\hat{i}\hat{a}^4$ and $\hat{a}^4\hat{i}$), situated
at the boundaries of the EPC, do not produce any additional restrictions on
the atomic displacements \textit{inside} this EPC. Indeed, they connect
with each other the atomic displacements in the adjacent extended primitive
cells, while the equivalence of the displacements of the atoms with the
same positions in the adjacent EPC is guaranteed by the translation
$\hat{a}^4$ which is the first generator of the considered subgroup
$[\hat{a}^4,\hat{i}]$.

On the other hand, the inversion $\hat{i}$ at the center of the EPC
(Fig.~\ref{fig2}), surviving in the subgroup $[\hat{a}^4,\hat{i}]$, demands
the displacements of the atoms, symmetrically situated with respect to this
center, to be equal in value, but opposite in sign:
\begin{equation}
x_1(t)=-x_4(t),\quad x_2(t)=-x_3(t).
\label{eq51}
\end{equation}
These relations between the atomic displacements are shown in
Fig.~\ref{fig2} by the appropriate arrows. As a consequence, the atomic
displacements, corresponding to the bush B$[\hat{a}^4,\hat{i}]$ with the
symmetry group $[\hat{a}^4,\hat{i}]$, can be written (for $N=12$) in the
following form:
\begin{equation}
\vec{X}(t)=\{x_1(t),x_2(t),-x_2(t),-x_1(t)~|~x_1(t),x_2(t),-x_2(t),
-x_1(t)~|~x_1(t),x_2(t),-x_2(t),-x_1(t)\}
\label{eq149}
\end{equation}

Since Eqs.~(\ref{eq51}) hold for an arbitrary moment $t$, we will drop the
argument $t$ and write down the \textit{vibrational bush in the $X$-space}
(configuration space) as follows:
\begin{equation}
\B[\hat{a}^4,\hat{i}]=|x_1,x_2,-x_2,-x_1|.
\label{eq52}
\end{equation}
Thus, we point out the atomic displacements in only one EPC.

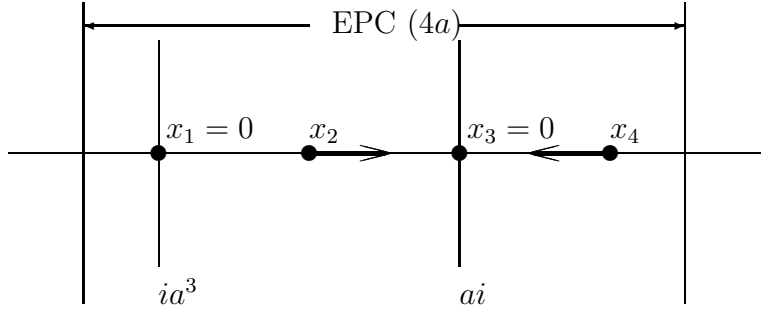
\begin{figure}
\centering
\setlength{\unitlength}{1mm}
\begin{picture}(100,40)(0,0)
\put(0,20){\line(1,0){100}}
\put(10,0){\line(0,1){40}}
\put(20,5){\line(0,1){30}}
\put(20,0){$ia^3$}
\put(20,20){\circle*{2}}
\put(21,22){$x_1=0$}
\put(40,20){\circle*{2}}
\put(40,22){$x_2$}
\put(60,5){\line(0,1){30}}
\put(60,0){$ai$}
\put(60,20){\circle*{2}}
\put(61,22){$x_3=0$}
\put(80,20){\circle*{2}}
\put(80,22){$x_4$}
\put(90,0){\line(0,1){40}}
\put(40,37){\vector(-1,0){30}}
\put(60,37){\vector(1,0){30}}
\put(43,36){EPC ($4a$)}
\linethickness{0.5mm}
%\put(40,20){\vector(1,0){10}}
\put(40,20){\rvec{10}}
%\put(80,20){\vector(-1,0){10}}
\put(80,20){\lvec{10}}
\end{picture}
\caption{\label{fig3} Vibrational bush corresponding to the subgroup
$G=[\hat{a}^4,\hat{a}\hat{i}]$.}
\end{figure}

2. For the subgroup
\begin{equation}
[\hat{a}^4,\hat{a}\hat{i}]\equiv\{\hat{E},\hat{a}^4,\hat{a}^8,\hat{a}\hat{i},
\hat{a}^5\hat{i}\equiv\hat{i}\hat{a}^7,\hat{a}^9\hat{i}\equiv\hat{i}\hat{a}^3\},
\label{eq53}
\end{equation}
we obtain the diagram depicted in Fig.~\ref{fig3}. In this diagram, one can
see two symmetry elements, $\hat{i}\hat{a}^3$ and $\hat{a}\hat{i}$, which
pass, respectively, through the first and third atoms of the represented
EPC. Therefore, the displacements of these atoms must satisfy the relations
$x_1(t)=-x_1(t)$, $x_3(t)=-x_3(t)$. In turn, this means that $x_1(t)\equiv
0$, $x_3(t)\equiv 0$, i.e.\ the atoms located on the inversion elements
(which are equivalent, in the one-dimensional space, to the mirror planes!)
must be \textit{immovable}.

On the other hand, the symmetry element $\hat{a}\hat{i}$ is situated in the
middle between the second and fourth atoms of the EPC and, therefore, their
possible displacements must satisfy the equation $x_2(t)=-x_4(t)$. Thus, we
have the following displacement pattern for the bush
B$[\hat{a}^4,\hat{a}\hat{i}]$
\begin{equation}
\vec{X}(t)=\{0,x(t),0,-x(t)~|~0,x(t),0,-x(t)~|~0,x(t),0,-x(t)\}
\label{eq150a}
\end{equation}
with \textit{only one} independent dynamical variable $x(t)$. We can write
this \textit{one-dimensional} vibrational bush in the abbreviated form:
\[\B[\hat{a}^4,\hat{a}\hat{i}]=|0,x,0,-x|.\]

3. The diagram of the symmetry elements of the subgroup
\[[\hat{a}^4,\hat{a}^2\hat{i}]\equiv\{\hat{E},\hat{a}^4,\hat{a}^8,
\hat{a}^2\hat{i},\hat{a}^6\hat{i}\equiv\hat{i}\hat{a}^6,
\hat{a}^{10}\hat{i}\equiv\hat{i}\hat{a}^2\}\]
is depicted in Fig.~\ref{fig4}.
\begin{figure}
\centering
\setlength{\unitlength}{1mm}
\begin{picture}(100,40)(0,0)
\put(0,20){\line(1,0){100}}
\put(10,0){\line(0,1){40}}
\put(20,20){\circle*{2}}
\put(20,22){$x_1$}
\put(30,5){\line(0,1){30}}
\put(30,0){$ia^2$}
\put(40,20){\circle*{2}}
\put(40,22){$x_2$}
\put(60,20){\circle*{2}}
\put(60,22){$x_3$}
\put(70,5){\line(0,1){30}}
\put(70,0){$a^2i$}
\put(80,20){\circle*{2}}
\put(80,22){$x_4$}
\put(90,0){\line(0,1){40}}
\put(40,37){\vector(-1,0){30}}
\put(60,37){\vector(1,0){30}}
\put(43,36){EPC ($4a$)}
\linethickness{0.5mm}
%\put(20,20){\vector(1,0){7}}
\put(20,20){\rvec{7}}
%\put(40,20){\vector(-1,0){7}}
\put(40,20){\lvec{7}}
%\put(60,20){\vector(1,0){9}}
\put(60,20){\rvec{9}}
%\put(80,20){\vector(-1,0){9}}
\put(80,20){\lvec{9}}
\end{picture}
\caption{\label{fig4} Vibrational bush corresponding to the subgroup
$G=[\hat{a}^4,\hat{a}^2\hat{i}]$.}
\end{figure}
There are two elements, $\hat{i}\hat{a}^2$ and $\hat{a}^2\hat{i}$ surviving
in the transition from the full dihedral group $D=[\hat{a},\hat{i}]$ to the
considered subgroup $[\hat{a}^4,\hat{a}^2\hat{i}]$. These elements are
located between two first and two last atoms of the depicted EPC.
Therefore, $x_1(t)=-x_2(t)$, $x_3(t)=-x_4(t)$ and the corresponding
displacement pattern is
\begin{equation}
\vec{X}(t)=\{x_1(t),-x_1(t),x_2(t),-x_2(t)~|~x_1(t),-x_1(t),x_2(t),
-x_2(t)~|~x_1(t),-x_1(t),x_2(t),-x_2(t)\}.
\label{eq57}
\end{equation}

There are two degrees of freedom in~(\ref{eq57}), $x_1(t)$ and $x_2(t)$.
Thus, the \textit{two-dimensional} bush with the symmetry group
$[\hat{a}^4,\hat{a}^2\hat{i}]$ can be written as
\[\B[\hat{a}^4,\hat{a}^2\hat{i}]=|x_1,-x_1,x_2,-x_2|.\]

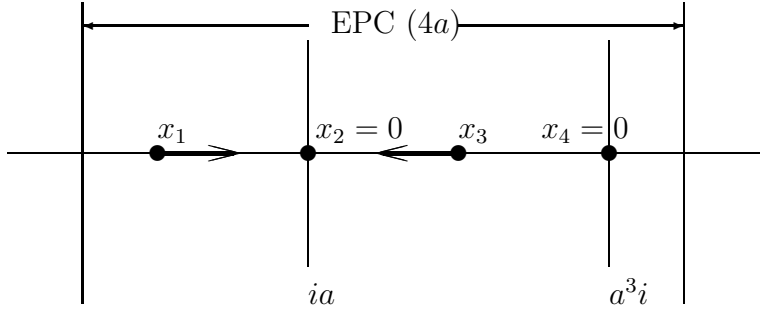
\begin{figure}
\centering
\setlength{\unitlength}{1mm}
\begin{picture}(100,40)(0,0)
\put(0,20){\line(1,0){100}}
\put(10,0){\line(0,1){40}}
\put(20,20){\circle*{2}}
\put(20,22){$x_1$}
\put(40,5){\line(0,1){30}}
\put(40,0){$ia$}
\put(40,20){\circle*{2}}
\put(41,22){$x_2=0$}
\put(60,20){\circle*{2}}
\put(60,22){$x_3$}
\put(80,5){\line(0,1){30}}
\put(80,0){$a^3i$}
\put(80,20){\circle*{2}}
\put(71,22){$x_4=0$}
\put(90,0){\line(0,1){40}}
\put(40,37){\vector(-1,0){30}}
\put(60,37){\vector(1,0){30}}
\put(43,36){EPC ($4a$)}
\linethickness{0.5mm}
%\put(20,20){\vector(1,0){10}}
\put(20,20){\rvec{10}}
%\put(60,20){\vector(-1,0){10}}
\put(60,20){\lvec{10}}
\end{picture}
\caption{\label{fig5} Vibrational bush corresponding to the subgroup
$G=[\hat{a}^4,\hat{a}^3\hat{i}]$.}
\end{figure}

4. The diagram of the symmetry elements for the last subgroup of the
list~(\ref{eq50})
\[[\hat{a}^4,\hat{a}^3\hat{i}]\equiv\{\hat{E},\hat{a}^4,\hat{a}^8,
\hat{a}^3\hat{i},\hat{a}^7\hat{i}\equiv\hat{i}\hat{a}^5,
\hat{a}^{11}\hat{i}\equiv\hat{i}\hat{a}\}\]
is depicted in Fig.~\ref{fig5}.
We see from this diagram that $x_2(t)\equiv 0$, $x_4(t)\equiv 0$,
$x_1(t)=-x_3(t)$, and, therefore,
\begin{equation}
\vec{X}(t)=\{x(t),0,-x(t),0~|~x(t),0,-x(t),0~|~x(t),0,-x(t),0\}.
\label{eq58}
\end{equation}
The corresponding \textit{one-dimensional} bush can be written as follows
\[\B[\hat{a}^4,\hat{a}^2\hat{i}]=|x,0,-x,0|.\]

We have just considered all the subgroups of the dihedral group $D$, which
contain the common translational group $T_4=[\hat{a}^4]$, and the
corresponding vibrational bushes. In the same way, we can find the
subgroups of the group $D$ corresponding to the arbitrary translational
group $T_m=[\hat{a}^m]$\footnote{Such subgroup exists in the case $N\mod
m=0.$} and then obtain the bushes induced by these subgroups~\cite{FPU2a}.
We will not discuss this method any further in the present paper, because
all the bushes of modes (symmetric invariant manifolds) for the dihedral
group were recently obtained by Bob Rink in~\cite{BRink}. Nevertheless, let
us note that the above described crystallographic method for finding bushes
of vibrational modes seems to be more straightforward than that used
in~\cite{BRink}. Actually, the former method represents a variant of our
general method of the splitting of the space group orbits
(SO)~\cite{Comput,Acta}, adapted for the sufficiently simple case of the
dihedral group $D$. The SO method was developed for finding the so-called
complete condensate of primary and secondary order parameters in the
framework of the theory of phase transitions in crystals. It was applied to
such nontrivial space groups as $O_h^1$~\cite{Acta},
$O_h^7$~\cite{Crystal}, etc.

The notion of the complete condensate of order parameters is equivalent to
the notion of the bush of modes in the \textit{geometrical} (only
geometrical!) sense. Three different methods for finding the condensate of
order parameters were considered in~\cite{PhysStatSol,Comput,Acta}: the
``direct method'', which is similar to that by Bob Rink~\cite{BRink}, the
above mentioned SO method and the method based on the concept of the
irreducible representations of the appropriate symmetry groups. The last
method was used, in particular, for treating bushes of vibrational modes in
nonlinear chains in the paper~\cite{FPU1}.

Now let us consider the notion of the bush equivalence. There are two pairs
of equivalent bushes of modes induced by the subgroups from the
list~(\ref{eq50}). Indeed, shifting the displacement pattern~(\ref{eq150a})
of the bush B$[\hat{a}^4,\hat{a}\hat{i}]$ by the lattice spacing $a$, we
obtain the displacement pattern~(\ref{eq58}) of the bush
B$[\hat{a}^4,\hat{a}^3\hat{i}]$\footnote{We can imagine that our chain
represents a ring, because of the periodic boundary conditions.}. The
pattern~(\ref{eq149}) of the bush B$[\hat{a}^4,\hat{i}]$ can be transformed
into the pattern~(\ref{eq57}) of the bush B$[\hat{a}^4,\hat{a}^2\hat{i}]$
by the same shifting.

The importance of the notion of bush equivalence is brought about by the
fact that equivalent bushes represent \textit{identical dynamical systems}:
they are described by equivalent equations of motions (see below). It can
be proved, in the general case, that the \textit{conjugate subgroups} of
the parent symmetry group induce equivalent bushes\footnote{Sometimes, we
call equivalent bushes by the term ``dynamical domains''. The term
``domain'' is borrowed from the theory of phase transition in crystals.}.
Remember that two subgroups $G_1$ and $G_2$ of the same group $G$
($G_1\subset G$, $G_2\subset G$) are called conjugate to each other
($G_1\sim G_2$), if there exists at least one element $g_0$ of the group
$G$ which transfers $G_1$ into $G_2$ by the transformation
\begin{equation}
G_2=g_0^{-1}G_1g_0\quad(g_0\in G).
\label{eq60}
\end{equation}
It can be easy checked the following conjugations:
\begin{equation}
\left[\hat{a}^4,\hat{i}\right]\sim\left[\hat{a}^4,\hat{a}^2\hat{i}\right],\quad
\left[\hat{a}^4,\hat{a}\hat{i}\right]\sim\left[\hat{a}^4,\hat{a}^3\hat{i}\right].
\label{eq71a}
\end{equation}
For example, using Eqs.~(\ref{eq170a}), we obtain the relation
\[[\hat{a}^4,\hat{a}\hat{i}]=\hat{a}^{-1}\cdot[\hat{a}^4,\hat{a}^3\hat{i}]\cdot a,\]
which proves the equivalence of the subgroups $[\hat{a}^4,\hat{a}\hat{i}]$
and $[\hat{a}^4,\hat{a}^3\hat{i}]$, if we take into account the
definition~(\ref{eq60}) with $g_0=\hat{a}$. Thus, the both above considered
one-dimensional bushes B$[\hat{a}^4,\hat{a}\hat{i}]$ and
B$[\hat{a}^4,\hat{a}^3\hat{i}]$ turn out to be equivalent to each other, as
well as the both two-dimensional bushes, B$[\hat{a}^4,\hat{i}]$ and
B$[\hat{a}^4,\hat{a}^2\hat{i}]$. Moreover, it can be proved that for any
even $m$ and $k=0,1,2,\dotsc,m-1$, all bushes of the form
B$[\hat{a}^m,\hat{a}^k\hat{i}]$ with $k$ of the \textit{same evenness} are
equivalent. For odd $m$, all the bushes of this form are equivalent.

Let us summarize the above said about the construction of the vibrational
bushes for nonlinear monoatomic chains.

-- Each subgroup of the dihedral group singles out a certain bush of
vibrational modes.

-- Every subgroup is determined by one or two generators and contains the
translational subgroup $T_k=[\hat{a}^k]$ (it is trivial, $T_0$, for the
subgroups $[\hat{a}^m\hat{i}]$,  $m=0,1,2,\dotsc,N-1$).

-- The translational subgroup $T_k=[\hat{a}^k]$ determines the extended
primitive cell (EPC) for the vibrational state of the chain.

-- The symmetry elements $\hat{a}^m\hat{i}$, belonging to the second coset
of the dihedral group (see Eq.~(\ref{eq500})) and surviving at the symmetry
lowering $D\equiv[\hat{a},\hat{i}]\rightarrow[\hat{a}^k,\hat{a}^m\hat{i}]$,
produce certain restrictions on the displacements inside the EPC.

-- The conjugate subgroups induce the equivalent bushes of modes. The
number of the equivalent bushes of a given type is equal to the
\textit{index}\footnote{The index of the subgroup $G$ of the group $G_0$ is
equal to the ratio $\|G_0\|/\|G\|$, where $\|G_0\|$ and $\|G\|$ are the
orders of these two groups, $G_0$ and $G$.} of the corresponding subgroup
in the dihedral group.

In addition, note that there exist subgroups to which no vibrational bushes
correspond. In our case, the subgroup $[\hat{a}^2,\hat{a}\hat{i}]$ does not
induce the vibrational bush, because the EPC, corresponding to this
subgroup, contains only two atoms through which the inversion elements,
$\hat{i}\hat{a}$ and $\hat{a}\hat{i}$, pass and, therefore, these atoms
cannot move.

The bushes obtained with the aid of the described method represent
invariant manifolds determined in the \textit{configuration} space. Let us
now consider the bushes in the \textit{modal} space.

We consider the normal modes for the monoatomic chain in the form used
in~\cite{PoggiRuffo}:
\begin{equation}
\vec{\psi}_k=\left\{\frac{1}{\sqrt{N}}\left.\left[\sin\left(\frac{2\pi
k}{N}n\right) +\cos\left(\frac{2\pi
k}{N}n\right)\right]\right|n=1,2,\dotsc,N\right\}.
\label{eq70}
\end{equation}
Here the subscript $k$ refers to the mode, while the subscript $n$ refers
to the atom. The modes $\vec{\psi}_k$ ($k=0,1,2,\dotsc,N-1$) form an
orthonormal basis in the modal space and we can decompose the set of the
atomic displacements $\vec{X}(t)$, corresponding to a given bush, in this
basis
\begin{equation}
\vec{X}(t)=\sum_{k=0}^{N-1}\nu_k(t)\vec{\psi}_k
\label{eq71}
\end{equation}
For example, we obtain the following decompositions for the bushes
B$[\hat{a}^4,\hat{i}]$ and B$[\hat{a}^4,\hat{a}^2\hat{i}]$:
\begin{eqnarray}
\vec{X}(\B[\hat{a}^4,\hat{i}])&\equiv&\{x_1(t),x_2(t),-x_1(t),-x_2(t)|%
x_1(t),x_2(t),-x_1(t),-x_2(t)|x_1(t),x_2(t),-x_1(t),-x_2(t)\}\nonumber\\
&=&\mu(t)\vec{\psi}_{N/2}+\nu(t)\vec{\psi}_{3N/4},
\label{eq72}\\
\vec{X}(\B[\hat{a}^4,\hat{a}^2\hat{i}])&=&\tilde{\mu}(t)\vec{\psi}_{N/2}+\tilde{\nu}(t)\vec{\psi}_{N/4}.
\label{eq73}
\end{eqnarray}
Only vectors $\vec{\psi}_{N/2}$, $\vec{\psi}_{N/4}$ and $\vec{\psi}_{3N/4}$
from the complete basis~(\ref{eq70}) contribute to these two-dimensional
bushes:
\begin{eqnarray}
\vec{\psi}_{N/2}&=&\frac{1}{\sqrt{N}}(-1,1,-1,1,-1,1,-1,1,-1,1,-1,1),
\label{eq74}\\
\vec{\psi}_{N/4}&=&\frac{1}{\sqrt{N}}(1,-1,-1,1,1,-1,-1,1,1,-1,-1,1),
\label{eq74a}\\
\vec{\psi}_{3N/4}&=&\frac{1}{\sqrt{N}}(-1,-1,1,1,-1,-1,1,1,-1,-1,1,1).
\label{eq74b}
\end{eqnarray}
Remember that these bushes are equivalent to each other, i.e.\ they
represent the so-called ``dynamical domains''. For the bush
B$[\hat{a}^4,\hat{i}]$ we can find the following relations between the old
dynamical variables $x_1(t)$, $x_2(t)$ (relating to the configuration
space) and the new dynamical variables $\mu(t)$, $\nu(t)$ (relating to the
modal space):
\begin{equation}
\begin{array}{l}
\mu(t)=-\frac{\sqrt{N}}{2}[x_1(t)-x_2(t)],\\
\nu(t)=-\frac{\sqrt{N}}{2}[x_1(t)+x_2(t)].
\end{array}
\label{eq75}
\end{equation}
Thus, each of the above bushes consists of two modes. One of these modes is
the \textit{root} mode ($\vec{\psi}_{3N/4}$ for the bush
B$[\hat{a}^4,\hat{i}]$ and $\vec{\psi}_{N/4}$ for the bush
B$[\hat{a}^4,\hat{a}^2\hat{i}]$), while the other mode $\vec{\psi}_{N/2}$
is the secondary mode. Indeed, we have already discussed (see Introduction)
that the symmetry of the secondary modes must be higher or equal to the
symmetry of the root mode. In our case, as it can be seen from
Eqs.~(\ref{eq74a}), the translational symmetry of the mode
$\vec{\psi}_{N/4}$ is $\hat{a}^4$ (acting by this element on~(\ref{eq74a})
we obtain the same displacement pattern), while the translational symmetry
of the mode $\vec{\psi}_{N/2}$ is $\hat{a}^2$, which is two times higher
than that of $\vec{\psi}_{N/4}$
(see~(\ref{eq74})-(\ref{eq74b}))\footnote{Note that the full symmetry of
the modes $\vec{\psi}_{N/4}$ and $\vec{\psi}_{N/2}$ are
$[\hat{a}^4,\hat{a}^2\hat{i}]$ and $[\hat{a}^2,\hat{i}]$, respectively.}.

One-dimensional (1D) and two-dimensional (2D) vibrational bushes for the
FPU chains in configuration and modal spaces can be seen in
Tables~\ref{table1},\ref{table2}. Note that in these tables, as well as in
Tables~\ref{table25},\ref{table3},\ref{table4} we list one-dimensional
bushes completely, but two-dimensional bushes partially: only those 2D
bushes are given whose stability properties are studied in the present
paper (see Sec.~6). The complete list of 2D bushes in the FPU chains can be
found in Appendix.

\begin{table}
\centering
\caption{\label{table1} Representation of 1D and 2D bushes of vibrational
modes for the FPU chains in the configuration space (classification by the
dihedral group~$D$)}
\begin{tabular}{|l|c|c|}
\hline
Bush&Displacement pattern&Dimension\\
\hline
B$[a^2,i]$&$|x,-x|$&1\\
\hline
B$[a^3]$&$|x_1,x_2,x_3|^*$&2\\
\hline
B$[a^3,i]$&$|x,0,-x|$&1\\
B$[a^3,ai]$&$|0,x,-x|$&1\\
B$[a^3,a^2i]$&$|x,-x,0|$&1\\
\hline
B$[a^4,i]$&$|x_1,x_2,-x_2,-x_1|$&2\\
B$[a^4,a^2i]$&$|x_1,-x_1,x_2,-x_2|$&2\\
\hline
B$[a^4,ai]$&$|0,x,0,-x|$&1\\
B$[a^4,a^3i]$&$|x,0,-x,0|$&1\\
\hline
B$[a^6,ai]$&$|0,x_1,x_2,0,-x_2,-x_1|$&2\\
B$[a^6,a^3i]$&$|x_1,0,-x_1,x_2,0,-x_2|$&2\\
B$[a^6,a^5i]$&$|x_1,x_2,0,-x_2,-x_1,0|$&2\\
\hline
\end{tabular}

\raggedright
\small{$^*$ Here $x_3$ is not an independent variable: its value must be
determined by the condition of \textit{immobility} of the mass center of
the chain, $x_1+x_2+x_3=0$.}
\end{table}

\begin{table}
\centering
\caption{\label{table2} Representation of 1D and 2D bushes of vibrational
modes for the FPU chains in the modal space (classification by the dihedral
group~$D$). Here
$\varepsilon_1=\sin\left(\frac{\pi}{12}\right)=\frac{\sqrt{2}(\sqrt{3}-1)}{4}$,
$\varepsilon_2=\cos\left(\frac{\pi}{12}\right)=\frac{\sqrt{2}(\sqrt{3}+1)}{4}$,
$\varepsilon_3=\frac{\sqrt{2}}{2}$}
\begin{tabular}{|l|c|}
\hline
Bush&Representation in modal space\\
\hline
B$[a^2,i]$&$\nu\vec{\psi}_{N/2}$\\
\hline
B$[a^3,i]$&$\nu(\varepsilon_1\vec{\psi}_{N/3}+\varepsilon_2\vec{\psi}_{2N/3})$\\
B$[a^3,ai]$&$\nu(\varepsilon_2\vec{\psi}_{N/3}+\varepsilon_1\vec{\psi}_{2N/3})$\\
B$[a^3,a^2i]$&$\nu\varepsilon_3(\vec{\psi}_{N/3}-\vec{\psi}_{2N/3})$\\
\hline
B$[a^4,i]$&$\nu_1\vec{\psi}_{3N/4}-\nu_2\vec{\psi}_{N/2}$\\
B$[a^4,a^2i]$&$\nu_1\vec{\psi}_{N/4}+\nu_2\vec{\psi}_{N/2}$\\
\hline
B$[a^4,ai]$&$\nu\varepsilon_3(\vec{\psi}_{N/4}+\vec{\psi}_{3N/4})$\\
B$[a^4,a^3i]$&$\nu\varepsilon_3(\vec{\psi}_{N/4}-\vec{\psi}_{3N/4})$\\
\hline
B$[a^6,ai]$&$\nu_1(\varepsilon_1\vec{\psi}_{N/6}+\varepsilon_2\vec{\psi}_{5N/6})+\nu_2(-\varepsilon_2\vec{\psi}_{N/3}-\varepsilon_1\vec{\psi}_{2N/3})$\\
B$[a^6,a^3i]$&$\nu_1(\varepsilon_2\vec{\psi}_{N/6}+\varepsilon_1\vec{\psi}_{5N/6})+\nu_2(\varepsilon_1\vec{\psi}_{N/3}+\varepsilon_2\vec{\psi}_{2N/3})$\\
B$[a^6,a^5i]$&$\nu_1\varepsilon_3(\vec{\psi}_{N/6}-\vec{\psi}_{5N/6})+\nu_2\varepsilon_3(\vec{\psi}_{N/3}-\vec{\psi}_{2N/3})$\\
\hline
\end{tabular}
\end{table}

\begin{table}
\centering
\caption{\label{table25} Additional 1D and 2D bushes, for the FPU-$\beta$
chain (these bushes were obtained in~\cite{BRink})}
\begin{tabular}{|l|c|c|c|}
\hline
~~Bush&Dim.&Displacement pattern&Representation in the modal space\\
\hline
\begin{tabular}{l}B$[a^3,iu]$\\B$[a^3,aiu]$\\B$[a^3,a^2iu]$\end{tabular} &
1 &
\begin{tabular}{l}$|x,-2x,x|$\\$|-2x,x,x|$\\$|x,x,-2x|$\end{tabular} &
\begin{tabular}{l}$\nu(\varepsilon_2\vec{\psi}_{N/3}-\varepsilon_1\vec{\psi}_{2N/3})$\\
$\nu(\varepsilon_1\vec{\psi}_{N/3}-\varepsilon_2\vec{\psi}_{2N/3})$\\
$\nu\varepsilon_3(\vec{\psi}_{N/3}+\vec{\psi}_{2N/3})$\end{tabular}\\
\hline
\begin{tabular}{l}B$[a^4,iu]$\\B$[a^4,a^2iu]$\end{tabular} &
1 &
\begin{tabular}{l}$|x,-x,-x,x|$\\$|x,x,-x,-x|$\end{tabular} &
\begin{tabular}{l}$\nu\vec{\psi}_{N/4}$\\$\nu\vec{\psi}_{3N/4}$\end{tabular}\\
\hline
\begin{tabular}{l}B$[a^6,ai,a^3u]$\\B$[a^6,a^3i,a^3u]$\\B$[a^6,a^5i,a^3u]$\end{tabular} &
1 &
\begin{tabular}{l}$|0,x,x,0,-x,-x|$\\$|x,0,-x,-x,0,x|$\\$|x,x,0,-x,-x,0|$\end{tabular} &
\begin{tabular}{l}$\nu(\varepsilon_1\vec{\psi}_{N/6}+\varepsilon_2\vec{\psi}_{5N/6})$\\
$\nu(\varepsilon_2\vec{\psi}_{N/6}+\varepsilon_1\vec{\psi}_{5N/6})$\\
$\nu\varepsilon_3(\vec{\psi}_{N/6}-\vec{\psi}_{5N/6})$\end{tabular}\\
\hline
\begin{tabular}{l}B$[a^4,aiu]$\\B$[a^4,a^3iu]$\end{tabular} &
2 &
\begin{tabular}{l}$|x_1,x_2,x_3,x_2|^{**}$\\$|x_2,x_1,x_2,x_3|^{**}$\end{tabular} &
\begin{tabular}{l}$\nu_1\varepsilon_3(\vec{\psi}_{N/4}-\vec{\psi}_{3N/4})+\nu_2\vec{\psi}_{N/2}$\\
$\nu_1\varepsilon_3(\vec{\psi}_{N/4}+\vec{\psi}_{3N/4})+\nu_2\vec{\psi}_{N/2}$\end{tabular}\\
\hline
\begin{tabular}{l}B$[a^4,a^2u]$\end{tabular} &
2 &
\begin{tabular}{l}$|x_1,x_2,-x_1,-x_2|$\end{tabular} &
\begin{tabular}{l}$\nu_1\vec{\psi}_{N/4}+\nu_2\vec{\psi}_{3N/4}$\end{tabular}\\
\hline
\begin{tabular}{l}B$[a^6,iu]$\\B$[a^6,a^2iu]$\\B$[a^6,a^4iu]$\end{tabular} &
2 &
\begin{tabular}{l}$|x_1,x_2,x_3,x_3,x_2,x_1|^{**}$\\$|x_1,x_1,x_2,x_3,x_3,x_2|^{**}$\\$|x_1,x_2,x_2,x_1,x_3,x_3|^{**}$\end{tabular} &
\begin{tabular}{l}$\nu_1(\varepsilon_2\vec{\psi}_{N/6}+\varepsilon_1\vec{\psi}_{5N/6})+\nu_2(\varepsilon_2\vec{\psi}_{N/3}-\varepsilon_1\vec{\psi}_{2N/3})$\\
$\nu_1\varepsilon_3(\vec{\psi}_{N/6}-\vec{\psi}_{5N/6})+\nu_2\varepsilon_3(\vec{\psi}_{N/3}+\vec{\psi}_{2N/3})$\\
$\nu_1(\varepsilon_1\vec{\psi}_{N/6}+\varepsilon_2\vec{\psi}_{5N/6})+\nu_2(\varepsilon_1\vec{\psi}_{N/3}-\varepsilon_2\vec{\psi}_{2N/3})$\end{tabular}\\
\hline
\begin{tabular}{l}B$[a^6,i,a^3u]$\\B$[a^6,a^2i,a^3u]$\\B$[a^6,a^4i,a^3u]$\end{tabular} &
2 &
\begin{tabular}{l}$|x_1,x_2,x_1,-x_1,-x_2,-x_1|$\\$|x_1,-x_1,x_2,-x_1,x_1,-x_2|$\\$|x_1,x_2,-x_2,-x_1,-x_2,x_2|$\end{tabular} &
\begin{tabular}{l}$\nu_1(-\varepsilon_1\vec{\psi}_{N/6}+\varepsilon_2\vec{\psi}_{5N/6})+\nu_2\vec{\psi}_{N/2}$\\
$\nu_1\varepsilon_3(\vec{\psi}_{N/6}+\vec{\psi}_{5N/6})-\nu_2\vec{\psi}_{N/2}$\\
$\nu_1(\varepsilon_2\vec{\psi}_{N/6}-\varepsilon_1\vec{\psi}_{5N/6})+\nu_2\vec{\psi}_{N/2}$\end{tabular}\\
\hline
\end{tabular}

\raggedright
\small{$^{**}$ See footnote to Table~\ref{table1}.}
\end{table}

\section{Dynamical description of the bushes of modes}

Up to this point, we've been considering bushes as geometrical objects. In
the present section, we address their dynamical aspect. For this purpose,
one would need to obtain the equations of motion for the appropriate
dynamical variables (in the configuration space or in the modal space). It
can be done with the aid of the Lagrange method. This method is very useful
for the mechanical systems with some constraints on the natural dynamical
variables (for example, on the displacements of the individual particles).
Note that we deal now with precisely this case, and the additional
constraints on the dynamical variables brought about by the
\textit{symmetry} causes. Indeed, the displacement pattern $\vec{X}(t)$,
corresponding to a given bush B$[G]$, is determined by the condition of its
invariance under the action of the bush symmetry group $G$:
$\hat{G}\vec{X}(t)=\vec{X}(t)$. Namely this condition permits us to
introduce the generalized coordinates which represent dynamical variables
of the reduced Hamiltonian system associated with the considered bush of
modes.

We have already used the Lagrange method for the derivation of the bush
dynamical equations in the paper~\cite{Okta} devoted to nonlinear
vibrations of the octahedral mechanical systems with Lennard-Jones
potential. Now we apply this method to the FPU chains.

The Hamiltonian of these nonlinear chains can be written as follows:
\begin{equation}
H=T+V=\frac{1}{2}\sum_{n=1}^N \dot{x}_n^2+\frac{1}{2}\sum_{n=1}^N
(x_{n+1}-x_n)^2 +\frac{\gamma}{p}\sum_{n=1}^N (x_{n+1}-x_n)^p.
\label{eq80}
\end{equation}
Here $p=3$, $\gamma=\alpha$ for the FPU-$\alpha$ model, and $p=4$,
$\gamma=\beta$ for the FPU-$\beta$ model\footnote{The constants $\alpha$
and $\beta$ can be removed from the Hamiltonian~(\ref{eq80}) by a trivial
scaling, but we keep them for the clarity of consideration.}. $T$ and $V$
are the kinetic and potential energy, respectively. We also assume the
periodic boundary conditions~(\ref{eq1}) to be valid. Let us consider the
set of atomic displacements corresponding to the two-dimensional bush
B$[\hat{a}^4,\hat{i}]$
\begin{equation}
\vec{X}(t)=\{x,y,-y,-x|x,y,-y,-x|x,y,-y,-x|\cdots\},
\label{eq81}
\end{equation}
where we rename $x_1(t)$ and $x_2(t)$ from Eq.~(\ref{eq149}) as $x(t)$ and
$y(t)$, respectively. Substituting the atomic displacements from
Eq.~(\ref{eq81}) into the Hamiltonian~(\ref{eq80}) corresponding to the
FPU-$\alpha$ chain we obtain the following expressions for the kinetic
energy $T$ and the potential energy $V$:
\begin{equation}
T=\frac{N}{4}(\dot{x}^2+\dot{y}^2),
\label{eq82}
\end{equation}
\begin{equation}
V=\frac{N}{4}(3x^2-2xy+3y^2)+\frac{N\alpha}{2}(x^3+x^2y-xy^2-y^3).
\label{eq83}
\end{equation}
These equations are valid for the arbitrary FPU-$\alpha$ chain for which
$N\mod 4=0$. The size of the extended primitive cell for the vibrational
state (\ref{eq81}) is equal to $4a$ and, therefore, we can restrict
ourselves calculating the energies $T$ and $V$, by summing over only one
EPC.

The Lagrange equations can be written as
\begin{equation}
\frac{d}{dt}\left(\frac{\partial
L}{\partial\dot{x}_j}\right)-\frac{\partial L}{\partial x_j}=0,
\label{eq84}
\end{equation}
where $L=T-V$. The subscript $j$ is used here for numbering the dynamical
variables (in our case, $x_1(t)=x(t)$, $x_2(t)=y(t)$). Using
Eqs.~(\ref{eq82}) and~(\ref{eq83}), we obtain from~(\ref{eq84}) the
following dynamical equations for the considered bush
B$[\hat{a}^4,\hat{i}]$:
\begin{equation}
\left\{\begin{array}{lll}
\ddot{x}+(3x-y)+\alpha(3x^2+2xy-y^2)&=&0,\\
\ddot{y}+(3y-x)+\alpha(x^2-2xy-3y^2)&=&0.
\end{array}\right.
\label{eq85}
\end{equation}
Let us emphasize that these equations do not depend on the number $N$ of
the particles in the chain (but the relation $N\mod 4=0$ must hold).

Eqs.~(\ref{eq85}) are written in terms of the atomic displacements $x(t)$
and $y(t)$. From these equations, it is easy to obtain the dynamical
equations for the bush in terms of the normal modes $\mu(t)$ and $\nu(t)$.
Using the relations~(\ref{eq75}) between the old and new variables, we find
the following equations for the bush B$[\hat{a}^4,\hat{i}]$ in the modal
space
\begin{eqnarray}
\ddot{\mu}+4\mu-\frac{4\alpha}{\sqrt{N}}\nu^2=0,
\label{eq86}\\
\ddot{\nu}+2\nu-\frac{8\alpha}{\sqrt{N}}\mu\nu=0.
\label{eq86b}
\end{eqnarray}
The Hamiltonian for the bush B$[\hat{a}^4,\hat{i}]$, considered as a
two-dimensional dynamical system, can be written in the modal space as
follows:
\begin{equation}
H[\hat{a}^4,\hat{i}]=\frac{1}{2}(\dot{\mu}^2+\dot{\nu}^2)+(2\mu^2+\nu^2)-%
\frac{4\alpha}{\sqrt{N}}\mu\nu^2.
\label{eq87}
\end{equation}

We already emphasized that identical dynamical equations correspond to
equivalent bushes of modes. Moreover, the dynamical equations of the given
\textit{type} are often associated with many bushes of very
\textit{different} symmetry groups (these equations differ from each other
by numerical values of their pertinent parameters only). Because of this
reason, we can say that such bushes belong to the same class of ``dynamical
universality''~\cite{DAN1,DAN2}. As an example, let us point out that all
one-dimensional bushes for the FPU-$\beta$ models belong to one and the
same class of dynamical universality. Indeed, the dynamics of all these
bushes is described by the Duffing equation $\ddot{\mu}+A\mu+B\mu^3=0$, but
with different values of the constants $A$ and $B$.

Dynamical equations for all one-dimensional and two-dimensional bushes of
vibrational modes for the FPU-chains are presented in
Tables~\ref{table3},\ref{table4}. In the first column, we present the
complete sets of equivalent bushes, to which the same dynamical equations
(see column 2 and 3) correspond.

Table~\ref{table4} describes the ``additional'' bushes\footnote{They are
additional with respect to the bushes of normal modes associated with the
dihedral group.}, which were found in~\cite{BRink}. They are brought about
by the evenness of the potential of the FPU-$\beta$ model and the
additional symmetry operator $\hat{u}$ is used in the description of these
bushes. This operator changes the signs of all atomic displacements:
$\hat{u}\vec{X}(t)=-\vec{X}(t)$. Note that the operator $\hat{u}$ is
contained in the description of the bushes only in combinations with some
other symmetry operators, $\hat{i}$, $\hat{a}^k$ or their products.

\begin{table}
\centering
\caption{\label{table3} Dynamical equations for 1D and 2D bushes in the FPU
chains, induced by the subgroups of the dihedral group}
\begin{tabular}{|l|c|c|}
\hline
~~Bush&Equations for the FPU-$\alpha$ chain&Equations for the FPU-$\beta$ chain\\
\hline
\begin{tabular}{l}B$[a^2,i]$\end{tabular} &
$\ddot{\nu}+4\nu=0$ &
$\ddot{\nu}+4\nu=-16\frac{\beta}{N}\nu^3$\\
\hline
\begin{tabular}{l}B$[a^3,i]$\\B$[a^3,ai]$\\B$[a^3,a^2i]$\end{tabular} &
$\ddot{\nu}+3\nu=\frac{3\sqrt{6}}{2}\frac{\alpha}{\sqrt{N}}\nu^2$&
$\ddot{\nu}+3\nu=-\frac{27}{2}\frac{\beta}{N}\nu^3$\\
\hline
\begin{tabular}{l}B$[a^4,ai]$\\B$[a^4,a^3i]$\end{tabular} &
$\ddot{\nu}+2\nu=0$ &
$\ddot{\nu}+2\nu=-4\frac{\beta}{N}\nu^3$\\
\hline
\begin{tabular}{l}B$[a^3]$\end{tabular} &
$\begin{array}{l}
\ddot{\nu}_1+3\nu_1=\frac{3\sqrt{3}}{2}\frac{\alpha}{\sqrt{N}}(-\nu_1^2-2\nu_1\nu_2+\nu_2^2)\\
\ddot{\nu}_2+3\nu_2=\frac{3\sqrt{3}}{2}\frac{\alpha}{\sqrt{N}}(-\nu_1^2+2\nu_1\nu_2+\nu_2^2)
\end{array}$ &
$\begin{array}{l}
\ddot{\nu}_1+3\nu_1=-\frac{27}{2}\frac{\beta}{N}\nu_1(\nu_1^2+\nu_2^2)\\
\ddot{\nu}_2+3\nu_2=-\frac{27}{2}\frac{\beta}{N}\nu_2(\nu_1^2+\nu_2^2)
\end{array}$\\
\hline
\begin{tabular}{l}B$[a^4,i]$\\B$[a^4,a^2i]$\end{tabular} &
$\begin{array}{l}
\ddot{\nu}_1+2\nu_1=-8\frac{\alpha}{\sqrt{N}}\nu_1\nu_2\\
\ddot{\nu}_2+4\nu_2=-4\frac{\alpha}{\sqrt{N}}\nu_1^2
\end{array}$ &
$\begin{array}{l}
\ddot{\nu}_1+2\nu_1=-8\frac{\beta}{N}\nu_1(\nu_1^2+3\nu_2^2)\\
\ddot{\nu}_2+4\nu_2=-8\frac{\beta}{N}\nu_2(3\nu_1^2+2\nu_2^2)
\end{array}$\\
\hline
\begin{tabular}{l}B$[a^6,ai]$\\B$[a^6,a^3i]$\\B$[a^6,a^5i]$\end{tabular} &
$\begin{array}{l}
\ddot{\nu}_1+\nu_1=-\sqrt{6}\frac{\alpha}{\sqrt{N}}\nu_1\nu_2\\
\ddot{\nu}_2+3\nu_2=\frac{\sqrt{6}}{2}\frac{\alpha}{\sqrt{N}}(-\nu_1^2+3\nu_2^2)
\end{array}$ &
$\begin{array}{l}
\ddot{\nu}_1+\nu_1=-\frac{3}{2}\frac{\beta}{N}\nu_1(\nu_1^2+3\nu_2^2)\\
\ddot{\nu}_2+3\nu_2=-\frac{9}{2}\frac{\beta}{N}\nu_2(\nu_1^2+3\nu_2^2)
\end{array}$\\
\hline
\end{tabular}
\end{table}

\begin{table}
\centering
\caption{\label{table4} Dynamical equations for the additional 1D and 2D
bushes, for the FPU-$\beta$ chain}
\begin{tabular}{|l|c|}
\hline
~~Bush&Equations\\
\hline
\begin{tabular}{l}B$[a^3,iu]$\\B$[a^3,aiu]$\\B$[a^3,a^2iu]$\end{tabular} &
$\ddot{\nu}+3\nu=-\frac{27}{2}\frac{\beta}{N}\nu^3$\\
\hline
\begin{tabular}{l}B$[a^4,iu]$\\B$[a^4,a^2iu]$\end{tabular} &
$\ddot{\nu}+2\nu=-8\frac{\beta}{N}\nu^3$\\
\hline
\begin{tabular}{l}B$[a^6,ai,a^3u]$\\B$[a^6,a^3i,a^3u]$\\B$[a^6,a^5i,a^3u]$\end{tabular} &
$\ddot{\nu}+\nu=-\frac{3}{2}\frac{\beta}{N}\nu^3$\\
\hline
\begin{tabular}{l}B$[a^4,aiu]$\\B$[a^4,a^3iu]$\end{tabular} &
$\begin{array}{l}
\ddot{\nu}_1+2\nu_1=-4\frac{\beta}{N}\nu_1(\nu_1^2+6\nu_2^2)\\
\ddot{\nu}_2+4\nu_2=-8\frac{\beta}{N}\nu_2(3\nu_1^2+2\nu_2^2)
\end{array}$\\
\hline
\begin{tabular}{l}B$[a^4,a^2u]$\end{tabular} &
$\begin{array}{l}
\ddot{\nu}_1+2\nu_1=-8\frac{\beta}{N}\nu_1^3\\
\ddot{\nu}_2+4\nu_2=-8\frac{\beta}{N}\nu_2^3
\end{array}$\\
\hline
\begin{tabular}{l}B$[a^6,iu]$\\B$[a^6,a^2iu]$\\B$[a^6,a^4iu]$\end{tabular} &
$\begin{array}{l}
\ddot{\nu}_1+\nu_1=-\frac{3}{2}\frac{\beta}{N}\nu_1(\nu_1^2+9\nu_2^2)\\
\ddot{\nu}_2+3\nu_2=-\frac{27}{2}\frac{\beta}{N}\nu_2(\nu_1^2+\nu_2^2)
\end{array}$\\
\hline
\begin{tabular}{l}B$[a^6,i,a^3u]$\\B$[a^6,a^2i,a^3u]$\\B$[a^6,a^4i,a^3u]$\end{tabular} &
$\begin{array}{l}
\ddot{\nu}_1+\nu_1=-\frac{\beta}{N}\nu_1(\frac{3}{2}\nu_1^2+3\sqrt{2}\nu_1\nu_2+12\nu_2^2)\\
\ddot{\nu}_2+4\nu_2=-\frac{\beta}{N}(\sqrt{2}\nu_1^3+12\nu_1^2\nu_2+16\nu_2^3)
\end{array}$\\
\hline
\end{tabular}
\end{table}

\section{Stability of bushes of normal modes}

The term ``stability'' is often used in various senses. Let us explain what
we mean by this term in the present paper.

In general case, stability of the bushes of modes was discussed
in~\cite{DAN1,PhysD}, while their stability in the FPU-$\alpha$ chain is
considered in~\cite{FPU1}. As well as in these papers, we will discuss here
the stability of a given bush of normal modes with respect of its
interactions\footnote{There is an essential difference between the
interactions of the modes which belong and which do not belong to a given
bush: we speak about ``force interaction'' in the former case and about
`parametric interaction'' in the last case~\cite{PhysD}.} with the modes
which \textit{do not belong} to this bush. Let us illustrate this idea with
the following example.

For the FPU-$\alpha$ chain, the two-dimensional bush B$[\hat{a}^4,\hat{i}]$
is described by Eqs.~(\ref{eq86}),(\ref{eq86b}). These equations allow a
solution of the form
\begin{equation}
\mu(t)\neq 0,\quad \nu(t)\equiv 0.
\label{eq90}
\end{equation}
(The vibrational regime of this type can be excited by imposing the
appropriate initial conditions: $\mu(t_0)=\mu_0\neq 0$, $\dot{\mu}(t_0)=0$,
$\nu(t_0)=0$, $\dot{\nu}(t_0)=0$). Substitution of the
solution~(\ref{eq90}) into~(\ref{eq86}) produces the dynamical equation of
the one-dimensional bush B$[\hat{a}^2,\hat{i}]$ (see Table~\ref{table4})
consisting of only one mode $\mu(t)$
\begin{equation}
\ddot{\mu}+4\mu=0
\label{eq92}
\end{equation}
with the trivial solution\footnote{We can consider the initial phase in the
solution~(\ref{eq93}) to be equal to zero~\cite{FPU1}.}
\begin{equation}
\mu(t)=\mu_0\cos(2t).
\label{eq93}
\end{equation}
Substituting~(\ref{eq93}) into Eq.~(\ref{eq86b}), we obtain
\begin{equation}
\ddot{\nu}+\left[2-\frac{8\alpha\mu_0}{\sqrt{N}}\cos(2t)\right]\nu=0.
\label{eq94}
\end{equation}
This equation can be easily transformed into the standard form of the
Mathieu equation
\begin{equation}
\ddot{\nu}+[a-2q\cos(2t)]\nu=0.
\label{eq95}
\end{equation}

On the other hand, there exist domains of stable and unstable motion of the
Mathieu equation~(\ref{eq95}) in the $a-q$ plane of its intrinsic
parameters, $a$ and $q$. The one-dimensional bush B$[\hat{a}^2,\hat{i}]$ is
stable for the sufficiently small amplitudes $\mu_0$ of the mode $\mu(t)$,
but it becomes unstable as the result of the increasing of this amplitude.
This phenomenon, similar to the well-known parametric resonance, takes
place for those $\mu_0$ which get into the domains of unstable motion of
the Mathieu-type equation~(\ref{eq94}). The loss of stability of the
dynamical regime~(\ref{eq90}) (the bush B$[\hat{a}^2,\hat{i}]$) manifests
itself in the appearance of the mode $\nu(t)$ which was identically equal
to zero for the vibrational state~(\ref{eq90}). As a result, the dimension
of the original one-dimensional bush B$[\hat{a}^2,\hat{i}]$ increases and
this bush transforms into the two-dimensional bush B$[\hat{a}^4,\hat{i}]$.
This transformation is accompanied by the breaking of the symmetry of the
vibrational state (the symmetry of the bush B$[\hat{a}^2,\hat{i}]$ is twice
larger than that of the bush B$[\hat{a}^4,\hat{i}]$). In general case, we
consider a given bush as a stable dynamical object, if the complete
collection of its modes (and, therefore, its dimension) does not change in
time. All other modes in the system, according to the definition of the
bush as the full collection of active modes, possess zero amplitudes, they
are ``sleeping'' modes. If we increase intensity of bush vibrations, some
sleeping modes, because of the parametric interactions with the active
modes, can lose their stability and become excited. In this situation, we
speak about the \textit{loss of stability} of the \textit{original bush},
since the dimension of the vibrational state (the number of active modes)
becomes larger, while its symmetry becomes lower. As a consequence of the
stability loss, the original bush transforms into another bush of higher
dimension.

As it was described above, the loss of stability of the bush
B$[\hat{a}^2,\hat{i}]$ with respect to its transformation into the bush
B$[\hat{a}^4,\hat{i}]$ can be investigated with the aid of the Mathieu
equation. But if our nonlinear chain contain $N$ particles, we must analyze
the stability of the bush B$[\hat{a}^2,\hat{i}]$ not only relatively to the
mode $\nu(t)$ of the bush B$[\hat{a}^4,\hat{i}]$, but with respect to
\textit{all the other} modes, too. This can be done by the Floquet method.

The stability of the bush B$[\hat{a}^2,\hat{i}]$\footnote{In the
paper~\cite{FPU1}, this bush was denoted by the symbol B$[2a]$.} in the
FPU-$\alpha$ chain for arbitrary even $N$ was investigated by us
in~\cite{FPU1}. Let us recall some points of this investigation which are
needed for analyzing the stability of the other one-dimensional bushes.

In the modal space, the dynamical equations of the FPU-$\alpha$ chain can
be written as follows (see Eqs.~(27) from the paper~\cite{FPU1}):
\begin{eqnarray}
\ddot{\nu}_j+\omega_j^2\nu_j=-\frac{\alpha}{\sqrt{N}}\sum_{k=0}^{N-1}\nu_k&&\left[
(\nu_{j+k}+\nu_{-j-k})\left(2\sin\frac{2\pi}{N}k+\sin\frac{2\pi}{N}j\right)\right.\nonumber\\
&&\;\left.+(\nu_{j-k}-\nu_{-j+k})\left(2\sin\frac{2\pi}{N}k-\sin\frac{2\pi}{N}j\right)\right],\label{eq100}
\end{eqnarray}
\[\omega_j^2=4\sin^2\left(\frac{\pi j}{N}\right),\quad j=0,1,2,\cdots,N-1.\]
Here $\nu_j(t)$ are the time-dependent coefficients in front of the basis
vectors $\vec{\psi}_j$~(\ref{eq70}) in the decomposition~(\ref{eq71}). For
brevity, we call $\nu_j(t)$ by the term ``mode'', in spite of the fact that
this term, rigorously speaking, corresponds to the product
$\nu_j(t)\vec{\psi}_j$. The numbers of all modes $\nu_j$ in~(\ref{eq100})
are assumed to be reduced to the interval $1\le i\le N-1$ by adding $\pm
N$, since these numbers are defined modulo $N$. Hereafter, the mode $\nu_0$
is excluded from the consideration of the \textit{vibrational} bushes
because it corresponds to the motion of the chain as a whole.

The bush B$[\hat{a}^2,\hat{i}]$ consists of only one mode $\nu_{N/2}(t)$
and looks as
\begin{equation}
\B[\hat{a}^2,\hat{i}]=\nu_{N/2}(t)\vec{\psi}_{N/2},
\label{eq102}
\end{equation}
where $\vec{\psi}_{N/2}$ is determined by Eq.~(\ref{eq74}). The dynamical
equation for the mode $\nu_{N/2}(t)$ can be obtained from~(\ref{eq100})
supposing that all the other modes are equal to zero: $\nu_j(t)\equiv 0$,
$j\neq \frac{N}{2}$. This equation reads
\begin{equation}
\ddot{\nu}_{N/2}+\omega_{N/2}^2\nu_{N/2}=0.
\label{eq104}
\end{equation}
Its solution can be written in the form
\begin{equation}
\nu_{N/2}(t)=A\cos(2\tau)
\label{eq105}
\end{equation}
(the initial phase in this solution, as well as in Eq.~(\ref{eq93}), may be
chosen equal to zero~\cite{FPU1}). \textit{Linearizing} the
system~(\ref{eq100}) near the exact solution~(\ref{eq102}) with respect of
all the modes $\nu_j$ ($j\neq \frac{N}{2}$), we obtain the following
approximate equations
\begin{equation}
\ddot{\nu}_j+\omega_j^2\nu_j=-\frac{8\alpha}{\sqrt{N}}\sin\left(\frac{2\pi j}{N}\right)\nu_{N/2}\nu_{N/2-j},
\label{eq106}
\end{equation}
\[\omega_j^2=4\sin^2\left(\frac{\pi j}{N}\right),\quad j=1,2,\cdots,\frac{N}{2}-1,\frac{N}{2}+1,\cdots,N-1.\]
The system~(\ref{eq106}) splits into a number of subsystems containing one
or two equations only. Indeed, the mode $\nu_j$ in~(\ref{eq106}) is
connected with the mode $\nu_{\tilde{j}}$, where $\tilde{j}=\frac{N}{2}-j$,
and vice versa, the mode $\nu_{\tilde{j}}$ is connected with
$\nu_{N/2-\tilde{j}}\equiv \nu_j$. Therefore, we have for
$j=1,2,\dotsc,\frac{N}{2}-1$ (see Eqs.~(\ref{eq106}) the following pairs of
linearized equations which are \textit{independent} from all the other
equations:
\begin{equation}
\begin{array}{lll}
\ddot{\nu}_j&+4\sin^2\left(\frac{\pi j}{N}\right)\nu_j&=-\gamma\sin\left(\frac{2\pi j}{N}\right)\nu_{N/2-j}\cos(2\tau),\\
\ddot{\nu}_{N/2-j}&+4\cos^2\left(\frac{\pi j}{N}\right)\nu_{N/2-j}&=-\gamma\sin\left(\frac{2\pi j}{N}\right)\nu_{j}\cos(2\tau),
\end{array}\label{eq107}
\end{equation}
\[\gamma=\frac{8\alpha A}{\sqrt{N}},\quad j=1,2,\cdots,\frac{N}{4}-1.\]
Here we replaced $\nu_{N/2}(t)$ with $A\cos(2\tau)$ from Eq.~(\ref{eq105}).
For $j=\frac{N}{2}+1,\dotsc,N-1$, we obtain equations equivalent to
Eqs.~(\ref{eq107})~\cite{FPU1}.

If $j=\frac{N}{2}-j$ and, therefore, $j=\frac{N}{4}$, Eqs.~(\ref{eq107})
reduce to a pair of identical equations of the form
\begin{equation}
\ddot{\nu}_{N/4}+2\nu_{N/4}=-\gamma\nu_{N/4}\cos(2\tau),
\label{eq108}
\end{equation}
which is the Mathieu equation. Actually, this is the above considered case,
describing the transformation of the bush B$[\hat{a}^2,\hat{i}]$ with the
mode $\mu(t)\equiv\nu_{N/2}(t)$ into the two-dimensional bush
B$[\hat{a}^4,\hat{i}]$ containing the modes $\mu(t)\equiv\nu_{N/2}(t)$ and
$\nu(t)\equiv\nu_{3N/4}(t)$.

It is convenient to rewrite Eqs.~(\ref{eq107}) as follows:
\begin{equation}
\begin{array}{lcl}
\ddot{x}+4\sin^2(k)x&=&\gamma\sin(2k)y\cos(2\tau)\\
\ddot{y}+4\cos^2(k)y&=&\gamma\sin(2k)x\cos(2\tau),
\end{array}
\label{eq109}
\end{equation}
where $k=\frac{\pi j}{N}$, $x(t)=\nu_j(t)$ and $y(t)=\nu_{N/2-j}(t)$. Thus,
studying the loss of stability of the bush B$[\hat{a}^2,\hat{i}]$, brought
about by its interactions with the modes $\nu_{N/4}$ (and $\nu_{3N/4}$), is
reduced to analyzing the Mathieu equation~(\ref{eq108}), while the loss of
its stability with respect to all other modes reduces to analyzing the
equations~(\ref{eq109}).

Eqs.~(\ref{eq109}) are linear differential equations with periodic
coefficients and, therefore, they can be studied with the aid of the
Floquet theory. The system~(\ref{eq109}) is very interesting, and we would
like to cite the following fragment of the text of our paper~\cite{FPU1},
devoted to its properties:

``Our computation of the multiplicators for the system~(\ref{eq109}) as
eigenvalues of the monodromic matrix reveals a surprising fact! Indeed, it
seems that the critical value $\gamma_c$ of the constant $\gamma$
from~(\ref{eq107}), corresponding to the loss of stability of the bush
B$[\hat{a}^2,\hat{i}]$, must depend on the mode number $j$, since the
coefficients of these equations depend explicitly on $\frac{j}{N}$.
However, this is not true. We found that $\gamma_c$ does not depend on
$\frac{j}{N}$, at least up to $10^{-5}$, and coincides with that of the
Mathieu equation~(\ref{eq108}):
\begin{equation}
\gamma_c=2.42332.
\label{eq110}
\end{equation}
Then from Eq.~(\ref{eq107}) we obtain
\begin{equation}
\alpha A_c=0.30292\sqrt{N}.
\label{eq111}
\end{equation}
This nontrivial fact means that the original bush B$[\hat{a}^2,\hat{i}]$
loses its stability with respect to all modes $\nu_j$ ($j\neq \frac{N}{2}$)
\textit{simultaneously}, i.e.\ for the same value $A_c$ of the amplitude of
the mode $\nu_{N/2}(t)$. In other words, all modes of the $N$-particle
FPU-$\alpha$ chain are excited parametrically because of interactions with
B$[\hat{a}^2,\hat{i}]$ when $\alpha A$ reaches its critical value $\alpha
A_c=0.30292\sqrt{N}$, and, therefore, the bush B$[\hat{a}^2,\hat{i}]$
transforms at once into the bush B$[\hat{a}^N]$ of trivial symmetry''.

One may think that the above-mentioned fact means that there exists such
transformation of Eqs.~(\ref{eq107}) which removes the dependence of these
equations on the parameter $\frac{j}{N}$ (or the dependence on $k$ of
Eqs.~(\ref{eq109})). Nevertheless, this doesn't turn out to be correct.
Indeed, in Fig.~\ref{fig6}, we show (in white color) not only the first
stability domain for the different modes $\nu_k$ for $0\leq k\leq\pi$
($k=\frac{\pi j}{N}$), but also the second stability domain. The former
domain looks as a white horizontal band near the $k$-axis, while the latter
domain looks as two white triangles above this band (the vertical axis
corresponds to the value $A$ of the amplitude of the $\nu_{N/2}(t)$ mode of
the bush B$[\hat{a}^2,\hat{i}]$). This means that there is no dependence of
the excitation conditions for the modes $\nu_j$ ($j\neq\frac{N}{2}$) on the
number $j$ for the first stability domain, but such a dependence is very
obvious for the second stability domain.

Since we use similar stability diagrams for all one-dimensional bushes, let
us describe their structure in more detail. In Fig.~\ref{fig6}, each point
$(A,k)$ determines a certain value of the root mode amplitude $A$ of the
bush B$[\hat{a}^2,\hat{i}]$ and a certain value $k=\frac{\pi j}{N}$, which
is connected with the index $j$ of a fixed mode of the FPU-$\alpha$ chain.
The black points $(A,k)$ correspond to the case where the mode
$j=k\frac{N}{\pi}$ becomes excited because of its parametric interaction
with the mode of the bush B$[\hat{a}^2,\hat{i}]$. The white color denotes
the opposite case: the corresponding mode $j$, being zero at the initial
instant, continues to be zero in spite of its interaction with the
considered bush. Dashed vertical lines correspond to the modes
($j=1,2,\dotsc,12$) for the FPU chain with $N=12$ particles. Considering
stability diagrams, we must make one remark concerning normalization of the
normal modes. In Eq.~(\ref{eq70}), there is the normalizing factor
$\frac{1}{\sqrt{N}}$ which leads to the relation
$\left|\vec{\psi}_k\right|^2=1$. Because of this factor, the absolute
values of the individual atomic displacements corresponding to a given mode
$\vec{\psi}_k$ tend to zero for $N\rightarrow\infty$. Such normalization of
the normal modes is not convenient for studying the bush stability. Indeed,
the critical value of atomic displacements, associated with the loss of
stability, say, for the bush B$[\hat{a}^2,\hat{i}]$ is equal to
$x_c=0.30292/\alpha$ (this result can be easy obtained from
Eq.~(\ref{eq111}) (see~\cite{FPU1}). This value $x_c$ does not depend on
the number $N$ of particles in the FPU chain. Because of this reason,
depicting stability diagrams of one-dimensional and two-dimensional bushes
in Figs.~\ref{fig26}-\ref{fig31}, we use different normalization of the
normal modes than that in Eq.~(\ref{eq70}), namely, we simply reject the
factor $\frac{1}{\sqrt{N}}$ which presents in Eq.~(\ref{eq70}). Then, the
maximal value of the atomic displacements corresponding to every normal
mode $\vec{\psi}_k$ turns out to be equal to $\sqrt{2}$.

We have just discussed the stability of the bush B$[\hat{a}^2,\hat{i}]$.
Now we intend to consider the stability of all the other one-dimensional
bushes in the FPU-$\alpha$ and FPU-$\beta$ chains. There are three such
bushes for \textit{every} monoatomic nonlinear chain with dihedral symmetry
group, but for the FPU-$\beta$ model, there exist three additional
one-dimensional bushes because of the evenness of its
potential~\cite{BRink}. For investigating stability of these bushes (three
for the FPU-$\alpha$ chain and six for the FPU-$\beta$ chain), we shall
also use the Floquet method.

There are two difficulties in implementing of this method for
one-dimensional bushes in the general case, as compared to that for the
above considered bush B$[\hat{a}^2,\hat{i}]$. Indeed, the dynamical
equations for the bushes B$[\hat{a}^2,\hat{i}]$ and
B$[\hat{a}^4,\hat{a}\hat{i}]$ in the FPU-$\alpha$ model are reduced to the
harmonic oscillator equation (with different frequencies), while the
dynamical equations for all the other bushes turn out to be
\textit{nonlinear}. As a consequence, the period $T$ of the oscillations,
described by the bush of such a type, depends on the amplitude $\mu_0$ of
these oscillations ($T=T(\mu_0)$). Therefore, we must construct the
monodromy matrix by integration of the linearized dynamical equations, for
the considered chain, over the period $T(\mu_0)$. This period must be found
for every value $\mu_0$ before obtaining the monodromy matrix.

The second difficulty is that the linearized dynamical system (near the
appropriate exact solution to the given bush) can be split into the
subsystems whose dimension (3 or 4) can be higher than that for the case of
the bush B$[\hat{a}^2,\hat{i}]$. As a result, we must deal, respectively,
with $6\times 6$ or $8\times 8$ monodromy matrices.

\section{Stability of one-dimensional bushes for the FPU chains} We will
consider the stability of the bushes for the FPU-$\alpha$ and FPU-$\beta$
models separately.

\subsection{FPU-$\alpha$ chain} The exact dynamical equations for the
FPU-$\alpha$ chain in the modal space are given by Eqs.~(\ref{eq100}).

We want to linearize these equations in the vicinity of the solution
described by a given one-dimensional bush. There are three one-dimensional
bushes associated with the subgroups of the dihedral group of the
monoatomic chains with the appropriate divisibility properties of the
number $N$ of the particles (see Table~\ref{table2}):
B$[\hat{a}^2,\hat{i}]$ for $N\mod 2=0$, B$[\hat{a}^3,\hat{i}]$ for $N\mod
3=0$, B$[\hat{a}^4,\hat{a}\hat{i}]$ for $N\mod 4=0$. The following root
modes correspond, respectively, to them: $\nu(t)\vec{\psi}_{N/2}$,
$\nu(t)[\varepsilon_1\vec{\psi}_{N/3}+\varepsilon_2\vec{\psi}_{2N/3}]$,
$\nu(t)[\vec\psi_{N/4}+\vec\psi_{3N/4}]$. Note that the root modes for the
bushes B$[\hat{a}^3,\hat{i}]$ and B$[\hat{a}^4,\hat{a}\hat{i}]$ are the
certain superpositions of the \textit{conjugate modes}\footnote{According
to~\cite{FPU1}, we call two \textit{real} modes $\vec{\psi}_j$ and
$\vec{\psi}_{N-j}$ by the term ``conjugate'' since normal modes
$\vec{\phi}_j$ and $\vec{\phi}_{N-j}$, corresponding to them in the
\textit{complex} form, are complex conjugate. The frequencies of the
conjugate normal modes are equal.} with time-independent coefficients
$\left[\varepsilon_1=\sin\left(\frac{\pi}{12}\right),
\varepsilon_2=\cos\left(\frac{\pi}{12}\right)\right]$.

Let us consider the bush B$[\hat{a}^3,\hat{a}^2\hat{i}]$\footnote{This
bush, being equivalent to the bush B$[\hat{a}^3,\hat{i}]$, turns out to be
more suitable for the next calculations.}. Linearizing the FPU-$\alpha$
dynamical equations near the solutions
$\vec{X}_0(t)=\nu(t)\varepsilon_3[\vec{\psi}_{N/3}-\vec{\psi}_{2N/3}]$ of
this bush $\left[\varepsilon_3=\frac{\sqrt{2}}{2}\right]$, we must consider
the amplitudes $\nu_{N/3}(t)=\varepsilon_3\nu(t)$ and
$\nu_{2N/3}(t)=-\varepsilon_3\nu(t)$ in Eqs.~(\ref{eq100}) to be much
greater than those of all other modes or, more precisely, $\nu_j(t)$ with
$j\neq\frac{N}{3},\frac{2N}{3}$ must be treated as the infinitesimal
values. Then we retain in Eqs.~(\ref{eq100}) only linear terms with respect
to the modes $\nu_j(t)$ ($j\neq\frac{N}{3},\frac{2N}{3}$) which do not
contribute to the given bush. The result of this computational procedure
can be written as follows
\begin{eqnarray}
\ddot\nu_j+\omega_j^2\nu_j=-\frac{\alpha\sqrt 6}{\sqrt N}\nu(t)
&&\left[\left(1+\sqrt 3\sin\frac{2\pi j}{N}-\cos\frac{2\pi j}{N}\right)\nu_{\frac{N}{3}+j}+\right.\nonumber\\
&&\;\left.\left(1-\sqrt 3\sin\frac{2\pi j}{N}-\cos\frac{2\pi j}{N}\right)\nu _{\frac{2N}{3}+j}\right],\label{eq5182}
\end{eqnarray}
\[j=1,2,\cdots,N-1.\]
It can be easy seen from this form of linearized dynamical system that, for
every fixed $j$, the modes $\nu_j$, $\nu_{N/3+j}$ and $\nu_{2N/3+j}$, being
coupled by Eq.~(\ref{eq5182}), are decoupled from all other modes.
Therefore, we obtain for $j=1,2,\cdots,\frac{N}{3}-1$ the following
$3\times 3$ systems of ordinary differential equations with
\textit{periodic coefficients} proportional to the function $\nu(t)$:
\begin{equation}
\begin{array}{l}
\ddot\nu_j+\omega_j^2\nu_j=-\frac{\alpha\sqrt 6}{\sqrt N}\nu(t)\left[ {\left( {1 - 2\cos \left( {\frac{{2\pi j}}{N} + \frac{\pi }{3}} \right)} \right)\nu _{\frac{N}{3} + j}  + \left( {1 - 2\sin \left( {\frac{{2\pi j}}{N} + \frac{\pi }{6}} \right)} \right)\nu _{\frac{{2N}}{3} + j} } \right], \\
\ddot\nu_{\frac{N}{3}+j}+\omega_{\frac{N}{3}+j}^2\nu_{\frac{N}{3}+j}=-\frac{\alpha\sqrt 6}{\sqrt N}\nu(t)\left[ {\left( {1 - 2\cos \left( {\frac{{2\pi j}}{N} + \frac{\pi }{3}} \right)} \right)\nu _j  + \left( {1 + 2\cos \frac{{2\pi j}}{N}} \right)\nu _{\frac{{2N}}{3} + j} } \right], \\
\ddot\nu_{\frac{2N}{3}+j}+\omega_{\frac{2N}{3}+j}^2\nu_{\frac{2N}{3}+j}=-\frac{\alpha\sqrt 6}{\sqrt N}\nu(t)\left[ {\left( {1 - 2\sin \left( {\frac{{2\pi j}}{N} + \frac{\pi }{6}} \right)} \right)\nu _j  + \left( {1 + 2\cos \frac{{2\pi j}}{N}} \right)\nu _{\frac{N}{3} + j} } \right]. \\
\end{array}
\label{eq5183}
\end{equation}
In turn, the function $\nu(t)$ is the solution to the dynamical equation
\[\ddot{\nu}+3\nu=\frac{3\sqrt{6}}{2}\frac{\alpha}{\sqrt{N}}\nu^2\]
of the considered bush B$[\hat{a}^3,\hat{a}^2\hat{i}]$. Thus, the stability
of the bush B$[\hat{a}^3,\hat{a}^2\hat{i}]$ can be investigated by the
Floquet method applied to the systems~(\ref{eq5183}) for
$j=1,2,\cdots,\frac{N}{3}-1$.

Similarly to the above said, we can obtain the $4\times 4$ systems of
differential equations for investigation of stability of the bush
B$[\hat{a}^4,\hat{a}\hat{i}]$. All systems of such a type, which must
subjected to the further Floquet analysis, are given for the FPU-$\alpha$
chain in Table~\ref{table5}.

\subsection{FPU-$\beta$ chain}
We have obtained the linearized dynamical equations for one-dimensional
bushes in the FPU-$\alpha$ chain by linearizing the \textit{exact}
nonlinear equations~(\ref{eq100}) defined in the modal space. This method,
namely, obtaining the exact equations in the modal space with their
subsequent linearization, is not the simplest method for our purpose.
Indeed, we can, firstly, linearize the dynamical equations in the
\textit{configuration} space and only then transform the obtained equations
to the modal space. In this manner, we can obtain the linearized equations
in the modal space for different nonlinear chains, in particular, for the
FPU-$\beta$ chain. Moreover, we want to find the dynamical equations
linearized near the solution
\begin{equation}
\vec{X}_0(t)=\nu(t)\vec{c},
\label{eq400}
\end{equation}
which describes the exact nonlinear regime corresponding to the considered
one-dimensional bush. Here $\vec{c}$ is a constant vector which determines
the directions of displacements of all particles in the chain, while
$\nu(t)$ is a time-dependent function satisfied the dynamical equation of a
given 1D bush
\begin{equation}
\ddot\nu+\omega^2\nu=F(\nu).
\label{eq401}
\end{equation}
Here $F(\nu)$ is a certain nonlinear function.

Let us then suppose that
\begin{equation}
\vec{X}(t)=\vec{X}_0(t)+\vec{\delta}(t)
\label{eq410}
\end{equation}
with $\vec{\delta}(t)=\left\{\delta_1(t),\cdots,\delta_N(t)\right\}$ being
an infinitesimal vector, and substitute $\vec{X}(t)$ in this form into the
dynamical equations of the nonlinear chain in the \textit{configuration
space}. For the chain with dynamical equations\footnote{For the
FPU-$\alpha$ and FPU-$\beta$ chains, $m$ is equal to $2$ and $3$,
respectively.}
\begin{equation}
\ddot{x}_k=(x_{k+1}+x_{k-1}-2x_k)+\gamma\left[(x_{k+1}-x_k)^m-(x_k-x_{k-1})^m\right],\quad k=1,2,\cdots,N,
\label{eq415}
\end{equation}
and with periodic boundary conditions $x_{N+1}(t)=x_1(t)$, $x_0(t)=x_N(t)$,
after linearization with respect to $\vec{\delta}(t)$, we obtain
\begin{equation}
\ddot{\vec{\delta}}=B(t)\vec{\delta}.
\label{eq420}
\end{equation}
Here time-periodic matrix $B(t)$ is almost three-diagonal (see
Fig.~\ref{fig50}) and can be determined by the equations
\[\begin{array}{l}
b_{j,j}=-2-\tilde\gamma(t)\left[(c_{j+1}-c_j)^{m-1}+(c_j-c_{j-1})^{m-1}\right],\\
b_{j,j-1}=1+\tilde\gamma(t)(c_j-c_{j-1})^{m-1},\\
b_{j,j+1}=1+\tilde\gamma(t)(c_{j+1}-c_j)^{m-1},
\end{array}\]
where $\tilde\gamma(t)=\gamma m\nu^{m-1}(t)$.

\begin{figure}
\centering
\begin{tabular}{|c|c|c|c|c|c|c|c|}
\hline
*&*&~~~~& & & & &*\\
\hline
*&*&*&~~~~& & & & \\
\hline
 &*&*&*&~~~~& & & \\
\hline
 & &*&*&*&~~~~& & \\
\hline
 & & &*&*&*&~~~~& \\
\hline
 & & & &*&*&*&~~~~\\
\hline
~~~~& & & & &*&*&*\\
\hline
*&~~~~& & & & &*&*\\
\hline
\end{tabular}
\caption{\label{fig50} The structure of the matrix $B$ from Eq.~\ref{eq420}
(for $N=8$).}
\end{figure}

The vector Eq.~(\ref{eq420}) represents the linearized dynamical equations
of the considered chain~(\ref{eq415}) in the configuration space. Now, we
must transform these equations to the \textit{modal} space. Let
\begin{equation}
\vec{\delta}(t)=\sum_{j=1}^{N-1}\nu_j(t)\vec{\psi}_j,
\label{eq430}
\end{equation}
where the basis vectors $\vec{\psi}_j$ are determined by Eq.~(\ref{eq70}).
Substituting $\vec{\delta}(t)$ from Eq.~(\ref{eq430}) into
Eq.~(\ref{eq420}) and taking into account that these basis vectors are
orthonormal, one can obtain the final dynamical equations of the chain,
linearized near the solution~(\ref{eq400}) to a given one-dimensional bush.
It must be noted, that this transformation, being simple in principle, is
practically very cumbersome. We performed it with the aid of the
mathematical packet \texttt{MAPLE}. The results of this procedure are
listed for all one-dimensional bushes in the FPU-$\alpha$ and FPU-$\beta$
chains in Table~\ref{table5} and Table~\ref{table6}, respectively.

As can be seen from Table~\ref{table6}, the full systems of linearized
differential equations corresponding to different one-dimensional bushes in
the FPU-$\beta$ chain split into certain subsystems whose dimensionalities
are equal to $1$, $2$, $3$.\footnote{Let us note that splitting of the
general dynamical system, linearized near a given bush, into independent
subsystems of small dimensionalities, is brought about by
\textit{symmetry-related} causes. Some general results concerning such
splitting will be considered elsewhere.}

These systems contain time-periodic coefficients proportional to
$\nu^2(t)$, where $\nu(t)$ is the solution to the Duffing equation
\[\ddot{\nu}+\omega^2\nu=\gamma\nu^3.\]
The coefficients $\omega^2$ and $\gamma$ are, naturally, different for
different bushes.

Using Tables~\ref{table5} and~\ref{table6}, we can investigate the
stability of all one-dimensional bushes, with the aid of the Floquet
method, for the FPU-$\alpha$ and FPU-$\beta$ models, respectively.

\begin{figure}
\centering
\setlength{\unitlength}{1mm}
\begin{picture}(130,45)(0,0)
\put(5,5){\includegraphics[width=120mm,height=36mm]{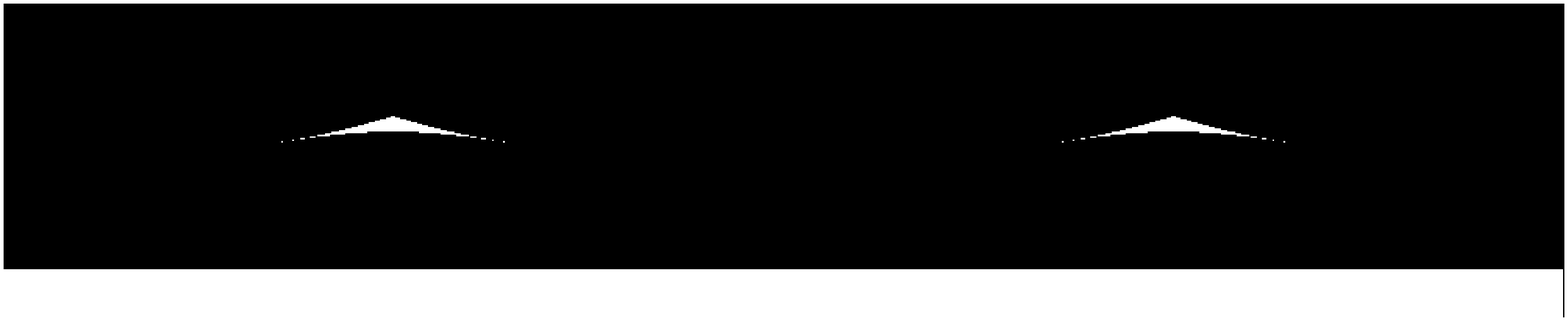}}
\drawgridB
\put(65,5){\line(0,1){36}}% 6
\end{picture}
\caption{\label{fig6} Regions of stability (white color) of different modes
of the FPU-$\alpha$ chain, interacting parametrically with the
one-dimensional bush B$[a^2,i]$.}
\end{figure}

In our bush stability diagrams (see Figs.~\ref{fig6}-\ref{fig14}), we
depict regions of stability (white color) and instability (black color) for
individual modes $j=1,2,3,\dotsc,N-1$ of the FPU chains. The loss of
stability by any mode, which does not belong to a given bush, leads to the
loss of stability by the entire bush. The fact that we can study stability
of individual modes allows us to investigate the bush stability not only
for the chains with finite number of particles, but also for the case of
thermodynamical (continuum) limit $N\rightarrow\infty$ (see a discussion in
the next section).

Practically, we study the eigenvalues of the monodromy matrices for a
certain subsets of sleeping modes. We choose as many points on the
horizontal $k$-axis ($k=\frac{\pi j}{N}$), as it is necessary for obtaining
sufficiently accurate picture. In some cases, the density of the analyzed
$k$-points corresponds to $N\sim 10^5$. Usually, we recognize instability
of a given set of sleeping modes, if there are eigenvalues of the
corresponding monodromy matrix with modules exceeding $1$ by $10^{-5}$.

In Table~\ref{table7}, for the case $N=12$, we present the thresholds of
the loss of stability for all one-dimensional bushes in the FPU-$\alpha$
and FPU-$\beta$ chains. In the stability diagrams
(Figs.~\ref{fig6}-\ref{fig14}), all possible modes for this case
($j=1,2,3,\dotsc,12$) are depicted by vertical dashed lines. Using these
diagrams, one can understand, in some sense, the \textit{cause} of the
given threshold values of bush stability for the FPU chains with $N=12$
(see below).

\subsection{Diagrams of stability for one-dimensional bushes in the
FPU-$\alpha$ chain}
First of all, we would like to emphasize, that studying the stability of
bushes of normal modes, we can restrict ourselves to only one specimen from
the set of ``dynamical domains'' which represent equivalent bushes (see the
previous section).

Now we consider the stability diagrams for the bushes in the FPU-$\alpha$
chain.

\subsubsection{Bush B$[\hat{a}^2,\hat{i}]$}
The stability diagram for this bush (see Fig.~\ref{fig6}) has already been
discussed. Let us recall that in all our diagrams, for one-dimensional
bushes, the stability regions are white, while unstable regions are black.

\subsubsection{Bush B$[\hat{a}^3,\hat{i}]$}
In the configuration and modal spaces this bush looks, respectively, as
follows:
\begin{equation}
\vec{X}(t)=\left\{x(t),0,-x(t)|x(t),0,-x(t)|\cdots\right\}=
\nu(t)\left\{\varepsilon_1\vec{\psi}_{N/3}+\varepsilon_2\vec{\psi}_{2N/3}\right\}.
\label{eq200a}
\end{equation}

\begin{figure}
\centering
\setlength{\unitlength}{1mm}
\begin{picture}(130,45)(0,0)
\put(5,5){\includegraphics[width=120mm,height=36mm]{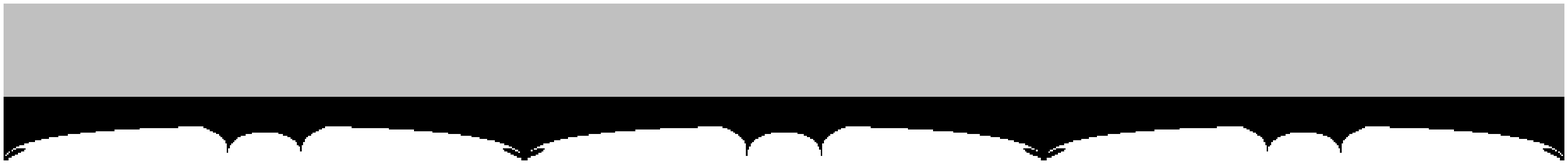}}
\drawgridA
\put(45,5){\line(0,1){36}}% 4
\put(85,5){\line(0,1){36}}% 8
\end{picture}
\caption{\label{fig7} Regions of stability (white color) of different modes
of the FPU-$\alpha$ chain, interacting parametrically with the
one-dimensional bush B$[a^3,i]$.}
\end{figure}

The stability diagram for the bush B$[\hat{a}^3,\hat{i}]$ is presented in
Fig.~\ref{fig7}. The domain depicted in grey color corresponds to the
disruption of the FPU-$\alpha$ chain\footnote{Obviously, this phenomenon is
impossible for the FPU-$\beta$ chain.} (atoms escape from the potential
wells in which the vibrational motion can occur). Thus, we must distinguish
between the stability of the chain itself and the stability of a given
\textit{vibrational} regime, i.e., of a bush of modes, in this chain.

Let us imagine a horizontal line in Fig.~\ref{fig7} corresponding to a
given value $A$ of the bush root mode. This line can partially pass through
the white and partially through the black regions of the considered
stability diagram. Actually, the parts of this line belonging to the black
(unstable) region represent the set of modes which are excited because of
the interactions with the bush root mode at the fixed level of its
amplitude ($A$). If $A$ is very small, our horizontal line $A=const$
intersects the \textit{narrow} unstable regions near the mode numbers $j=0,
\frac{N}{3}, \frac{2N}{3}$ and $N$.

Let us suppose that $N=12$ and $A<<1$. Then it is easy to see from
Fig.~\ref{fig7} that only two modes, corresponding to $j=\frac{N}{3}$ and $j=\frac{2N}{3}$,
are excited in our mechanical system\footnote{Note that the mode $\nu_0$
(and the equivalent to it mode $\nu_{12}$) must be excluded because of the
momentum conservation law~\cite{FPU1}.}. This fact is consistent with the
stability of the bush B$[\hat{a}^3,\hat{i}]$ for small values of its root
mode amplitude. Indeed, according to Eq.~(\ref{eq200a}), this bush contains
only two (conjugate) modes $\nu_{N/3}(t)$ and $\nu_{2N/3}(t)$ with the
certain relation between their amplitudes:
\begin{equation}
\nu_{N/3}(t)=\varepsilon_1\nu(t),\quad \nu_{2N/3}(t)=\varepsilon_2\nu(t)
\label{eq201a}
\end{equation}
Moreover, we can find the \textit{threshold of the stability loss} of the
bush B$[\hat{a}^3,\hat{i}]$ directly from the diagram in Fig.~\ref{fig7},
taking into account that all the possible modes for the case $N=12$ are
marked by vertical dashed lines. This threshold is the minimal vertical
distance from the horizontal coordinate axis to the black (unstable) region
for $j=1,2,3,5,6,7,9,10,11$. The stability thresholds for all
one-dimensional bushes in the FPU-$\alpha$ and FPU-$\beta$ chains for
$N=12$ are given in Table~\ref{table7}. From this table, we find that the
stability threshold for the bush B$[\hat{a}^3,\hat{i}]$ (the critical value
$x_c$ of the atomic displacements) is equal to $x_c=0.203$. The
corresponding critical values of the root mode amplitude ($A_c$) and the
energy ($E_c$) are equal to $A_c=0.166$, $E_c=0.047$.

Let us now increase the value $A$ of the root mode amplitude of the bush
B$[\hat{a}^3,\hat{i}]$. Such an increase leads to the broadening of the
unstable regions near the $j=0, \frac{N}{3}, \frac{2N}{3}, N$ which
intersects the line $A=const$ in Fig.~\ref{fig7}. This fact does not affect
the stability of the bush B$[\hat{a}^3,\hat{i}]$ for $N=12$, as long as
there are no modes $\nu_j$ with $j=1,2,\cdots,11$ (except the modes
$\nu_{N/3}$ and $\nu_{2N/3}$) which fall into the above-mentioned
broadening unstable intervals.

On the other hand, if $N\rightarrow\infty$ (continuum limit), the mode
density, i.e.\ the density of vertical dashed lines in Fig.~\ref{fig7},
increases and there appear narrow groups of modes near the values
$j=0,\frac{N}{3},\frac{2N}{3},N$ ($k=0,\frac{\pi}{3},\frac{2\pi}{3},\pi$).
This case corresponds to the onset of the modulation instability of the
considered bush of modes. Note that we then have a bush of \textit{narrow
wave packets} which gradually collapses on a long time scale\footnote{We
will discuss this phenomenon elsewhere.}. Thus, we conclude that the bush
B$[\hat{a}^3,\hat{i}]$, being stable for small number of particles in the
chain ($N$), for a fixed value of its root mode amplitude, becomes unstable
in the continuum limit $N\rightarrow\infty$.

\subsubsection{Bush B$[\hat{a}^4,\hat{a}\hat{i}]$} The one-dimensional bush
B$[\hat{a}^4,\hat{a}\hat{i}]$ can be written as follows
\begin{equation}
\vec{X}(t)=\left\{0,x(t),0,-x(t)|0,x(t),0,-x(t)|\cdots\right\}=\frac{\sqrt{2}}{2}\nu(t)\left[\vec{\psi}_{N/4}+\vec{\psi}_{3N/4}\right].
\label{eq150b}
\end{equation}
Thus, it represents a superposition of two conjugate modes
$\vec{\psi}_{N/4}$ and $\vec{\psi}_{3N/4}$ with equal amplitudes
\begin{equation}
\nu_{N/4}(t)=\nu_{3N/4}(t)=\frac{\sqrt{2}}{2}\nu(t).
\label{eq151}
\end{equation}

According to the Table~\ref{table7}, the threshold\footnote{This is the
upper boundary of the first stability domain of the considered bush.} of
the stability loss for the bush B$[\hat{a}^4,\hat{a}\hat{i}]$ in the
FPU-$\alpha$ chain, in contrast to the similar thresholds for all other
bushes, is equal to \textit{zero}. In particular, the loss stability
threshold (LST) for the same bush in the FPU-$\beta$ chain is equal to
$1.161$ (for the case $N=12$). Let us discuss this unusual case in more
detail.

\begin{figure}
\centering
\setlength{\unitlength}{1mm}
\begin{picture}(130,45)(0,0)
\put(5,5){\includegraphics[width=120mm,height=36mm]{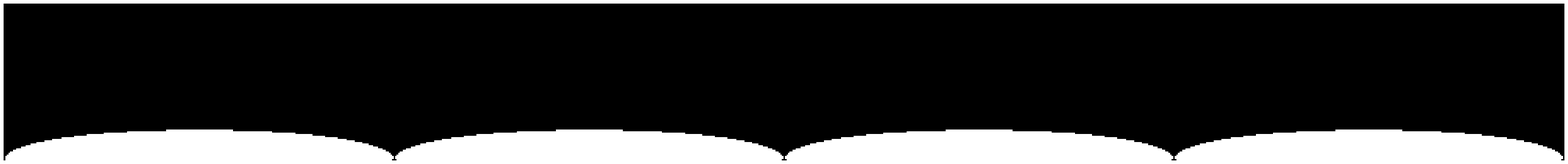}}
\drawgridA
\put(35,5){\line(0,1){36}}% 3
\put(95,5){\line(0,1){36}}% 9
\end{picture}
\caption{\label{fig8} Regions of stability (white color) of different modes
of the FPU-$\alpha$ chain, interacting parametrically with the
one-dimensional bush B$[a^4,ai]$.}
\end{figure}

In Fig.~\ref{fig8}, we see that besides the modes $\nu_{N/4}(t)$ and
$\nu_{3N/4}(t)$ contributing to B$[\hat{a}^4,\hat{a}\hat{i}]$, the mode
$\nu_{N/2}(t)$ is also excited for \textit{arbitrarily small} values of the
root mode amplitude of the considered bush. On the other hand, all the
results presented in Table~\ref{table7} and in Figs.~\ref{fig6}-\ref{fig14}
are obtained \textit{numerically} by the Floquet method and, partially, by
a straightforward integration of the dynamical equations of the FPU-chains.
Since numerical results are only approximate, it is desirable to prove
\textit{analytically} that the LST for the bush
B$[\hat{a}^4,\hat{a}\hat{i}]$ in the FPU-$\alpha$ chain is indeed equal to
zero exactly.

In the general case, the presence of three modes $\nu_{N/4}(t)$,
$\nu_{N/2}(t)$, $\nu_{3N/4}(t)$, in the dynamical regime, means that we
have the \textit{three-dimensional} bush B$[\hat{a}^4]$ (see~\cite{FPU2a})
with
\begin{equation}
\vec{X}(t)=\nu_{N/4}(t)\vec{\psi}_{N/4}+\nu_{N/2}(t)\vec{\psi}_{N/2}+\nu_{3N/4}(t)\vec{\psi}_{3N/4}.
\label{eq144}
\end{equation}
Therefore, one can suppose that the initially excited one-dimensional bush
B$[\hat{a}^4,\hat{a}\hat{i}]$ transforms spontaneously into the
three-dimensional bush B$[\hat{a}^4]$:
\begin{equation}
\B[\hat{a}^4,\hat{a}\hat{i}]\rightarrow\B[\hat{a}^4].
\label{eq160a}
\end{equation}
This transformation is accompanied by the lowering of the symmetry, because
of the loss of the invertion element $\hat{a}\hat{i}$. In the dynamical
picture of the above transformation, the new mode $\nu_{N/2}(t)$ appears
and the specific relation between the modes $\nu_{N/4}(t)$ and
$\nu_{3N/4}(t)$, presented in the bush B$[\hat{a}^4,\hat{a}\hat{i}]$, is
destroyed:
\[\nu_{N/4}(t)\neq\nu_{3N/4}(t).\]

Thus, we must now prove that the bush B$[\hat{a}^4,\hat{a}\hat{i}]$ loses
its stability with respect to the interaction exactly with the mode
$\nu_{N/2}$. So, let us exclude all the other modes of the FPU-$\alpha$
chain from our consideration and focus only on the
transformation~(\ref{eq160a}). Obviously, if we prove that the stability
threshold of the bush B$[\hat{a}^4,\hat{a}\hat{i}]$, for the
transformation~(\ref{eq160a}), is zero, then it is also zero with respect
to the interactions with \textit{all modes} of the chain.

Because of the above said, let us consider the dynamical equations of the
three-dimensional bush B$[\hat{a}^4]$ which can be written as follows
\begin{equation}
\begin{array}{l}
\ddot{\mu}_1+2\mu_1=-\frac{8\alpha}{N}\mu_1\mu_2,\\
\ddot{\mu}_2+4\mu_2=-\frac{4\alpha}{N}\left(\mu_1^2-\mu_3^2\right),\\
\ddot{\mu}_3+2\mu_3=\frac{8\alpha}{N}\mu_2\mu_3.
\end{array}
\label{eq155}
\end{equation}
Here we renamed the modes $\nu_{N/4}(t)$, $\nu_{N/2}(t)$ and
$\nu_{3N/4}(t)$ as $\mu_1(t)$, $\mu_2(t)$ and $\mu_3(t)$, respectively.

In the dynamical regime corresponding to the bush
B$[\hat{a}^4,\hat{a}\hat{i}]$ only one degree of freedom is excited (see
Eq.~(\ref{eq150b})). In our new notation it can be written as
\begin{equation}
\vec{X}(t)=\mu_1(t)\vec{\psi}_{N/4}+\mu_3(t)\vec{\psi}_{3N/4},
\label{eq170b}
\end{equation}
where $\mu_1(t)=\mu_3(t)$. Substituting $\mu_1(t)=\mu_3(t)$, $\mu_2(t)=0$
into Eqs.~(\ref{eq155}), we find that the first and third equations are
both reduced to the equation of the harmonic oscillator
\begin{equation}
\ddot{\mu}_1+2\mu_1=0,
\label{eq156}
\end{equation}
while both sides of the second equation turn out to be zero. Thus,
\begin{equation}
\mu_1^{(0)}(t)=A\cos(\sqrt{2}t+\phi_0)
\label{eq157}
\end{equation}
and we can linearize Eqs.~(\ref{eq155}) near the dynamical
regime~(\ref{eq157}). Note that one can assume $\phi_0=0$ in
Eq.~(\ref{eq157}), without loss of generality, since it is possible to
shift the time variable $t$ in the dynamical equations of the bush
B$[\hat{a}^4]$ by the appropriate value.

According to the general idea of linearization, let us suppose that
\begin{equation}
\begin{array}{lll}
\mu_1(t)&=&\mu_1^{(0)}(t)+\delta_1(t),\\
\mu_2(t)&=&\delta_2(t),\\
\mu_3(t)&=&\mu_1^{(0)}(t)+\delta_3(t).
\end{array}
\label{eq160b}
\end{equation}
Here $\delta_i$ ($i=1,2,3$) are small corrections to the original dynamical
regime~(\ref{eq170b}) $\left\{\mu_1^{(0)}(t),0,\mu_1^{(0)}(t)\right\}$
which is determined by the one-dimensional bush
B$[\hat{a}^4,\hat{a}\hat{i}]$. Let us now substitute~(\ref{eq160b}) into
Eqs.~(\ref{eq155}) and neglect all nonlinear in $\delta_i$ terms. Then we
have:
\begin{equation}
\begin{array}{lll}
\ddot{\delta}_1+2\delta_1&=&-8\alpha A\cos(\sqrt{2}t)\delta_2,\\
\ddot{\delta}_2+4\delta_2&=&-8\alpha A\cos(\sqrt{2}t)(\delta_1-\delta_3),\\
\ddot{\delta}_3+2\delta_3&=&8\alpha A\cos(\sqrt{2}t)\delta_2.
\end{array}
\label{eq162}
\end{equation}

Subtracting the third equation from the first equation and introducing new
variables $\gamma_1(t)=\delta_1(t)-\delta_3(t)$, $\gamma_2(t)=\delta_2(t)$,
we obtain finally
\begin{equation}
\begin{array}{lcl}
\ddot{\gamma}_1+2\gamma_1&=&-16\alpha A\cos(\sqrt{2}t)\gamma_2,\\
\ddot{\gamma}_2+4\gamma_2&=&-8\alpha A\cos(\sqrt{2}t)\gamma_1.
\end{array}
\label{eq163}
\end{equation}
Namely for these equations, we must find analytically the nontrivial
solution ($\gamma_1(t)\neq 0$, $\gamma_2(t)\neq 0$)

Note that the same system can be obtained from the linearized equations for
investigation of the stability of the bush B$[\hat{a}^4,\hat{a}^3\hat{i}]$
which is presented in Table~\ref{table5} for the FPU-$\alpha$ chain.
Indeed, let us assume $j=0$ in the last system from this table, keeping in
mind the relation $k=\frac{\pi j}{N}$. Then we have
\begin{equation}
\begin{array}{lllll}
\ddot{\nu}_0&&&=&0,\\
\ddot{\nu}_{N/4}&+&\omega_{N/4}^2\nu_{N/4}&=&-\frac{4\sqrt{2}\alpha}{\sqrt{N}}\mu(t)\nu_{N/2}(t),\\
\ddot{\nu}_{N/2}&+&\omega_{N/2}^2\nu_{N/2}&=&-\frac{4\sqrt{2}\alpha}{\sqrt{N}}\mu(t)\left[\nu_{N/4}(t)+\nu_{3N/4}(t)\right],\\
\ddot{\nu}_{3N/4}&+&\omega_{3N/4}^2\nu_{3N/4}&=&-\frac{4\sqrt{2}\alpha}{\sqrt{N}}\mu(t)\nu_{N/2}(t),
\end{array}
\label{eq164}
\end{equation}
where $\omega_j^2=4\sin^2(\frac{\pi j}{N})$, while
$\mu(t)=A\cos(\sqrt{2}t)$ is the solution to the equation
$\ddot{\mu}+2\mu=0$. Summing the second and forth equations of the
system~(\ref{eq164}) leads us again to Eqs.~(\ref{eq163}), where
$\gamma_1(t)=\nu_{N/4}(t)+\nu_{3N/4}(t)$, $\gamma_2(t)=\nu_{N/2}(t)$.

Thus, the analysis of the stability loss by the bush
B$[\hat{a}^4,\hat{a}\hat{i}]$ is reduced to the problem of parametric
excitation of the nontrivial solution to Eqs.~(\ref{eq163}). On the other
hand, it is well known that the onset of unstable motion for the Mathieu
equation can be investigated with the aid of different asymptotic methods
of nonlinear mechanics (averaging method, multiscale method, etc.).
In~\cite{FPU2b}, we use one of such a type methods, namely, the
normalization procedure based on the Poincare-Dulac theorem, for studying
the parametric excitation in the system~(\ref{eq163}). Using this
procedure, we remove, step by step, the so-called \textit{nonresonance}
terms of a given order to higher orders in the small parameter (in our
case, this is the root mode amplitude $A$) with the aid of the appropriate
nonlinear transformations of the dynamical variables $\gamma_i(t)$. In such
a way, the normalized system of differential equations with respect to new
variables $\tilde{\gamma}_i(t)$ is obtained and, as a rule, this system
turns out to be simpler than the original system.

In~\cite{FPU2b}, it was shown that the normalized equations up to $O(A^3)$,
obtained from Eqs.~(\ref{eq163}), can be solved exactly. The corresponding
solution for $\tilde{\gamma}_1(t)$ contains the growing exponent
$\exp(\sqrt{2} \alpha^2 A^2 t)$ which increases from zero for the
\textit{arbitrary small} value of the root mode amplitude $A$. For the old
variable $\gamma_1(t)$ this exponent, being decomposed into the power
series with respect to $A$, produces the unremovable secular term of the
form\footnote{The full approximate analytical solution to
Eqs.~(\ref{eq163}) represents the power series for $\gamma_1(t)$ and
$\gamma_2(t)$ with respect to the small parameter $A$. Since this solution
is rather cumbersome, we do not present it here.}
\[\alpha^2A^2\sqrt{2}t\sin(\sqrt{2}t+\delta).\]

From the above said, it is clear that the parametric excitation in the
system~(\ref{eq163}) takes place for the arbitrary small value of the root
mode amplitude $A$ and, therefore, the bush B$[\hat{a}^4,\hat{a}\hat{i}]$
possesses the stability domain of zero size, as it has already been
concluded directly from Fig.~\ref{fig8}.

\subsection{Diagrams of stability for one-dimensional bushes in the
FPU-$\beta$ chain}
\subsubsection{Bush B$[\hat{a}^2,\hat{i}]$}

\begin{figure}
\centering
\setlength{\unitlength}{1mm}
\begin{picture}(130,45)(0,0)
\put(5,5){\includegraphics[width=120mm,height=36mm]{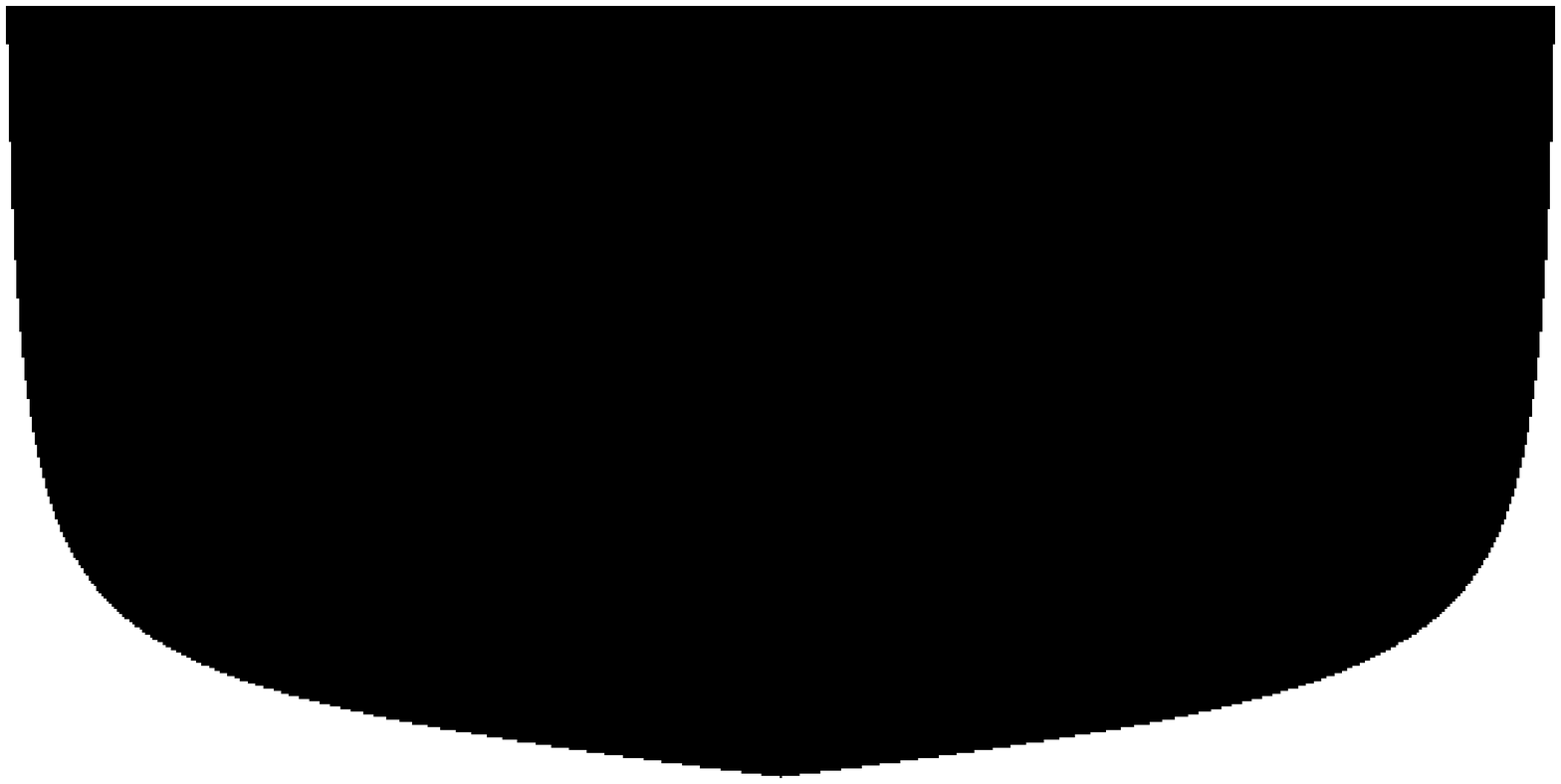}}
\drawgrid
\put(65,5){\line(0,1){36}}% 6
\end{picture}
\caption{\label{fig9} Regions of stability (white color) of different modes
of the FPU-$\beta$ chain, interacting parametrically with the
one-dimensional bush B$[a^2,i]$.}
\end{figure}

The stability diagram corresponding to the bush B$[\hat{a}^2,\hat{i}]$ in
the FPU-$\beta$ chain is presented in Fig.~\ref{fig9}. From this diagram,
one can see that for any \textit{finite} $N$ and for sufficiently small
root mode amplitude $A$, there exists, in our mechanical systems, only one
mode $\nu_{N/2}(t)$, which is the root (and single!) mode of the bush
B$[\hat{a}^2,\hat{i}]$. On the other hand, if we \textit{fix} arbitrary
small value of $A$ and then begin to increase the number of particles $N$,
such $N$ can be always found for which some other modes, close to the mode
$\nu_{N/2}(t)$, turn out to be excited (see the center of the
diagram~\ref{fig9}). In other words, the threshold of the loss of stability
of the considered bush decreases with increasing $N$. Thus, the bush
B$[\hat{a}^2,\hat{i}]$ becomes unstable in the continuum limit
$N\rightarrow\infty$.

Another interesting fact can be revealed with the aid of the
diagram~\ref{fig9}. Namely, there are certain intervals of mode numbers on
r.h.s.\ and l.h.s.\ of this diagram which cannot be excited because of the
interactions with the mode $\nu_{N/2}(t)$. Note that the both above
mentioned facts were found\footnote{The authors of this paper used the
direct integration of the dynamical equations of the FPU-$\beta$ chain in
the modal space, while we use the Floquet method.} and discussed, in rather
different terms, in the paper~\cite{PoggiRuffo}.

\subsubsection{Bush B$[\hat{a}^3,\hat{i}]$}

\begin{figure}
\centering
\setlength{\unitlength}{1mm}
\begin{picture}(130,45)(0,0)
\put(5,5){\includegraphics[width=120mm,height=36mm]{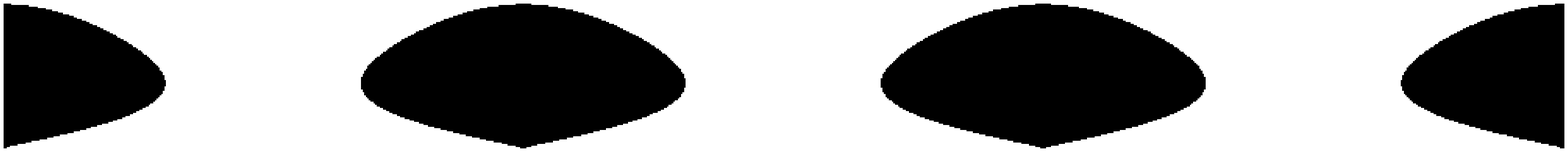}}
\drawgridB
\put(45,5){\line(0,1){36}}% 4
\put(85,5){\line(0,1){36}}% 8
\put( 5, 8.9){\dashbox{1}(120,0){}}% A1
\put(20, 9.1){$A_1$}
\put( 5,16.7){\dashbox{1}(120,0){}}% A2
\put(20,16.9){$A_2$}
\put( 5,21.5){\dashbox{1}(120,0){}}% A3
\put(20,21.7){$A_3$}
\end{picture}
\caption{\label{fig10} Regions of stability (white color) of different
modes of the FPU-$\beta$ chain, interacting parametrically with the
one-dimensional bush B$[a^3,i]$.}
\end{figure}

The form of the bush B$[\hat{a}^3,\hat{i}]$ in the configuration and modal
spaces was given in Eq.~(\ref{eq200a}) The corresponding stability diagram
is shown in Fig.~\ref{fig10}. From this diagram, we find that, for small
amplitudes, the bush B$[\hat{a}^3,\hat{i}]$ is not stable in the continuum
limit $N\rightarrow\infty$, but can be stable for finite $N$. For example,
for $N=12$ the stability takes place up to the value $A_1$ of the root
mode\footnote{This mode is a certain superposition,
$\varepsilon_1\vec{\psi}_{N/3}+\varepsilon_2\vec{\psi}_{2N/3}$, of the
conjugate modes $\vec{\psi}_{N/3}$ and $\vec{\psi}_{2N/3}$.} amplitude~$A$.
Note that three horizontal dashed lines are depicted in Fig.~\ref{fig10}:
the lower line is described by the equation $A=A_1$, while the middle and
upper lines correspond to the equations $A=A_2$ and $A=A_3$, respectively.

For $A_1<A<A_2$ all the modes, with the only exception of the modes with
$j=\frac{2N}{12}=\frac{N}{6}$, $j=\frac{6N}{12}=\frac{N}{2}$ and
$j=\frac{10N}{12}=\frac{5N}{6}$, are excited.\footnote{Note that only the
results of a \textit{linear} stability analysis are presented in our
stability diagrams: we find those modes, which are excited because of the
direct parametric excitation with the root mode of the considered bush. On
the other hand, if some new modes appear in the mechanical system, we must
also take into account their interactions with other modes (it is the
problem of nonlinear stability analysis).} Therefore, for $N=12$, the bush
with trivial symmetry is excited. Indeed, the excitation of the mode with
$j=\frac{N}{12}$ can lead, in principle, to excitation of all the modes
with $j=\frac{kN}{12}$ ($k=1,2,\cdots,11$).

It is very interesting that there exists also the \textit{upper} value
($A_3$) for the loss of stability of the bush B$[\hat{a}^3,\hat{i}]$.
Indeed, for $A>A_3$ (see Fig.~\ref{fig10}) this bush turns out to be stable
for \textit{all} values of the number ($N$) of particles in the chain.
Thus, the bush B$[\hat{a}^3,\hat{i}]$ becomes stable even in the continuum
limit, for such values of the root mode amplitude $A$.

\subsubsection{Bush B$[\hat{a}^4,\hat{a}\hat{i}]$}
\begin{figure}
\centering
\setlength{\unitlength}{1mm}
\begin{picture}(130,45)(0,0)
\put(5,5){\includegraphics[width=120mm,height=36mm]{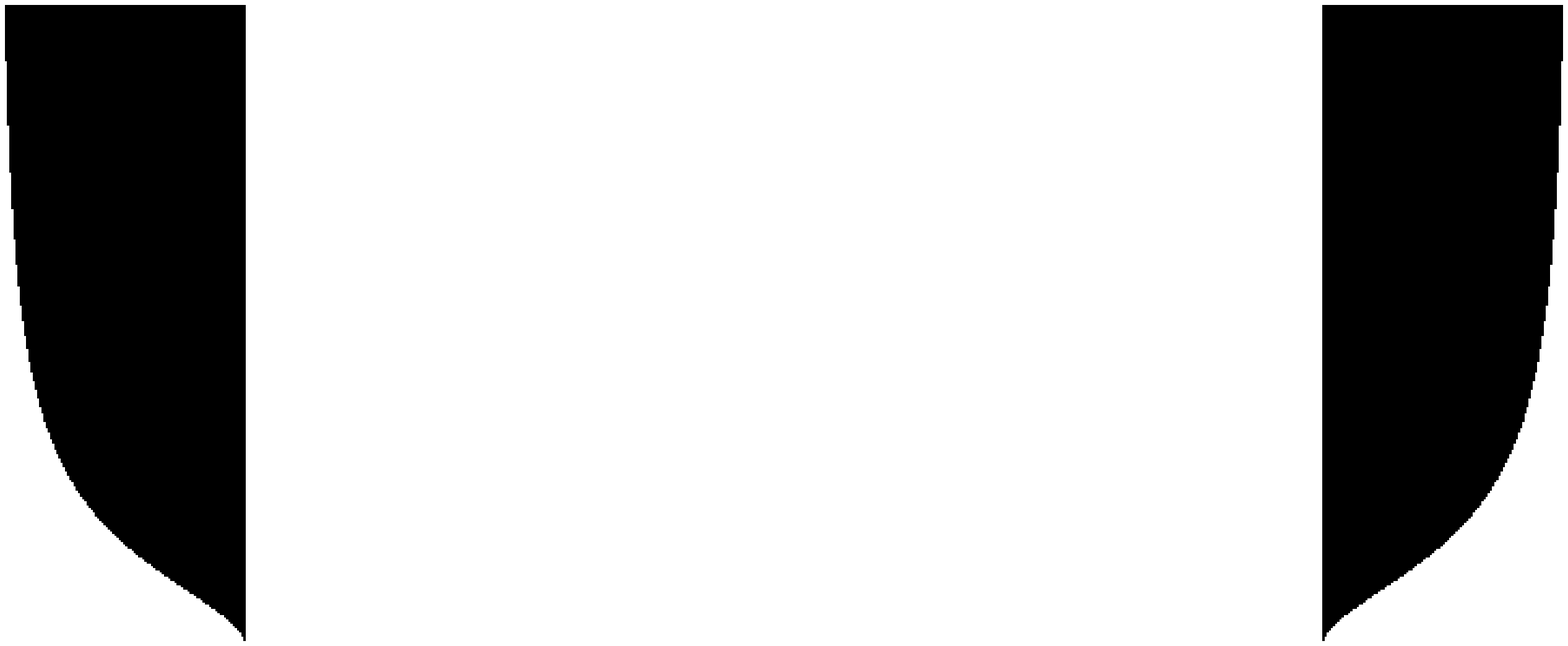}}
\drawgrid
\put(35,5){\line(0,1){36}}% 3
\put(95,5){\line(0,1){36}}% 9
\put( 5,15.0){\dashbox{1}(120,0){}}% A1
\put(20,15.2){$A_1$}
\end{picture}
\caption{\label{fig12} Regions of stability (white color) of different
modes of the FPU-$\beta$ chain, interacting parametrically with the
one-dimensional bush B$[a^4,ai]$.}
\end{figure}

\begin{figure}
\centering
\setlength{\unitlength}{1mm}
\begin{picture}(130,45)(0,0)
\put(5,5){\includegraphics[width=120mm,height=36mm]{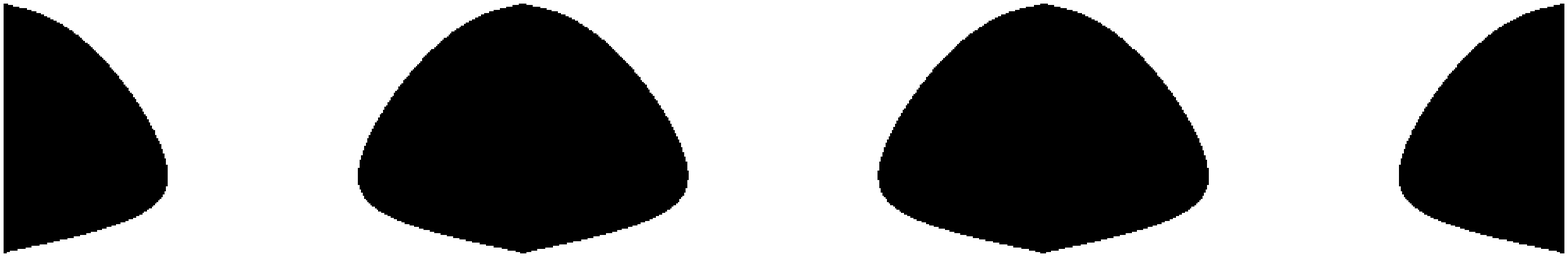}}
\drawgrid
\put(45,5){\line(0,1){36}}% 4
\put(85,5){\line(0,1){36}}% 8
\put( 5, 7.8){\dashbox{1}(120,0){}}% A1
\put(20, 8.0){$A_1$}
\put( 5,17.0){\dashbox{1}(120,0){}}% A2
\put(20,17.2){$A_2$}
\put( 5,24.2){\dashbox{1}(120,0){}}% A3
\put(20,24.4){$A_3$}
\end{picture}
\caption{\label{fig11} Regions of stability (white color) of different
modes of the FPU-$\beta$ chain, interacting parametrically with the
one-dimensional bush B$[a^3,iu]$.}
\end{figure}

The description of the bush B$[\hat{a}^4,\hat{a}\hat{i}]$ in the
configuration and modal spaces has already been given by Eq.~(\ref{eq150b}):
\[\vec{X}(t)=\left\{0,x,0,-x|0,x,0,-x|\dots\right\}=\varepsilon_3\nu(t)\left[\vec{\psi}_{N/4}+\vec{\psi}_{3N/4}\right].\]
The stability diagram for this bush for the FPU-$\beta$ chain is
represented in Fig.~\ref{fig12}. From this diagram, one can see that, for
any finite $N$, the bush B$[\hat{a}^4,\hat{a}\hat{i}]$ is stable for
sufficiently small root mode amplitudes $A$. For example, for $N=12$, this
bush loses its stability only for $A>A_1$ (see Fig.~\ref{fig12}). For very
large $N$, the bush B$[\hat{a}^4,\hat{a}\hat{i}]$ practically also remains
stable for sufficiently small $A$ because the intervals of the numbers of
the excited modes turn out to be very narrow (these intervals are situated
near $j=\frac{N}{4}$ and $j=\frac{3N}{4}$). In the continuum limit
$N\rightarrow\infty$, rigorously speaking, the bush
B$[\hat{a}^4,\hat{a}\hat{i}]$ appears to be unstable, but its instability
is weak.

We have just discussed the stability of all one-dimensional bushes for the
FPU-$\beta$ chain which are induced by the dihedral symmetry group. Now let
us consider the stability of three ``additional'' one-dimensional bushes
whose existence is brought about by the evenness of the potential of the
FPU-$\beta$ chain.

\subsubsection{Bush B$[\hat{a}^3,\hat{i}\hat{u}]$}
The description of this bush in configuration and modal spaces looks as
follows:
\begin{equation}
\vec{X}(t)=\left\{x,-2x,x|x,-2x,x|\dots\right\}=\nu(t)\left[\varepsilon_2\vec{\psi}_{N/3}-\varepsilon_1\vec{\psi}_{2N/3}\right].
\label{eq200b}
\end{equation}
The stability diagram of the bush B$[\hat{a}^3,\hat{i}\hat{u}]$ is given in
Fig.~\ref{fig11}. It seems similar to that of the above discussed bush
B$[\hat{a}^3,\hat{i}]$ (see Fig.~\ref{fig10}). Therefore, the discussion of
the stability properties of the bush B$[\hat{a}^3,\hat{i}\hat{u}]$ can be
practically reduced to that of the bush B$[\hat{a}^3,\hat{i}]$.

\subsubsection{Bush B$[\hat{a}^4,\hat{i}\hat{u}]$}
\begin{figure}
\centering
\setlength{\unitlength}{1mm}
\begin{picture}(130,45)(0,0)
\put(5,5){\includegraphics[width=120mm,height=36mm]{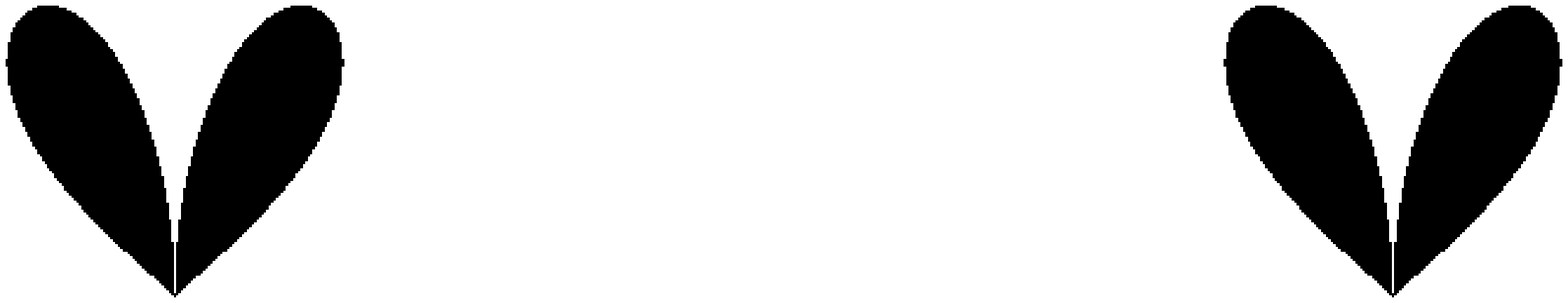}}
\drawgridB
\put(35,5){\line(0,1){36}}% 3
\put(95,5){\line(0,1){36}}% 9
\put( 5,26.0){\dashbox{1}(120,0){}}% A1
\put(20,26.2){$A_1$}
\end{picture}
\caption{\label{fig13} Regions of stability (white color) of different
modes of the FPU-$\beta$ chain, interacting parametrically with the
one-dimensional bush B$[a^4,iu]$.}
\end{figure}

\begin{figure}
\centering
\setlength{\unitlength}{1mm}
\begin{picture}(130,45)(0,0)
\put(5,5){\includegraphics[width=120mm,height=36mm]{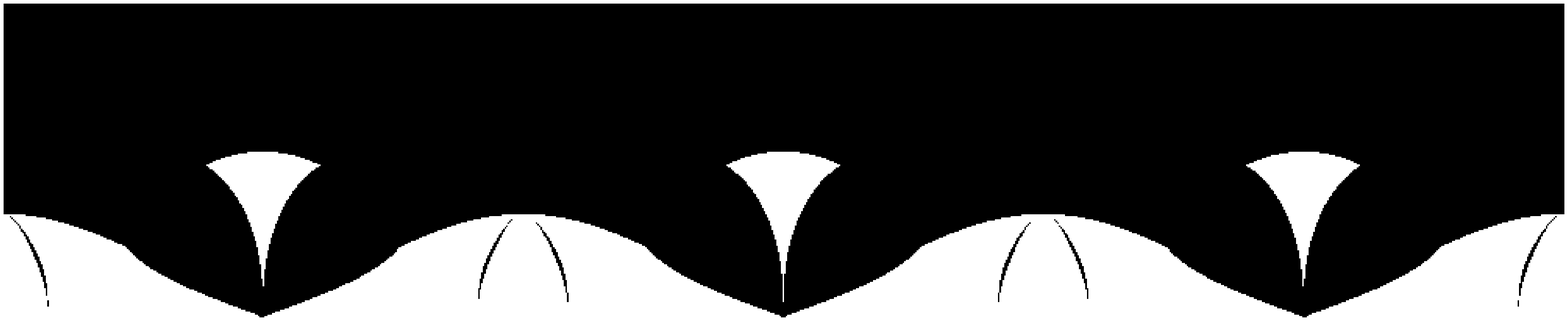}}
\drawgridB
\put( 25,5){\line(0,1){36}}% 2
\put(105,5){\line(0,1){36}}% 10
\end{picture}
\caption{\label{fig14} Regions of stability (white color) of different
modes of the FPU-$\beta$ chain, interacting parametrically with the
one-dimensional bush B$[a^6,ai,a^3u]$.}
\end{figure}

This bush can be described by the equation
\begin{equation}
\vec{X}(t)=\left\{x,-x,-x,x|x,-x,-x,x|\dots\right\}=\nu(t)\vec{\psi}_{N/4}.
\label{eq201b}
\end{equation}
The corresponding stability diagram is represented in Fig.~\ref{fig13} and
looks as ``rabbit ears''. For root mode amplitudes $A>A_1$ (see
Fig.~\ref{fig13}) this bush is stable for \textit{arbitrary} $N$. It is
interesting that for the case $N=12$ there are no excited modes, certainly,
besides the mode $\vec{\psi}_{N/4}$ (or the mode $\vec{\psi}_{3N/4}$ for
the equivalent bush B$[\hat{a}^4,\hat{a}^2\hat{i}\hat{u}]$ which looks in
the modal space as $\vec{X}(t)=\nu(t)\vec{\psi}_{3N/4}$). This is in
agreement with Table~\ref{table7} from which we find that this bush is
stable, \textit{at least}, up to the root mode amplitude $A<20$.

\subsubsection{Bush B$[\hat{a}^6,\hat{a}\hat{i},\hat{a}^3\hat{u}]$}
This bush can be described as follows:
\begin{equation}
\vec{X}(t)=\left\{0,x,x,0,-x,-x|0,x,x,0,-x,-x|\dots\right\}=\nu(t)\left[\varepsilon_1\vec{\psi}_{N/6}+\varepsilon_2\vec{\psi}_{5N/6}\right].
\label{eq202}
\end{equation}

One of the distinguishing features of the stability diagram for this bush
(see Fig.~\ref{fig14}) is the presence of three white regions in the
``black sea'' of instability. They correspond to the second stability zones
for some modes whose numbers fall into these white regions. Another
interesting feature of the stability diagram for the considered bush is the
very narrow unstable intervals in the ``white sea'' of stability (see the
black lines which resemble parabolas). We did not examine these unstable
parabolas in detail, because they seem to be not very essential for
consideration of the stability of the bush
B$[\hat{a}^6,\hat{a}\hat{i},\hat{a}^3\hat{u}]$. Indeed, they represent too
narrow regions of unstable movements.

\subsection{About numerical values of root mode amplitudes for the FPU
chains}
The FPU models of $\alpha$ and $\beta$ type, considered as exact mechanical
systems, do not contain the interparticle distance $a$. As a consequence,
one cannot estimate the reasonable values of the root mode amplitudes for
the above discussed bushes. This is an important problem because atomic
displacements in real crystals cannot, as a rule, exceed $10-15\%$ of the
equilibrium interparticle distance.

On the other hand, we can consider FPU system as a certain approximation to
more realistic models of one-dimensional crystals. Let us consider a
nonlinear monoatomic chain with the Lennard-Jones potential $U(r)$
describing the interaction between the nearest particles:
\begin{equation}
U(r)=\varepsilon\left[\left(\frac{\sigma}{r}\right)^{12}-\left(\frac{\sigma}{r}\right)^6\right].
\label{eq300}
\end{equation}
Here, $r$ is the interparticle distance, while $\varepsilon$ and $\sigma$
are certain constants.

Let us expand the force $f(r)$ acting on one particle, in Taylor series
near the equilibrium interparticle distance $a=\sigma\sqrt[6]{2}$ up to the
terms of the second order:
\begin{equation}
f(r)=-\frac{9\sqrt[3]{4}}{\sigma^2}(r-a)+\frac{189\sqrt{2}}{2\sigma^3}(r-a)^2.
\label{eq301}
\end{equation}
Then we can write the Newton equations of motion in the form
\begin{equation}
m\ddot{x}_i=\frac{9\sqrt[3]{4}\varepsilon}{\sigma^2}\left(-2x_i+x_{i+1}+x_{i-1}\right)+
\frac{189\sqrt{2}\varepsilon}{2\sigma^3}\left[\left(x_{i+1}-x_i\right)^2-\left(x_i-x_{i-1}\right)^2\right],
\label{eq302}
\end{equation}
where $m$ is the mass of one particle.

Introducing new dimensionless time and space variables
\begin{equation}
t'=\sqrt{\frac{9\sqrt[3]{4}\varepsilon}{m\sigma^2}}t,\quad
x'_n=\frac{21\sqrt[3]{2}}{a}x_n,
\label{eq303}
\end{equation}
we obtain the following equations:
\begin{equation}
\ddot{x}'_i=(-2x'_i+x'_{i+1}+x'_{i-1})+[(x'_{i+1}-x'_i)^2-(x'_i-x'_{i-1})^2].
\label{eq304}
\end{equation}
These equations precisely correspond to the FPU-$\alpha$ model considered
in the present paper.

On the other hand, the dimensionless variables $x'_i$, according to
Eq.~(\ref{eq303}), are proportional to the ratio of the real atomic
displacements $x_i$ and the equilibrium interparticle distance $a$. As it
has already been noted, for real crystal, $x_i/a$ must not exceed
$10-15\%$. The dimensionless displacements $x'_i$ from the FPU-$\alpha$
model~(\ref{eq304}) can be considerably larger than this value, because of
the coefficient $21\sqrt[3]{2}\approx 29.7$ before $x_i/a$. For example, if
$x_i\sim 0.1a$, then $x'_i\sim 3$.

Let us now recall that, studying the bush stability, we use such
normalization of the normal modes, that their amplitudes are of the same
order as the atomic displacements corresponding to them. Thus, the root
mode amplitudes $A$ at about $3$ seem to be rather reasonable. It is
precisely because of this fact that we depict the bush stability diagram
up to $A$ of such an order of magnitude.

\section{Stability of two-dimensional bushes for the FPU chains}
In the previous section, we considered the stability of one-dimensional
bushes with respect to increasing their root mode amplitudes. Such an
increasing leads to strengthening of the parametric interactions of the
root mode of a given bush with all the other modes which do not belong to
this bush and, finally, to a loss of its stability. Actually, the threshold
of the loss of stability of any one-dimensional bush depends only on its
\textit{total energy} (the energy of the initial excitation), but not on
the distribution of this energy between the kinetic and potential
counterparts~\cite{FPU2b}. In two last columns of Table~\ref{table7} we
give the energy thresholds of the loss of stability for all one-dimensional
bushes in the FPU chains. In contrast to one-dimensional bushes, the
stability of the \textit{multi-dimensional} bushes depends not only on the
energy of the initial excitation, but also on the complete set of the
initial conditions. Let us consider this problem in more detail.

The four-dimensional phase space corresponds to the two-dimensional bushes
considered below in the present paper. In Fig.~\ref{fig15}-\ref{fig24}, we
depict only certain planar sections of the four-dimensional stability
domains, aiming to show how nontrivial the boundary of these domains can
be.

The complete set of initial conditions for the dynamical equations of a
given two-dimensional bush determines the concrete values of
$\nu_1(0),\dot{\nu}_1(0),\nu_2(0),\dot{\nu}_2(0)$, where $\nu_i(t)$ and
$\dot{\nu}_i(t)$ ($i=1,2$) are the two modes of the considered bush and
their velocities, respectively. We depict the bush stability diagrams (see
Fig.~\ref{fig15}-\ref{fig31}) in certain planes which are chosen by fixing
two of the four above-mentioned initial values. Most frequently, we specify
$\dot{\nu}_1(0)=0,\dot{\nu}_2(0)=0$ and change $\nu_1(0),\nu_2(0)$ in some
intervals near their zero values. Therefore, the initial kinetic energy
assumed to be equal to zero and the dependence of bush stability on the
initial mode amplitudes $\nu_1(0)$ and $\nu_2(0)$ is studied.

The diagrams discussed in the previous section allow us to analyze
stability of one-dimensional bushes for the FPU chains with arbitrary
number ($N$) of particles. Unfortunately, we cannot obtain similar diagrams
for two-dimensional bushes. Stability diagrams considered in this section
depend on the number of particles and, therefore, they must be constructed
individually for each specific $N$. Because of this reason, we consider
stability of 2D bushes only for fixed $N$ (for the most diagrams $N=12$).
As a consequence, we do not discuss the stability properties of the 2D
bushes with translation symmetry $[\hat{a}^5]$, $[\hat{a}^8]$ and
$[\hat{a}^{10}]$ (the complete list of 2D bushes for the FPU chains can be
found in Appendix).

\begin{figure}
\centering
\setlength{\unitlength}{1mm}
\includegraphics[width=90mm]{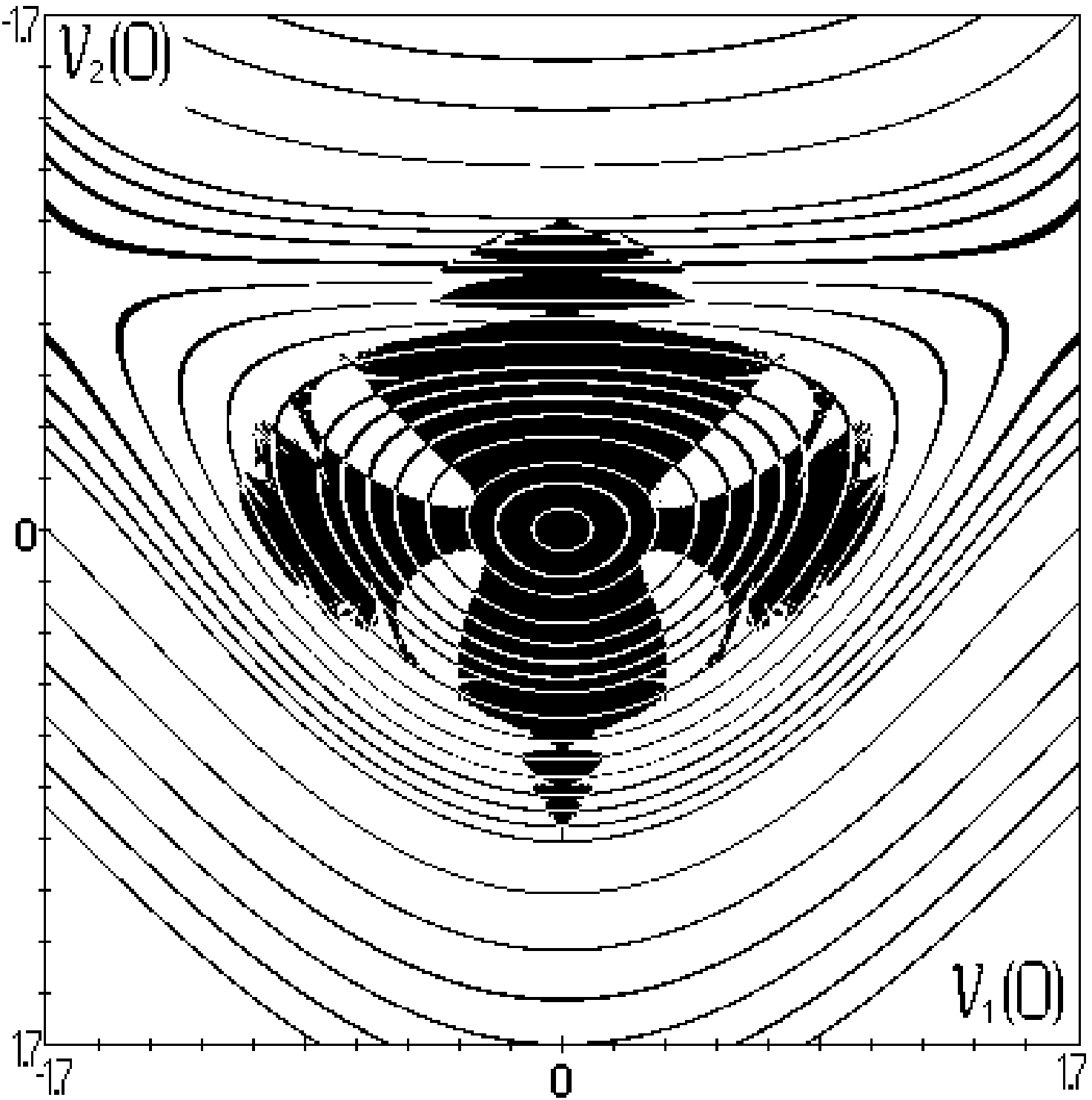}
\caption{\label{fig77} Stability diagram of the bush B$[a^4,i]$ in the
FPU-$\alpha$ chain with $N=12$ for the case $\dot\nu_1(0)=0$,
$\dot\nu_2(0)=0$.}
\end{figure}

Firstly, let us consider the stability of the two-dimensional bush
B$[\hat{a}^4,\hat{i}]$ in the FPU-$\alpha$ chain with $N=12$ particles. The
stability domain for this bush, in the plane of the initial values of its
both modes $[\nu_1(0),\nu_2(0)]$, is presented in \textit{black}
color\footnote{In stability diagrams for one-dimensional bushes, we use
white color for stability regions and black color for instability regions.
In contrast, for two-dimensional bushes, white color corresponds to
instability domains, while black color is used for stability domains. It
seems to us that this manner of depicting stability diagrams turns out to
be more expressive.} in the center of Fig.~\ref{fig77}. This domain,
resembling a beetle, is depicted on the background of the energy level
lines. The bush B$[\hat{a}^4,\hat{i}]$ loses its stability when we cross
the boundary of the black region in any direction. Moreover, it can be
shown that the considered bush transforms into \textit{different} bushes of
larger dimensions at different points of the boundary. Which bushes appear
near the various parts of the boundary of the stability domain of the
bush~B$[\hat{a}^4,\hat{i}]$ will be discussed elsewhere. Now we merely want
to emphasize that the loss of stability of this bush is not determined only
by the energy of the initial excitation (the boundary of stability domain
does not coincide with any energy level line!), but depends on the relation
between the values of its both modes $[\nu_1(0),\nu_2(0)]$, as well. Thus,
the stability domain of the bush B$[\hat{a}^4,\hat{i}]$  is strongly
anisotropic: the linear size of this domain in different directions, from
zero point $\nu_1(0)=0,\nu_2(0)=0$, turns out to be different.

The stability diagram of the bush B$[\hat{a}^4,\hat{i}]$ in the
FPU-$\alpha$ chain, depicted in Fig.~\ref{fig77}, was obtained by the
direct numerical experiment for the case $\dot\nu_1(0)=0,\dot\nu_2(0)=0$.
Therefore, initial kinetic energy was equal to zero and the total
excitation energy was purely potential.

\begin{figure}
\centering
\setlength{\unitlength}{1mm}
\includegraphics[width=90mm]{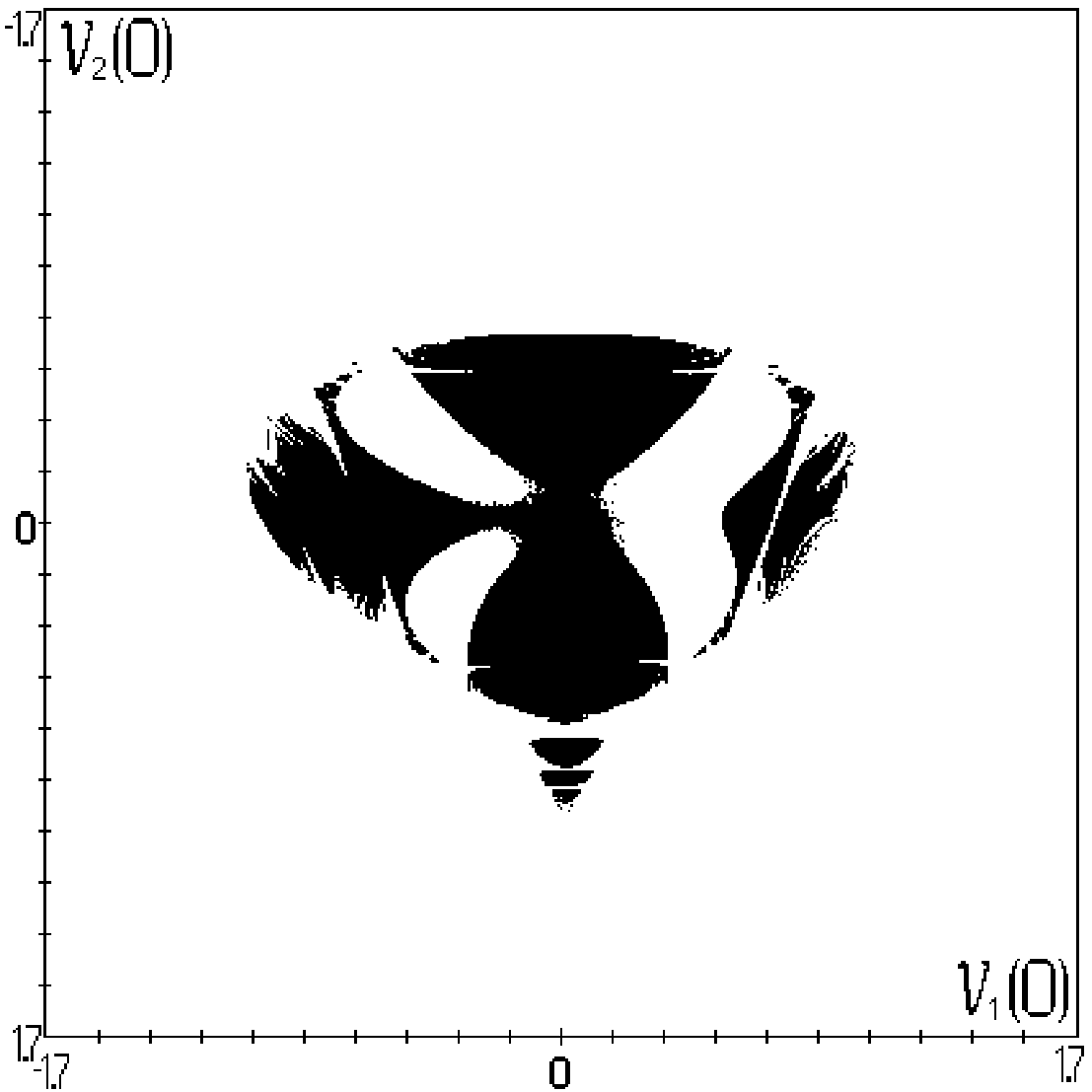}
\caption{\label{fig78} Stability diagram of the bush B$[a^4,i]$ in the
FPU-$\alpha$ chain with $N=12$ for the case $\dot\nu_1(0)=0.2$,
$\dot\nu_2(0)=0.1$.}
\end{figure}

\begin{figure}
\centering
\setlength{\unitlength}{1mm}
\includegraphics[width=60mm]{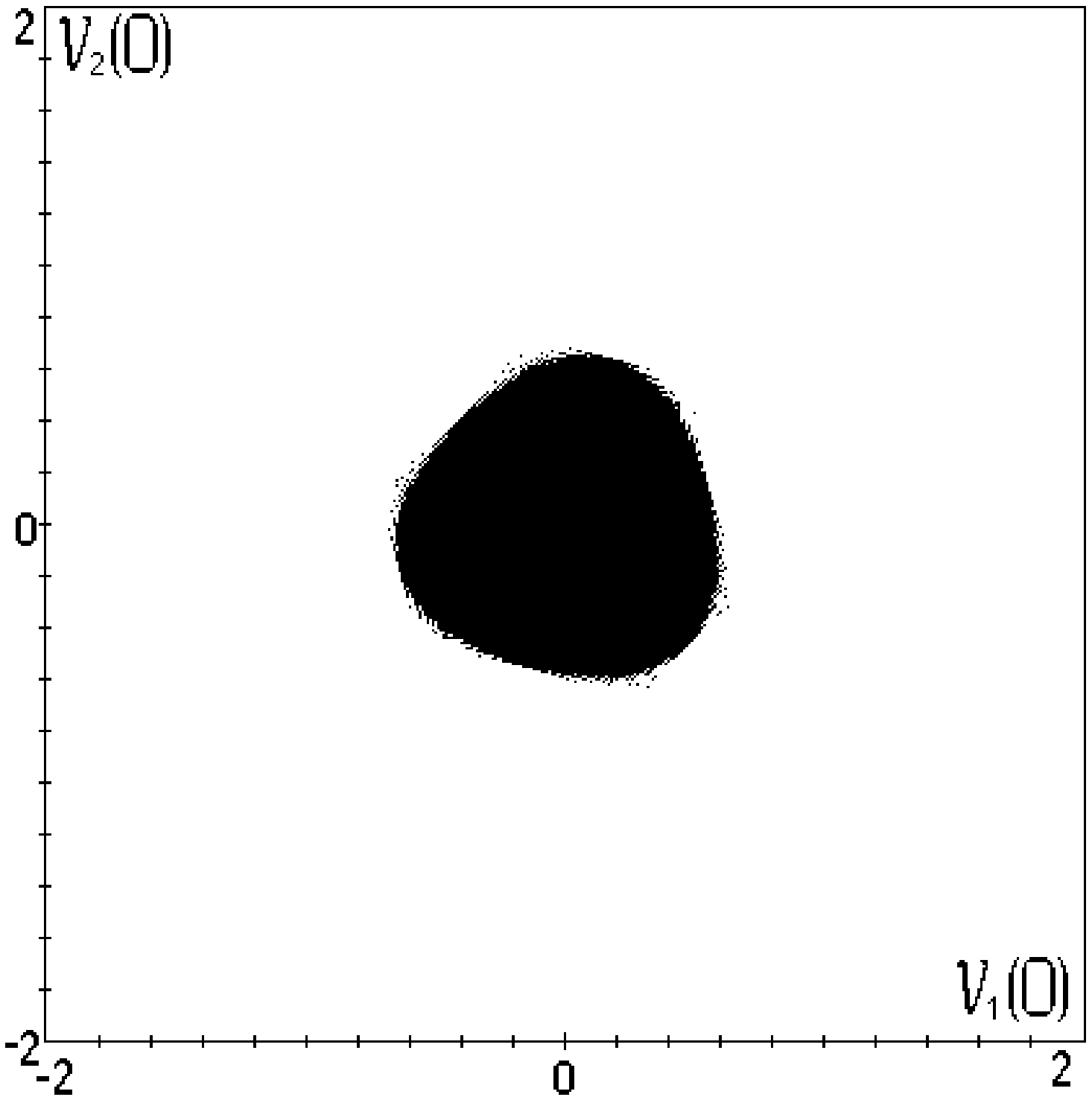}
\caption{\label{fig15} Stability diagram for the bush B$[a^3]$ in the
FPU-$\alpha$ chain.}
\end{figure}

\begin{figure}
\centering
\setlength{\unitlength}{1mm}
\includegraphics[width=60mm]{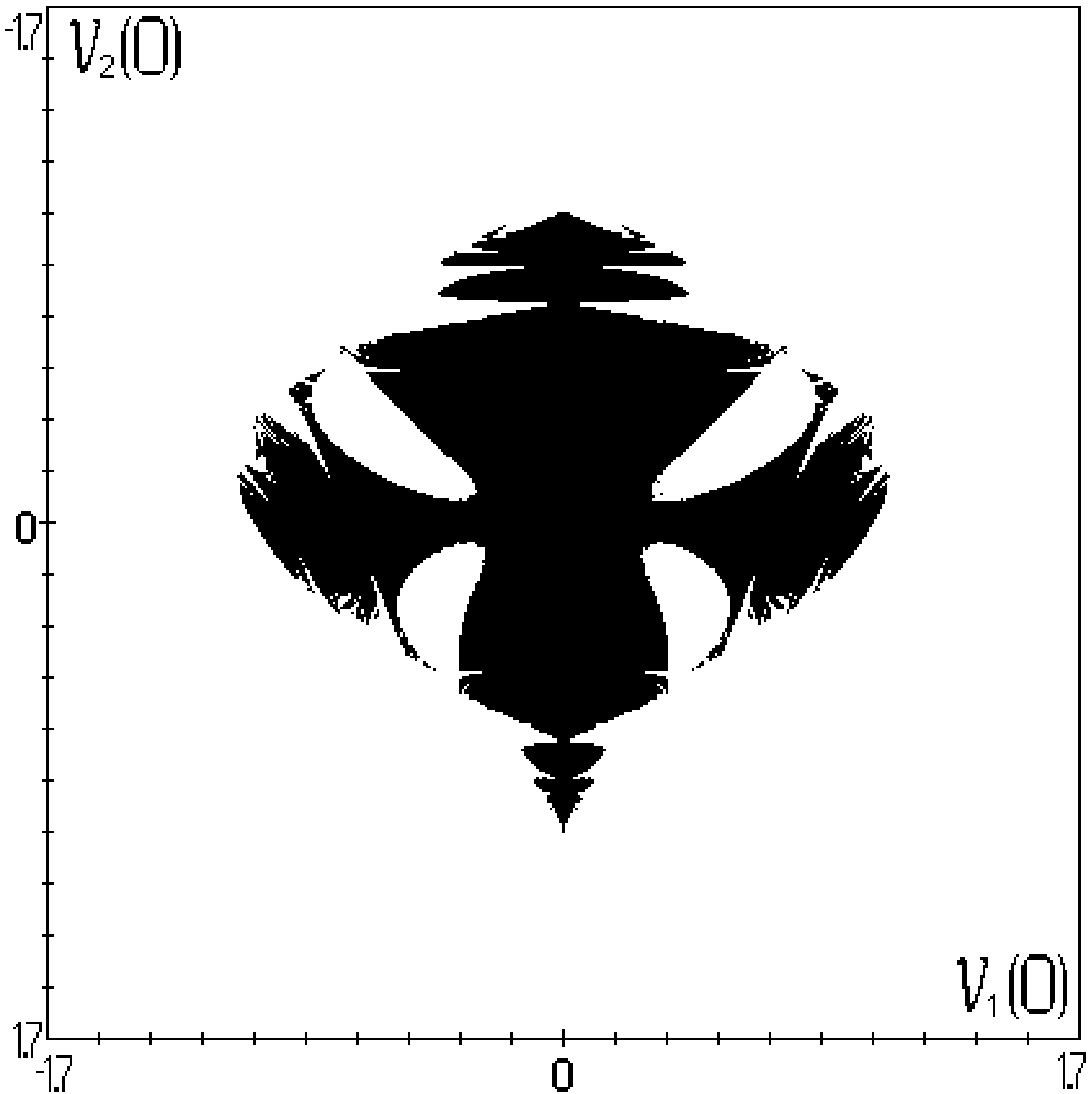}
\caption{\label{fig16} Stability diagram for the bush B$[a^4,i]$ in the
FPU-$\alpha$ chain.}
\end{figure}

\begin{figure}
\centering
\setlength{\unitlength}{1mm}
\includegraphics[width=60mm]{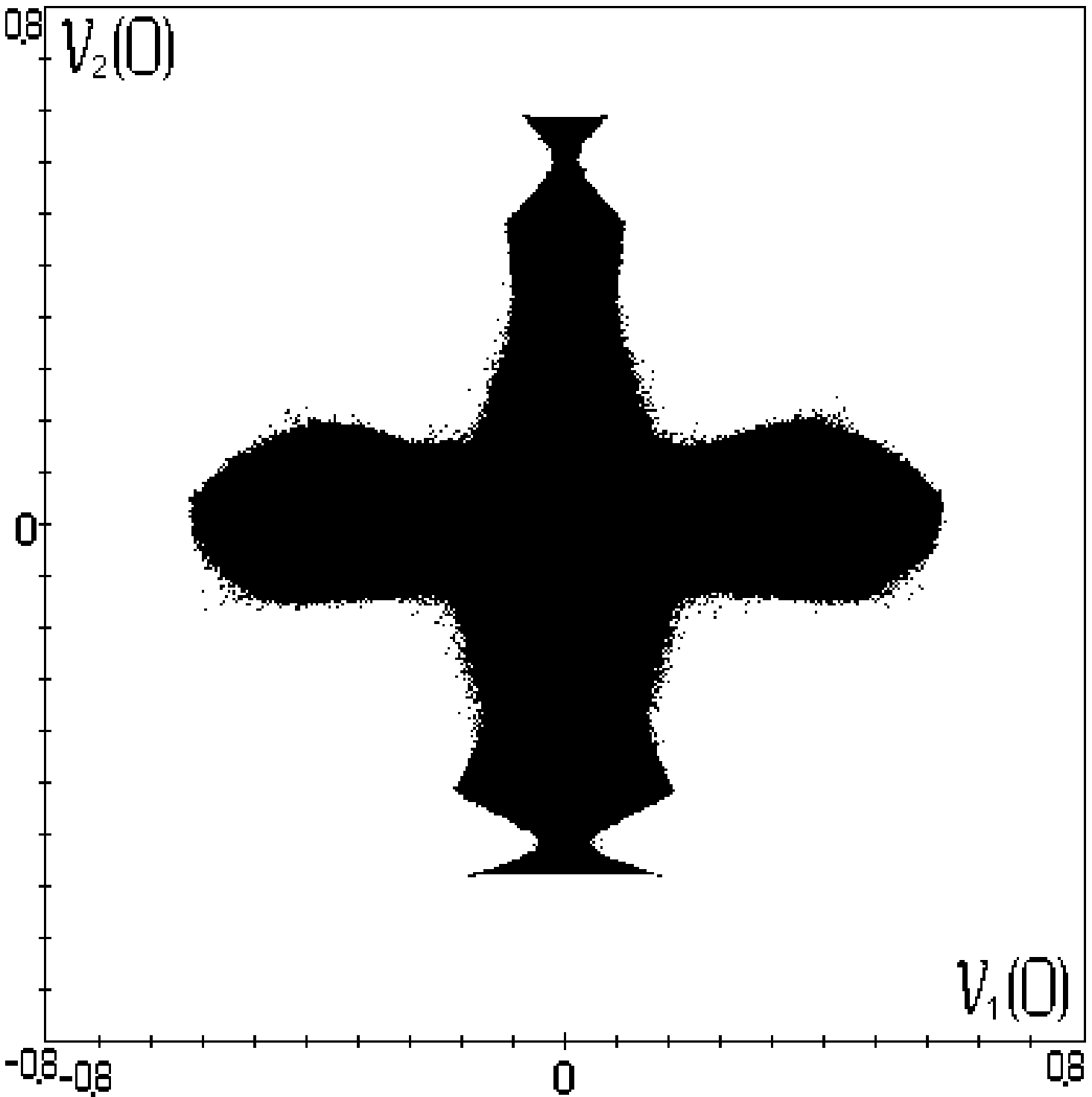}
\caption{\label{fig17} Stability diagram for the bush B$[a^6,ai]$ in the
FPU-$\alpha$ chain.}
\end{figure}

\begin{figure}
\centering
\setlength{\unitlength}{1mm}
\includegraphics[width=60mm]{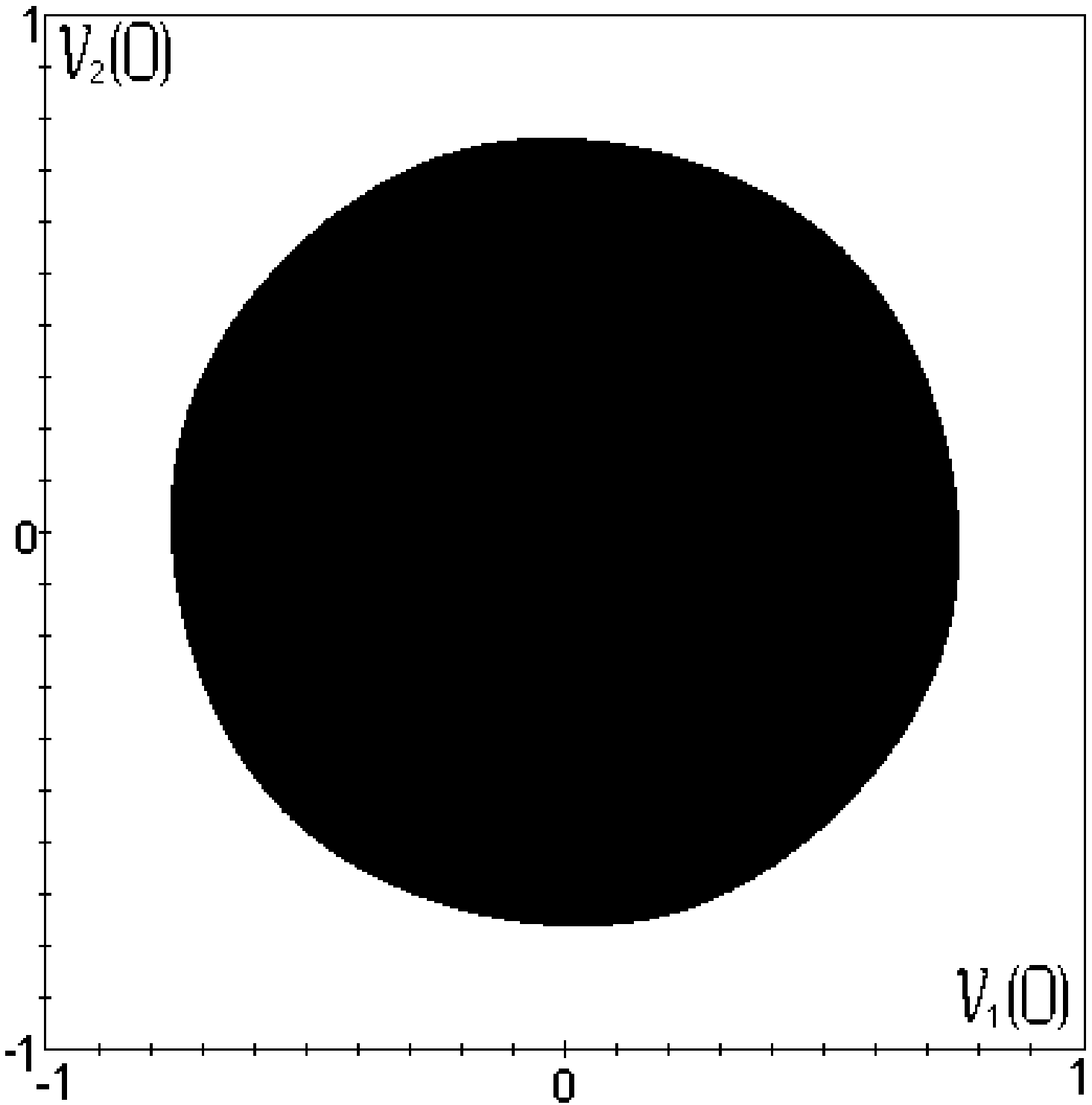}
\caption{\label{fig18} Stability diagram for the bush B$[a^3]$ in the
FPU-$\beta$ chain.}
\end{figure}

\begin{figure}
\centering
\setlength{\unitlength}{1mm}
\includegraphics[width=60mm]{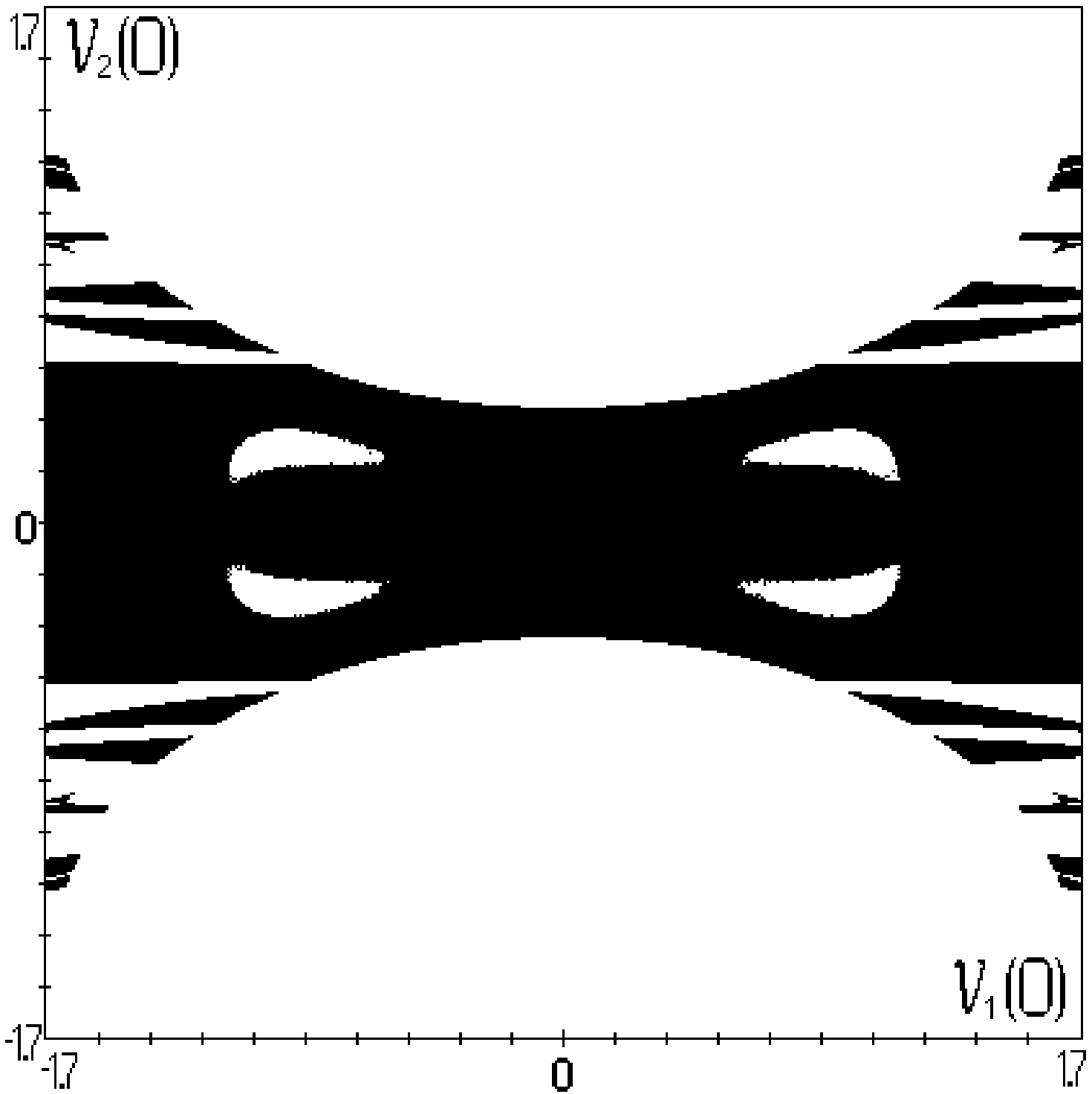}
\caption{\label{fig19} Stability diagram for the bush B$[a^4,i]$ in the
FPU-$\beta$ chain.}
\end{figure}

\begin{figure}
\centering
\setlength{\unitlength}{1mm}
\includegraphics[width=60mm]{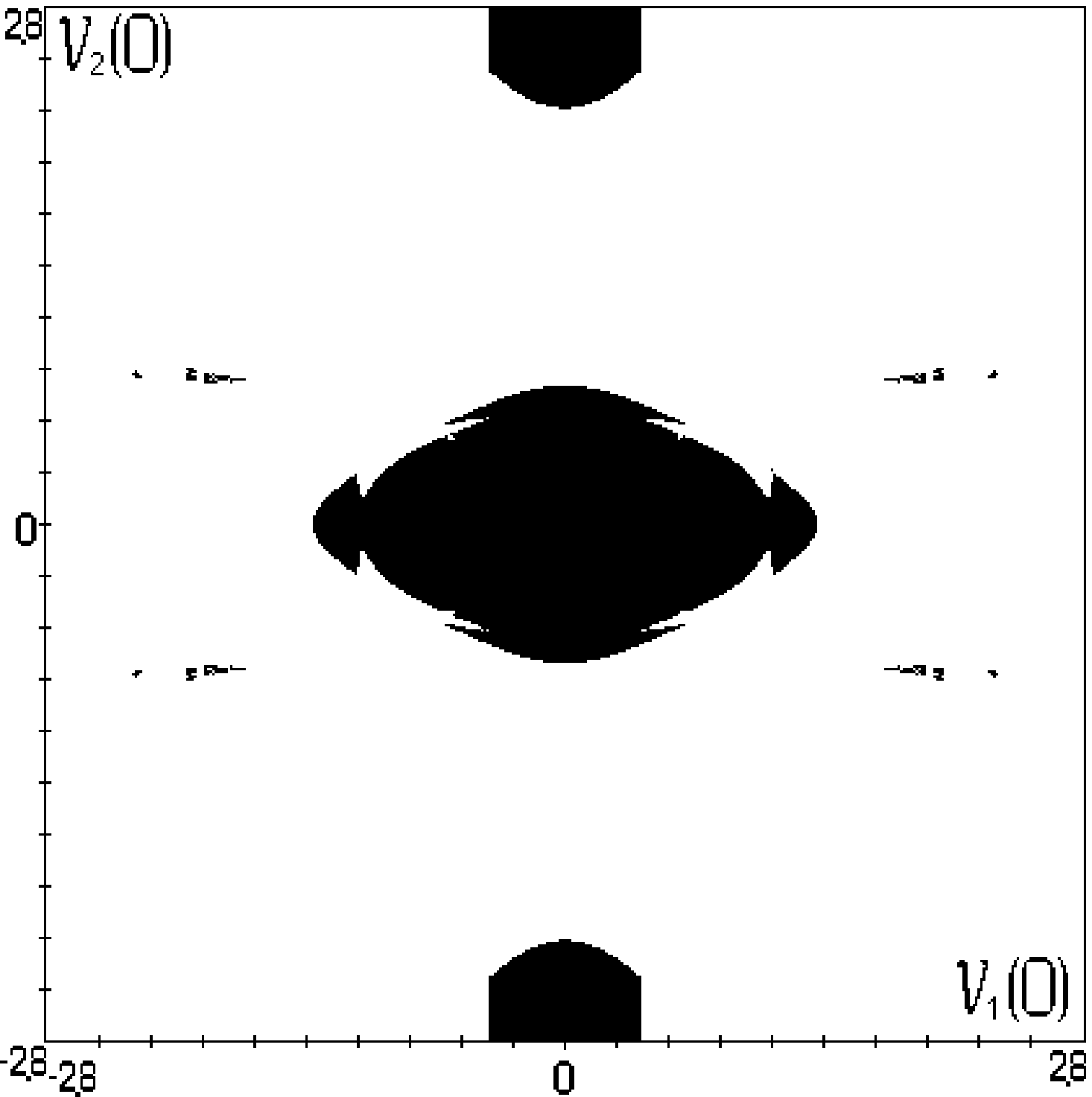}
\caption{\label{fig22} Stability diagram for the bush B$[a^6,ai]$ in the
FPU-$\beta$ chain.}
\end{figure}

\begin{figure}
\centering
\setlength{\unitlength}{1mm}
\includegraphics[width=60mm]{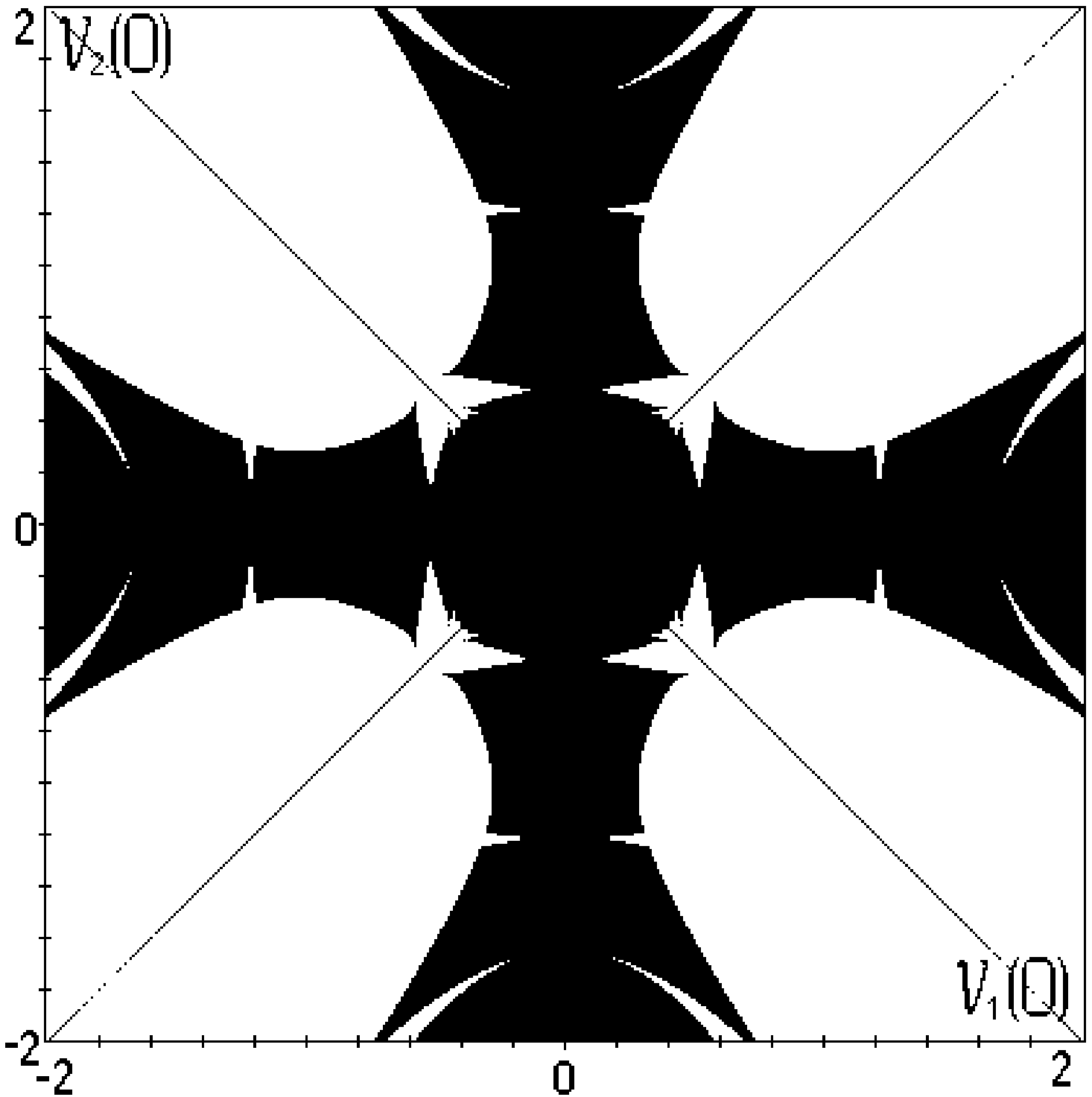}
\caption{\label{fig20} Stability diagram for the bush B$[a^4,a^2u]$ in the
FPU-$\beta$ chain.}
\end{figure}

\begin{figure}
\centering
\setlength{\unitlength}{1mm}
\includegraphics[width=60mm]{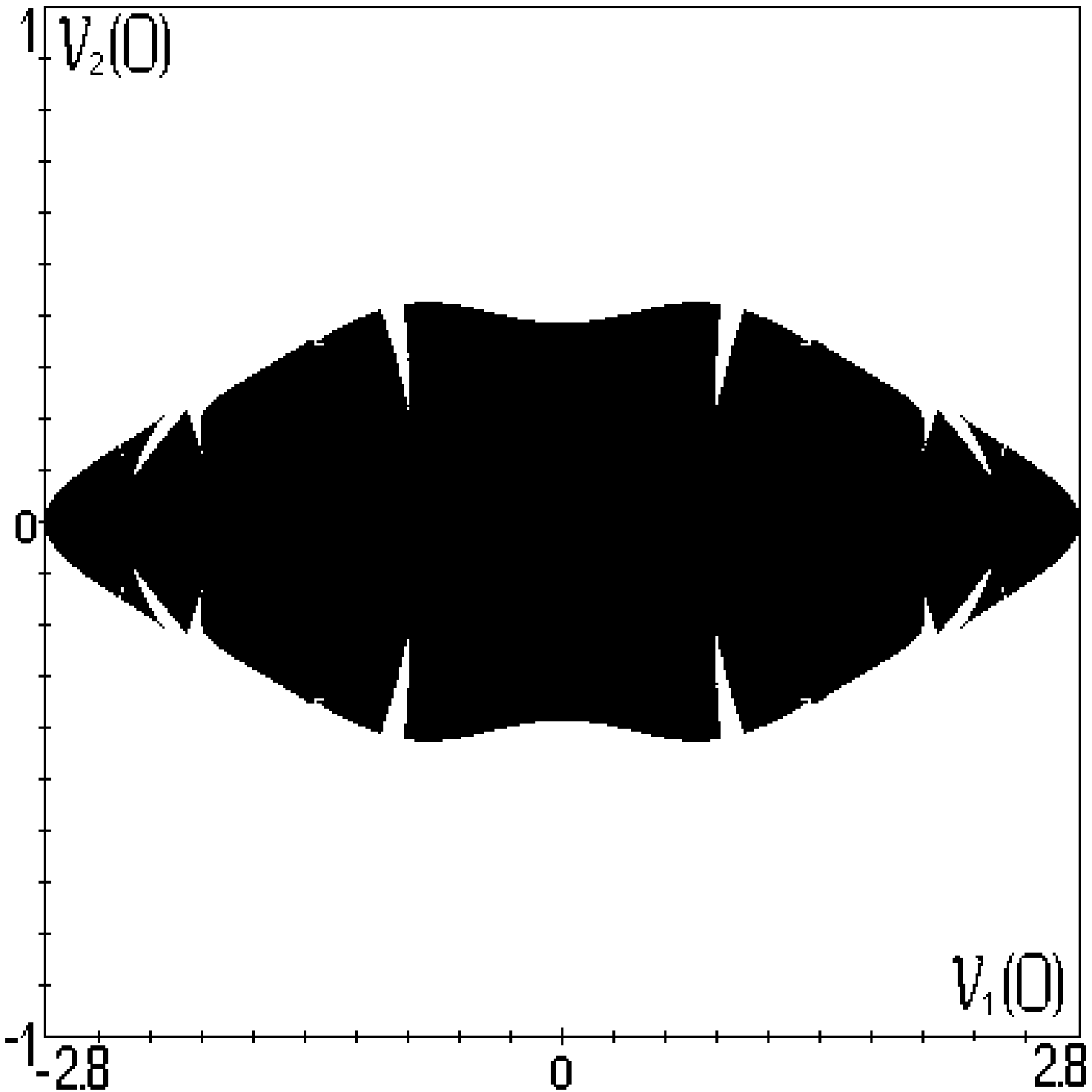}
\caption{\label{fig21} Stability diagram for the bush B$[a^4,aiu]$ in the
FPU-$\beta$ chain.}
\end{figure}

\begin{figure}\centering
\setlength{\unitlength}{1mm}
\includegraphics[width=60mm]{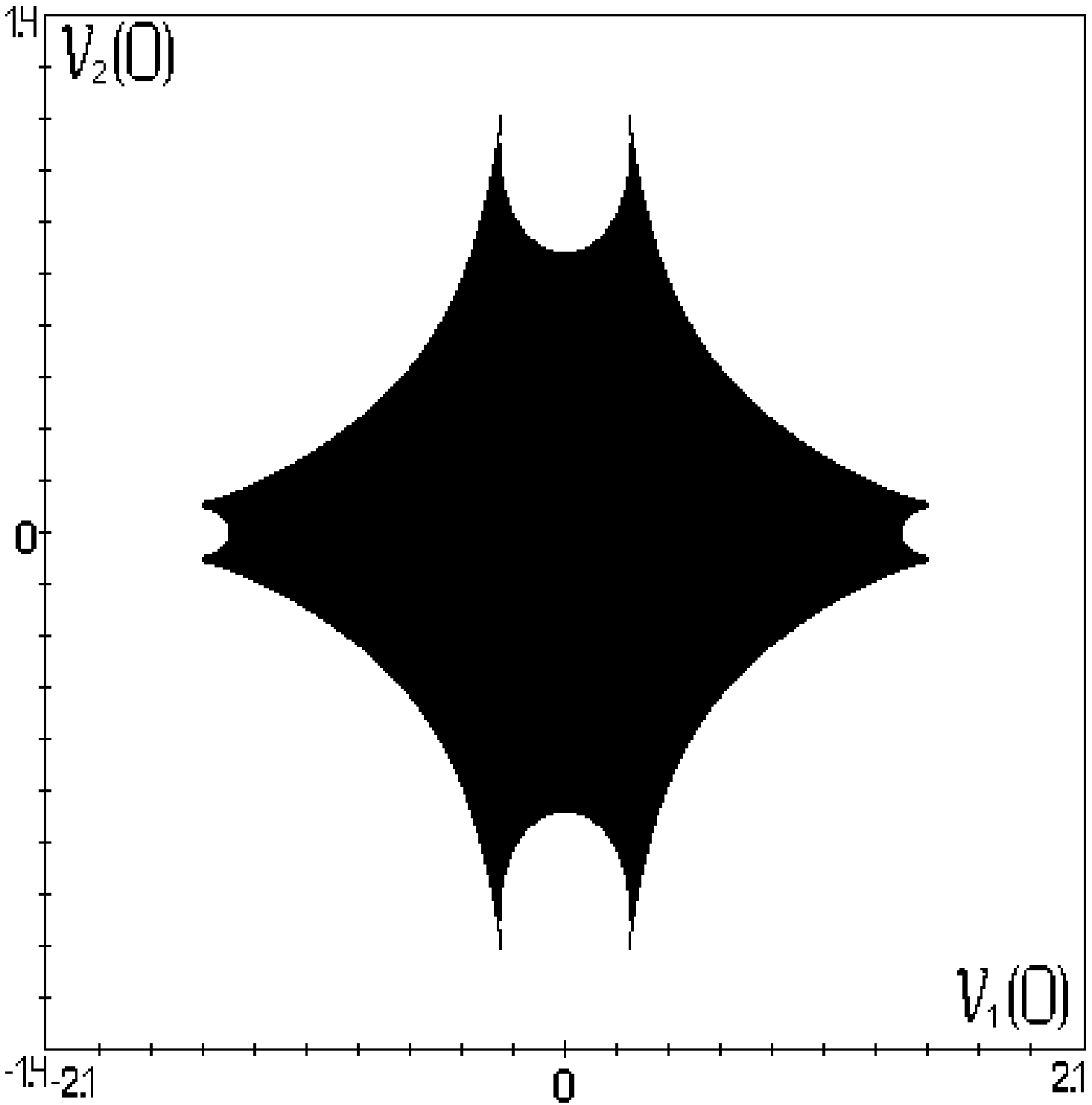}
\caption{\label{fig23} Stability diagram for the bush B$[a^6,iu]$ in the
FPU-$\beta$ chain.}
\end{figure}

\begin{figure}
\centering
\setlength{\unitlength}{1mm}
\includegraphics[width=60mm]{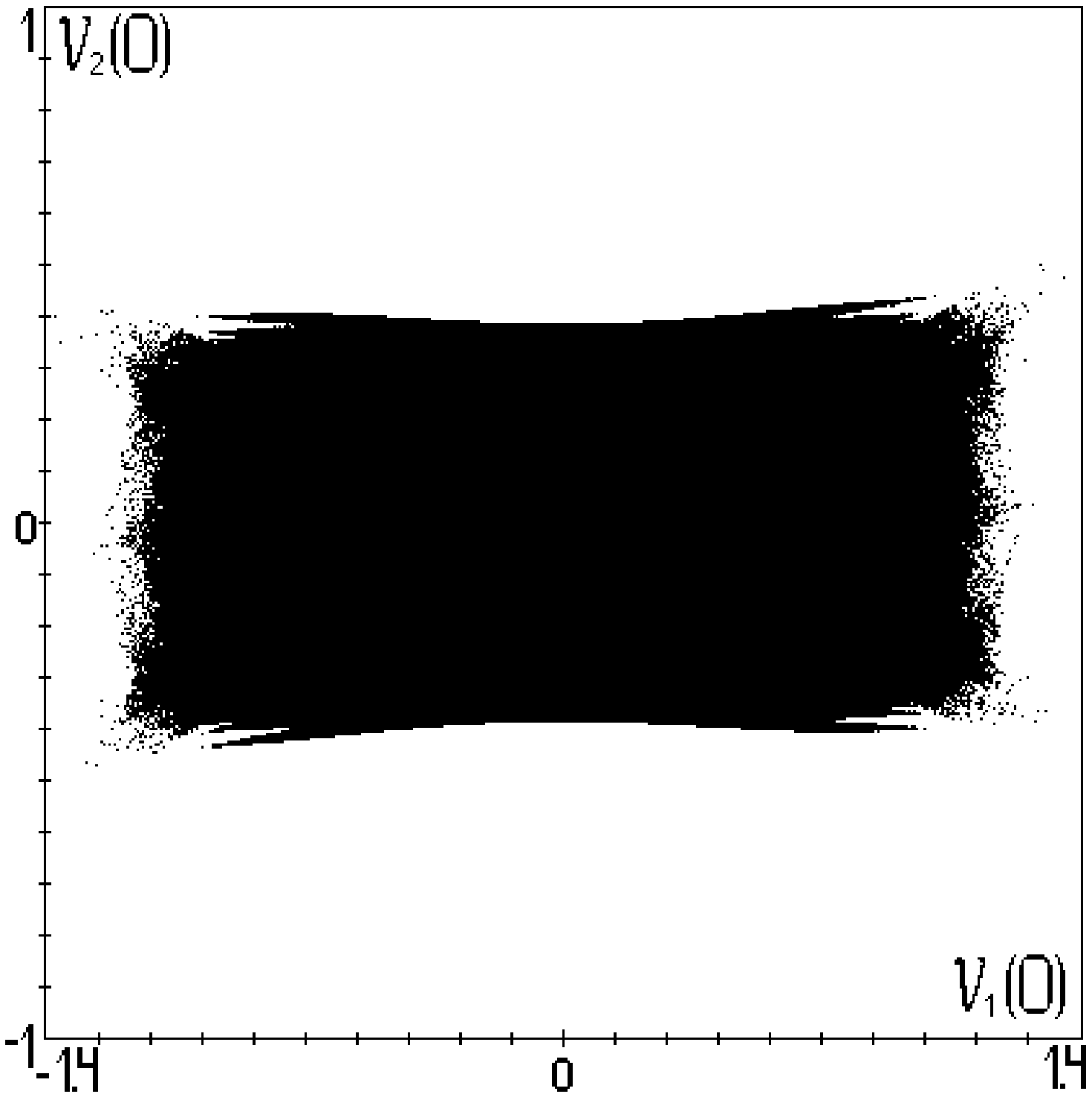}
\caption{\label{fig24} Stability diagram for the bush B$[a^6,i,a^3u]$ in
the FPU-$\beta$ chain.}
\end{figure}

\begin{figure}
\centering
\setlength{\unitlength}{1mm}
\includegraphics[width=60mm]{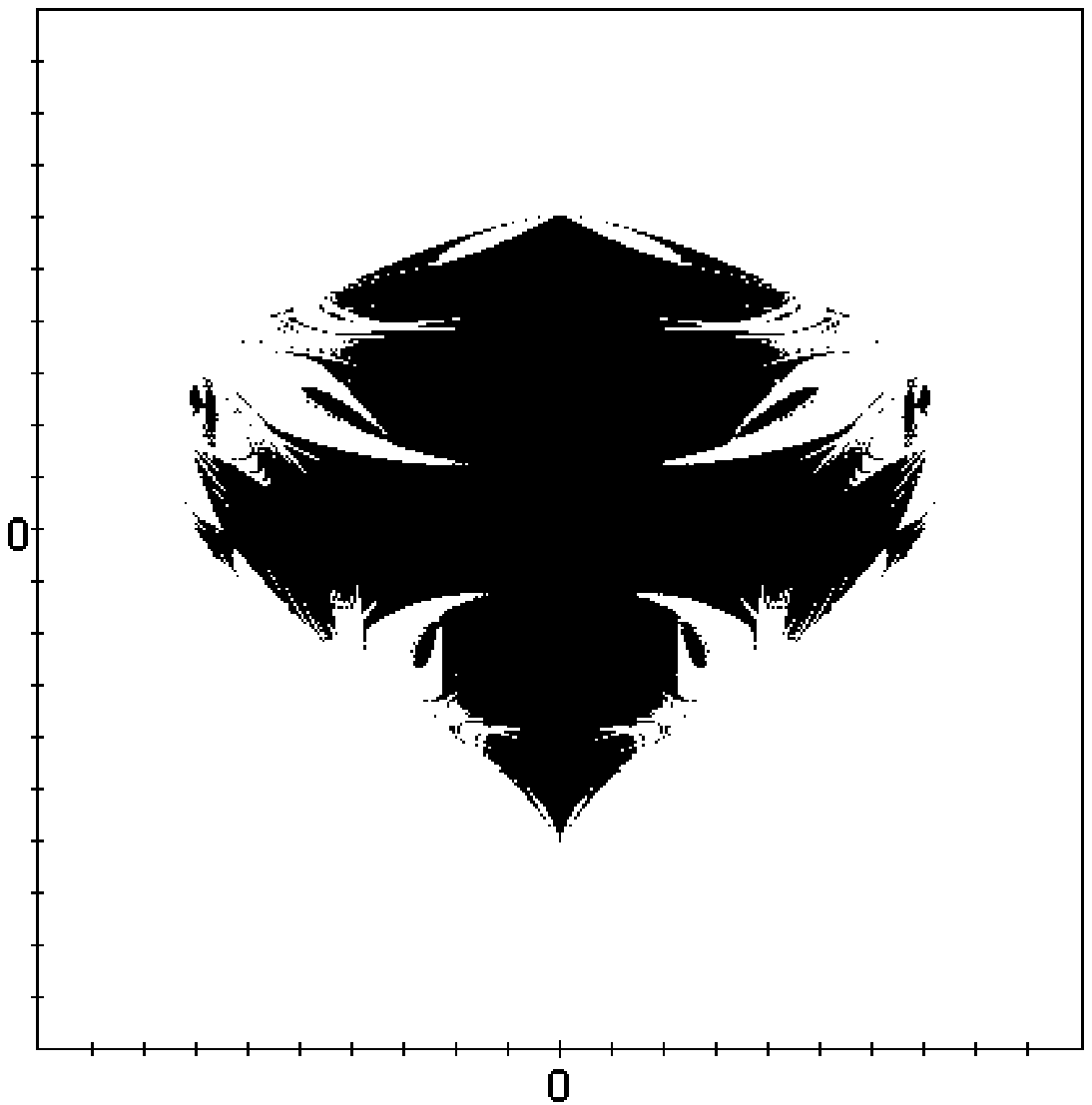}
\caption{\label{fig26} Bush B$[a^4,i]$ in FPU-$\alpha$ with $N=8$.}
\end{figure}

\begin{figure}
\centering
\setlength{\unitlength}{1mm}
\includegraphics[width=60mm]{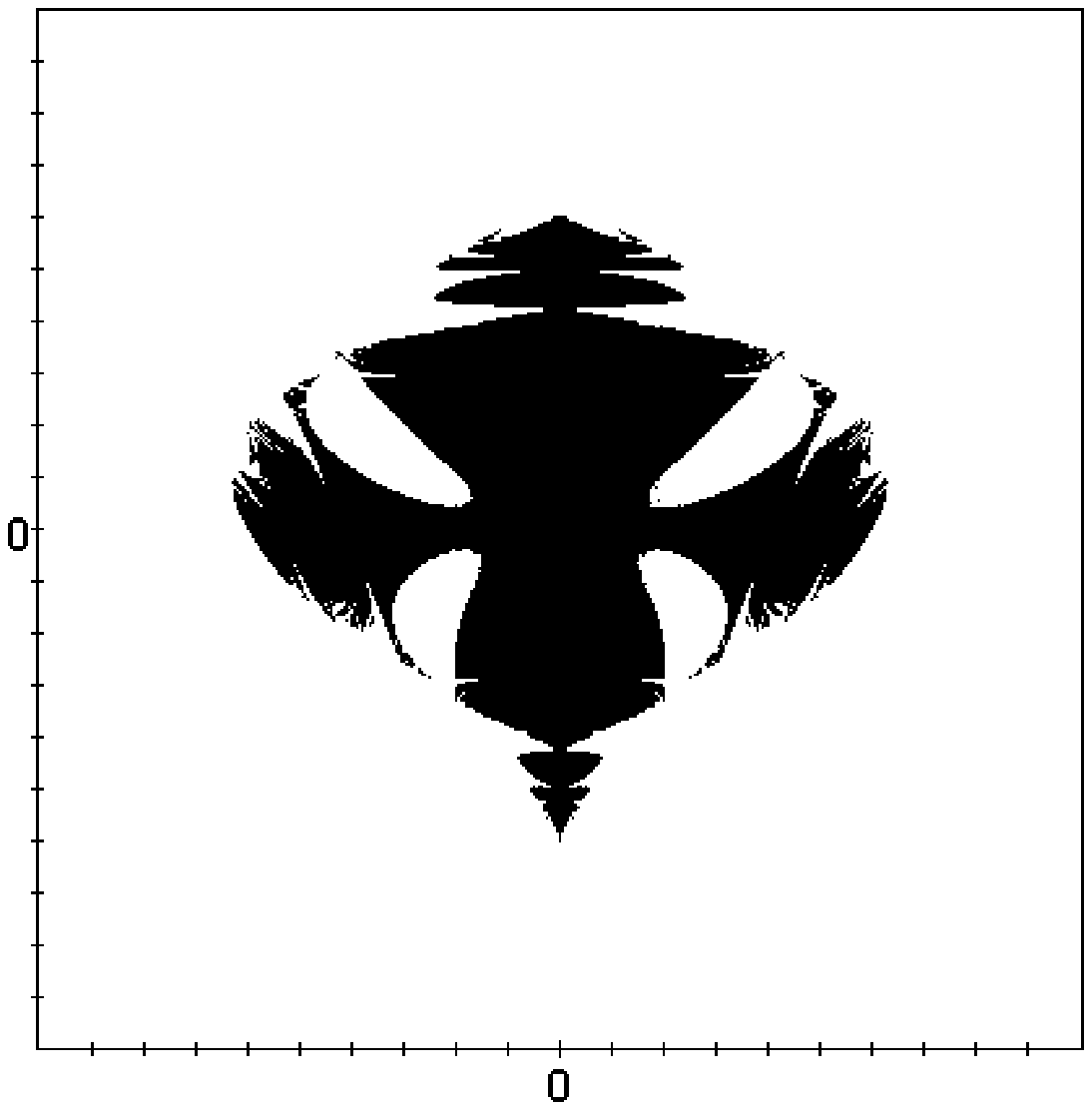}
\caption{\label{fig27} Bush B$[a^4,i]$ in FPU-$\alpha$ with $N=12$.}
\end{figure}

\begin{figure}
\centering
\setlength{\unitlength}{1mm}
\includegraphics[width=60mm]{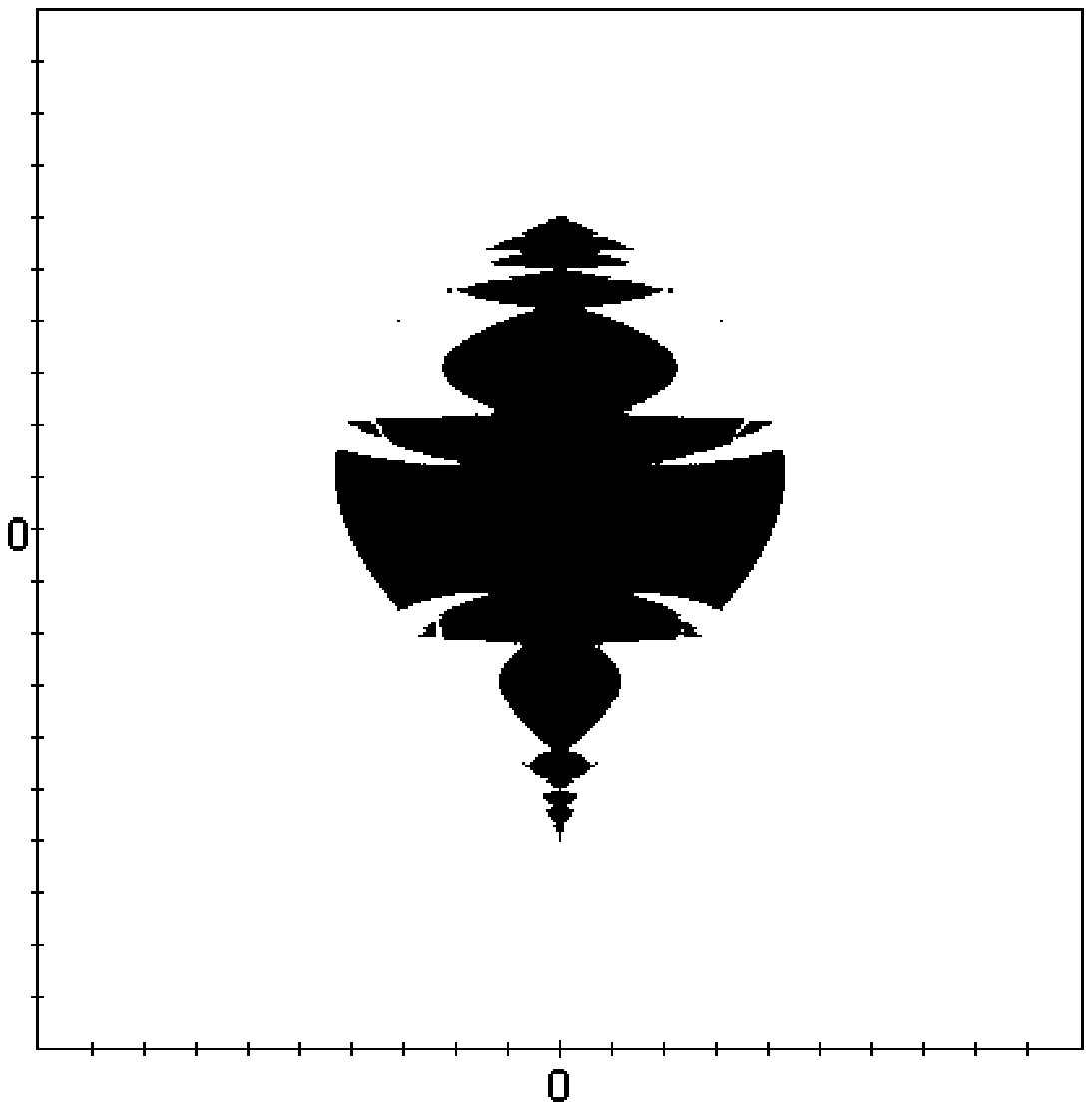}
\caption{\label{fig28} Bush B$[a^4,i]$ in FPU-$\alpha$ with $N=16$.}
\end{figure}

\begin{figure}
\centering
\setlength{\unitlength}{1mm}
\includegraphics[width=60mm]{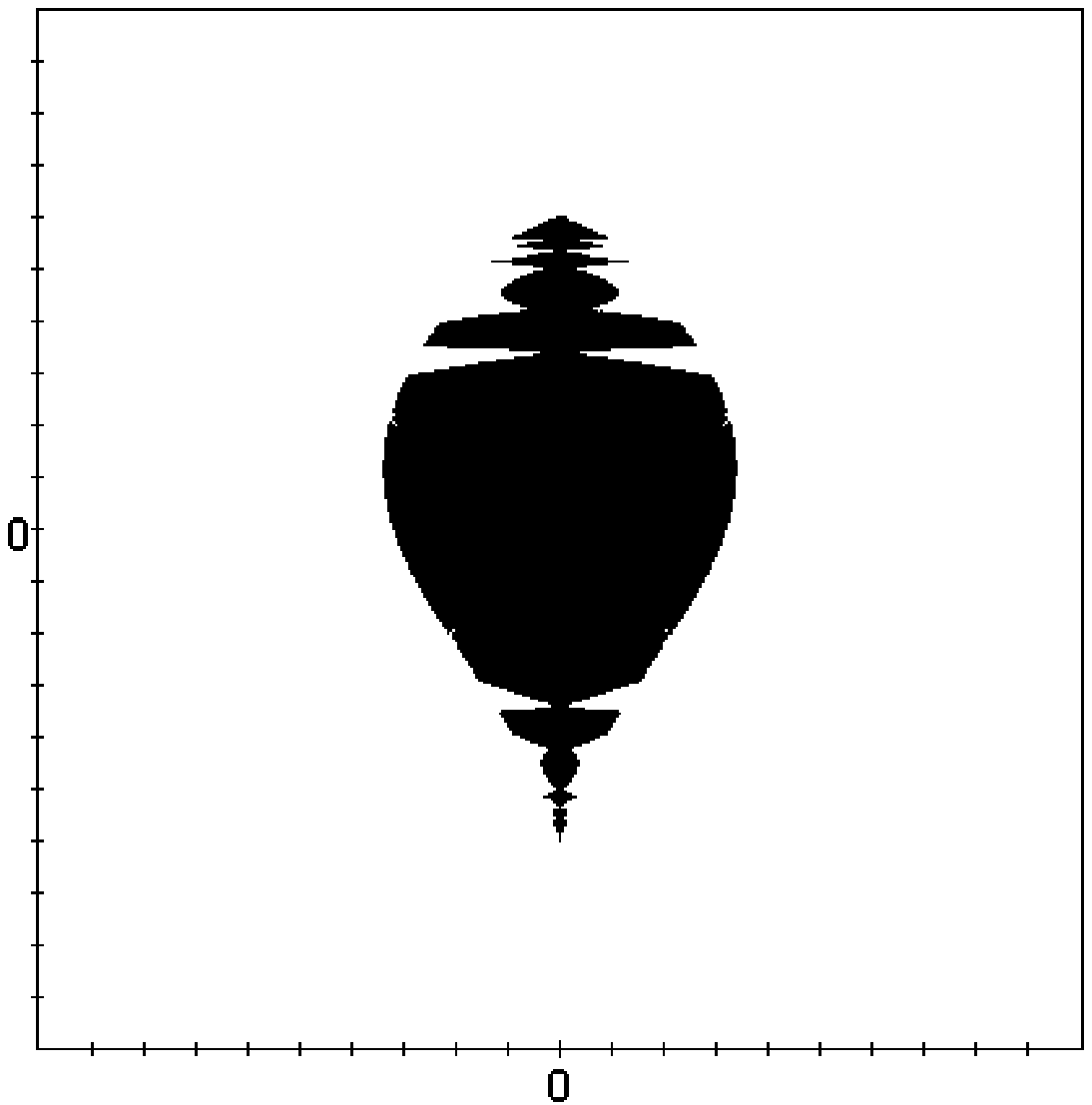}
\caption{\label{fig29} Bush B$[a^4,i]$ in FPU-$\alpha$ with $N=20$.}
\end{figure}

\begin{figure}
\centering
\setlength{\unitlength}{1mm}
\includegraphics[width=60mm]{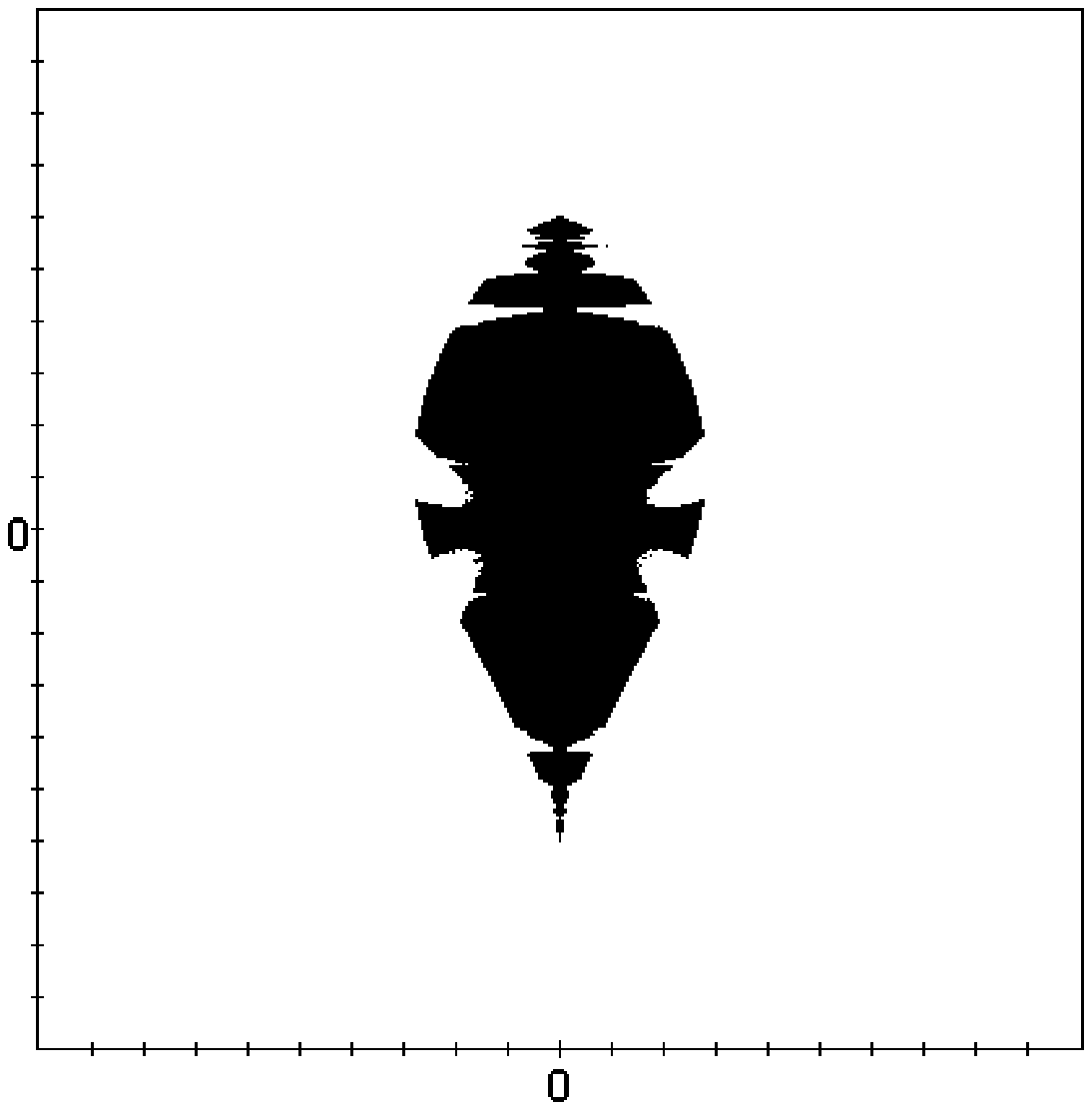}
\caption{\label{fig30} Bush B$[a^4,i]$ in FPU-$\alpha$ with $N=24$.}
\end{figure}

\begin{figure}
\centering
\setlength{\unitlength}{1mm}
\includegraphics[width=60mm]{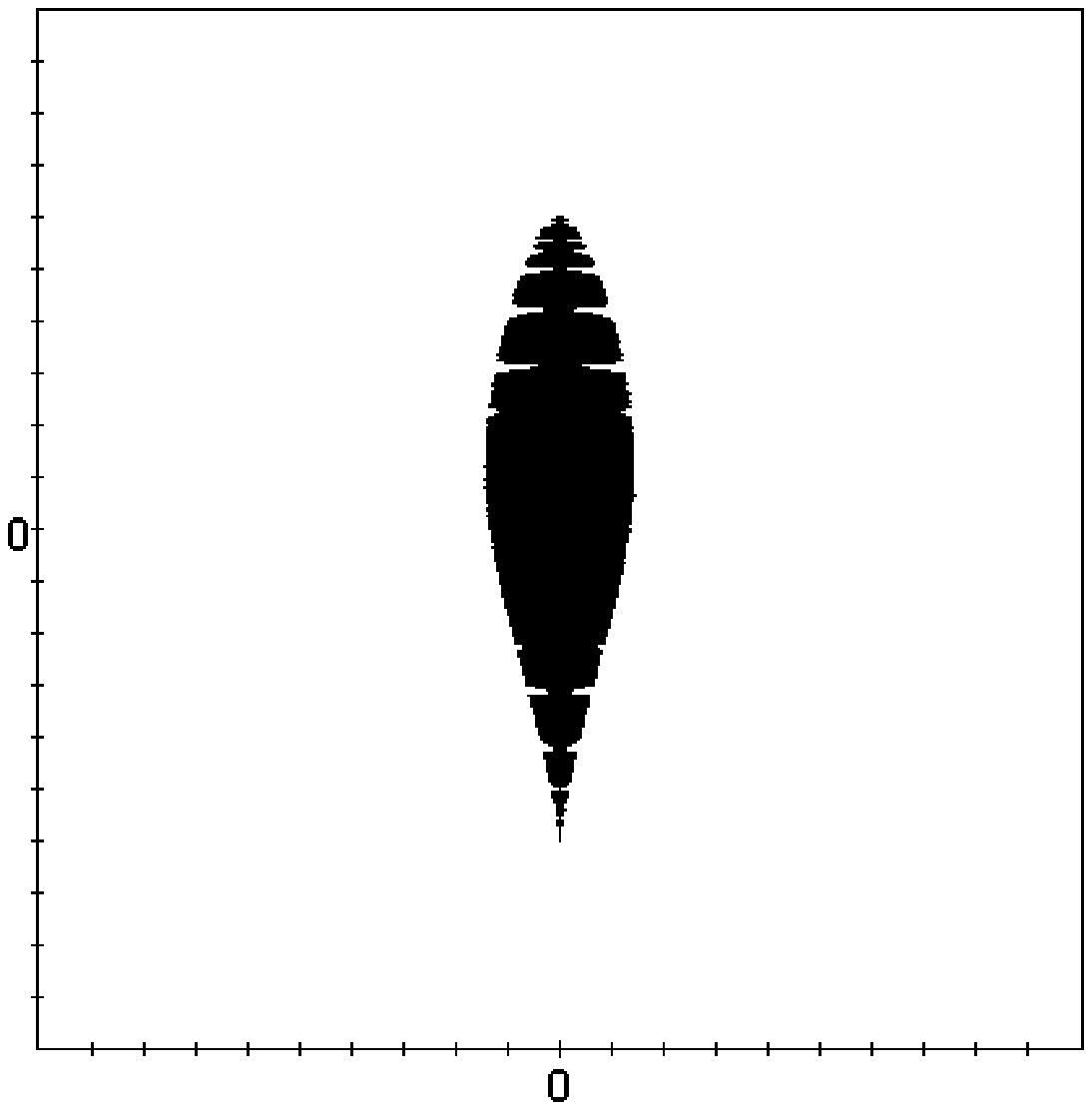}
\caption{\label{fig31} Bush B$[a^4,i]$ in FPU-$\alpha$ with $N=48$.}
\end{figure}

It is interesting to consider the case where the excitation energy is
distributed at the initial instant, in some a manner, between its kinetic
and potential counterparts. The appropriate stability diagram is depicted
in Fig.~\ref{fig78}. Actually, this is simply another section of the same
four-dimensional stability domain of the considered bush which is
determined by the equations $\dot\nu_1(0)=0.2,\dot\nu_2(0)=0.1$.

Comparing Fig.~\ref{fig77} and Fig.~\ref{fig78}, one can see the effect of
the ``wind blowing from the right to the left''. This effect leads to a
loss of the symmetry of the diagram in Fig.~\ref{fig77} with respect to the
vertical axis passing through the origin.

Thus, the threshold of the loss of stability of the bush
B$[\hat{a}^4,\hat{i}]$ in the FPU-$\alpha$ chain depend essentially not
only on the distribution of the initial energy between its two modes, but
also between the kinetic and potential counterparts. Let us repeat, once
more, that instead of dealing with the four-dimensional stability diagram,
we investigate different \textit{sections} of this domain.

The stability diagrams for two other two-dimensional bushes, B$[\hat{a}^3]$
and B$[\hat{a}^6,\hat{a}\hat{i}]$, in the FPU-$\alpha$ chain are depicted
in Fig.~\ref{fig15} and Fig.~\ref{fig17}, respectively. These diagrams, as
well as the diagrams in Fig.~\ref{fig18}-\ref{fig24} for the FPU-$\beta$
chain, correspond to the initial conditions $\dot\nu_1(0)=0$,
$\dot\nu_2(0)=0$, i.e.\ the kinetic energy of the initial excitation is
supposed to be equal to zero.

Let us emphasize that the stability diagrams for the \textit{same} bushes
in the FPU-$\alpha$ and FPU-$\beta$ chains look very different. As an
example, one can compare Fig.~\ref{fig16} and Fig.~\ref{fig19} for the bush
B$[\hat{a}^4,\hat{i}]$ in FPU-$\alpha$ and FPU-$\beta$ chains,
respectively. It is interesting to note that the last diagram
(Fig.~\ref{fig19}) demonstrates the presence of several regions of
instability (white color) in the stable sea (black color).

In Figs.~\ref{fig20}-\ref{fig24}, we represent the stability diagrams
corresponding to four additional two-dimensional bushes for the FPU-$\beta$
chain (as it has already been said, they are induced by the evenness of the
potential of this model).

The most nontrivial picture corresponds to the stability diagram of the
bush B$[\hat{a}^4,\hat{a}^2\hat{u}]$. It resembles a chivalrous cross with
two crossed swords (along the diagonals of the diagram). Note that these
swords describe, actually, the stability of two equivalent one-dimensional
bushes $\nu(t)[\vec{\psi}_3+\vec{\psi}_9]$ and
$\nu(t)[\vec{\psi}_3-\vec{\psi}_9]$, while the two-dimensional bush
B$[\hat{a}^4,\hat{a}^2\hat{u}]$ is
$\nu_1(t)\vec{\psi}_3+\nu_2(t)\vec{\psi}_9$.

For many bushes, the stability domains depend essentially on the number $N$
of particles in the chain. The area of the corresponding regions of
stability, in our diagrams, tends often to decrease with the increasing of
$N$. We demonstrate such a behavior in the cases $N=8,12,16,20,24,48$ for
the bush B$[\hat{a}^4,\hat{i}]$ in the FPU-$\alpha$ chain (see
Figs.~\ref{fig26}-\ref{fig31})\footnote{We do not discuss the continuum
limit $N\rightarrow\infty$ for two-dimensional bushes in the present
paper.}. Note that the stability domain of this bush shrinks only along the
horizontal axis $\nu_1(0)$ when $N\rightarrow\infty$. This is the
consequence of the fact that when $\nu_1(t)\rightarrow 0$ the bush
B$[\hat{a}^4,\hat{i}]=\nu_1(t)\vec{\psi}_{3N/4}-\nu_2(t)\vec{\psi}_{N/2}$
(or its equivalent form
B$[\hat{a}^4,\hat{a}^2\hat{i}]=\nu_1(t)\vec{\psi}_{N/4}+\nu_2(t)\vec{\psi}_{N/2}$)
tends to the one-dimensional bush
B$[\hat{a}^2,\hat{i}]=\nu_2(t)\vec{\psi}_{N/2}$. On the other hand, the
latter bush possesses, for $N\rightarrow\infty$, a finite one-dimensional
interval of stability with respect to increasing of the initial value
$\nu_2(0)$ of its single mode $\nu_2(t)$.

\section{Conclusion}

The bushes of normal modes as \textit{exact} nonlinear excitations in the
physical systems with discrete (point or space) symmetry groups were
introduced and discussed in our previous
papers~\cite{DAN1,DAN2,PhysD,Okta,NNM,ENOC3,C60}. In general case, bushes
can be rather complex, for example, the bushes in the $C_{60}$ fullerene
structure~\cite{C60}.

In~\cite{FPU1}, we considered bushes in monoatomic nonlinear chains which
represent the simplest systems with translational symmetry. Mainly, we
studied there the bushes induced by the group $T$ of pure translations and
discussed only in part those induced by the dihedral symmetry group $D$.
The stability of a few simplest bushes was also studied in~\cite{FPU1}.
Nevertheless, there are some more one-dimensional and two-dimensional
bushes in the FPU-chains, in particular, those obtained by Bob Rink
in~\cite{BRink}, whose stability still has not been investigated. On the
other hand, the problem of the bush stability is one of the most important
problems in the general theory of the bushes of normal modes. Because of
this, in the present paper, we discuss in detail the stability of all
one-dimensional and some two-dimensional bushes in both FPU-$\alpha$ and
FPU-$\beta$ chains.

Let us summarize the results obtained in the previous sections of this
paper.

1. A simple crystallographic method for finding bushes in nonlinear chains
is developed.

2. The stability of all one-dimensional bushes in the FPU-$\alpha$ (three
bushes) and in the FPU-$\beta$ (six bushes, including those obtained
in~\cite{BRink}) are investigated. The stability diagrams which can be used
for studying stability of one-dimensional bushes for the FPU chains with
any finite and infinite number of particles (continuum limit
$N\rightarrow\infty$) are presented. They were obtained by Floquet method.

3. The stability diagrams of different type, which represent certain
sections of four-dimensional stability domains, corresponding to some
two-dimensional bushes in the FPU-$\alpha$ (three bushes) and in the
FPU-$\beta$ (seven bushes), are found. These diagrams demonstrate
explicitly that the bush stability domains depend not only on the energy of
the initial excitation, but on the complete set of the initial conditions
in the bush phase space.

The above mentioned diagrams for one-dimensional and two-dimensional bushes
show how nontrivial the shape of the stability domains for the bushes of
normal modes can be.

In this paper, we consider only the main mechanism for the loss of bush
stability, wherein it is brought about by the parametric excitations of the
bush modes with those modes which do not belong to the given bush.
Nevertheless, one can consider some different reasons for a bush to lose
its stability, such as a finite temperature of the thermal bath, the
presence of impurities in the chain structure, etc. These problems, as well
as the methods for excitation of the bushes of normal modes in the
FPU-chains will be considered in a separate paper.

\section*{Acknowledgments}
We are very grateful to Prof.\ V.P.~Sakhnenko for his friendly support and
to O.E.~Evnin for his valuable help with the language corrections in the
text of this paper.

\clearpage

\section*{Appendix. Two-dimensional bushes of vibrational modes for the FPU
chains}
\begin{table}[hb]
\centering
\caption{\label{tableA1} Representation of 2D bushes of vibrational modes
for the FPU chains in the configuration space (classification by the
dihedral group~$D$)}
\begin{tabular}{|l|c|}
\hline
Bush&Displacement pattern\\
\hline
B$[a^3]$&$|x_1,x_2,x_3|$\\
\hline
B$[a^4,i]$&$|x_1,x_2,-x_2,-x_1|$\\
B$[a^4,a^2i]$&$|x_1,-x_1,x_2,-x_2|$\\
\hline
B$[a^5,i]$&$|x_1,x_2,0,-x_2,-x_1|$\\
B$[a^5,ai]$&$|0,x_1,x_2,-x_2,-x_1|$\\
B$[a^5,a^2i]$&$|x_1,-x_1,x_2,0,-x_2|$\\
B$[a^5,a^3i]$&$|x_1,0,-x_1,x_2,-x_2|$\\
B$[a^5,a^4i]$&$|x_1,x_2,-x_2,-x_1,0|$\\
\hline
B$[a^6,ai]$&$|0,x_1,x_2,0,-x_2,-x_1|$\\
B$[a^6,a^3i]$&$|x_1,0,-x_1,x_2,0,-x_2|$\\
B$[a^6,a^5i]$&$|x_1,x_2,0,-x_2,-x_1,0|$\\
\hline
\end{tabular}
\end{table}

\begin{table}[h]
\centering
\caption{\label{tableA2} Additional 2D bushes for the FPU-$\beta$ chain}
\begin{tabular}{|l|c|}
\hline
Bush&Displacement pattern\\
\hline
B$[a^4,aiu]$&$|x_1,x_2,x_3,x_2|$\\
B$[a^4,a^3iu]$&$|x_2,x_1,x_2,x_3|$\\
\hline
B$[a^4,a^2u]$&$|x_1,x_2,-x_1,-x_2|$\\
\hline
B$[a^5,iu]$&$|x_1,x_2,x_3,x_2,x_1|$\\
B$[a^5,aiu]$&$|x_1,x_2,x_3,x_3,x_2|$\\
B$[a^5,a^2iu]$&$|x_1,x_1,x_2,x_3,x_2|$\\
B$[a^5,a^3iu]$&$|x_1,x_2,x_1,x_3,x_3|$\\
B$[a^5,a^4iu]$&$|x_1,x_2,x_2,x_1,x_3|$\\
\hline
B$[a^6,iu]$&$|x_1,x_2,x_3,x_3,x_2,x_1|$\\
B$[a^6,a^2iu]$&$|x_1,x_1,x_2,x_3,x_3,x_2|$\\
B$[a^6,a^4iu]$&$|x_1,x_2,x_2,x_1,x_3,x_3|$\\
\hline
B$[a^6,i,a^3u]$&$|x_1,x_2,x_1,-x_1,-x_2,-x_1|$\\
B$[a^6,a^2i,a^3u]$&$|x_1,-x_1,x_2,-x_1,x_1,-x_2|$\\
B$[a^6,a^4i,a^3u]$&$|x_1,x_2,-x_2,-x_1,-x_2,x_2|$\\
\hline
B$[a^8,i,a^4u]$&$|x_1,x_2,x_2,x_1,-x_1,-x_2,-x_2,-x_1|$\\
B$[a^8,a^2i,a^4u]$&$|x_1,-x_1,x_2,x_2,-x_1,x_1,-x_2,-x_2|$\\
B$[a^8,a^4i,a^4u]$&$|x_1,x_2,-x_2,-x_1,-x_1,-x_2,x_2,x_1|$\\
B$[a^8,a^6i,a^4u]$&$|x_1,x_1,x_2,-x_2,-x_1,-x_1,-x_2,x_2|$\\
\hline
B$[a^8,ai,a^4u]$&$|0,x_1,x_2,x_1,0,-x_1,-x_2,-x_1|$\\
B$[a^8,a^3i,a^4u]$&$|x_1,0,-x_1,x_2,-x_1,0,x_1,-x_2|$\\
B$[a^8,a^5i,a^4u]$&$|x_1,x_2,0,-x_2,-x_1,-x_2,0,x_2|$\\
B$[a^8,a^7i,a^4u]$&$|x_1,x_2,x_1,0,-x_1,-x_2,-x_1,0|$\\
\hline
B$[a^{10},ai,a^5u]$&$|0,x_1,x_2,x_2,x_1,0,-x_1,-x_2,-x_2,-x_1|$\\
B$[a^{10},a^3i,a^5u]$&$|x_1,0,-x_1,x_2,x_2,-x_1,0,x_1,-x_2,-x_2|$\\
B$[a^{10},a^5i,a^5u]$&$|x_1,x_2,0,-x_2,-x_1,-x_1,-x_2,0,x_2,x_1|$\\
B$[a^{10},a^7i,a^5u]$&$|x_1,x_1,x_2,0,-x_2,-x_1,-x_1,-x_2,0,x_2|$\\
B$[a^{10},a^9i,a^5u]$&$|x_1,x_2,x_2,x_1,0,-x_1,-x_2,-x_2,-x_1,0|$\\
\hline
\end{tabular}
\end{table}

\clearpage

\begin{table}
\centering
\caption{\label{table5} Linearized dynamical equations for studying
stability of 1D bushes for the FPU-$\alpha$ chain. Here $\nu(t)$ is the
vibrational mode representing a given bush, while $k=\frac{\pi j}{N}$
determines the number $j$ of ``sleeping'' modes}
\begin{tabular}{|l|}
\hline
B$[a^2,i]$: $\ddot\nu+4\nu=0$\\
$\left\{\begin{array}{l}
\ddot\nu_k+4\nu_k\sin^2k=-\frac{8\alpha}{N}\nu\nu_{\frac{\pi}{2}-k}\sin 2k\\
\ddot\nu_{\frac{\pi}{2}-k}+4\nu_{\frac{\pi}{2}-k}\cos^2k=-\frac{8\alpha}{N}\nu\nu_k\sin 2k
\end{array}\right.$\\
\hline
B$[a^3,a^2i]$: $\ddot\nu+3\nu=\frac{3\alpha\sqrt{6}}{2\sqrt{N}}\nu^2$\\
$\left\{\begin{array}{l}
\ddot\nu_k+4\nu_k\sin^2k=-\frac{\alpha\sqrt{6}}{\sqrt{N}}\nu
\left[\left(1-2\cos\left(2k+\frac{\pi}{3}\right)\right)\nu_{\frac{\pi}{3}+k}+
\left(1-2\sin\left(2k+\frac{\pi}{6}\right)\right)\nu_{\frac{2\pi}{3}+k}\right]\\
\ddot\nu_{\frac{\pi}{3}+k}+4\nu_{\frac{\pi}{3}+k}\sin^2\left(k+\frac{\pi}{3}\right)=-\frac{\alpha\sqrt{6}}{\sqrt{N}}\nu
\left[\left(1-2\cos\left(2k+\frac{\pi}{3}\right)\right)\nu_k+
\left(1+2\cos 2k\right)\nu_{\frac{2\pi}{3}+k}\right]\\
\ddot\nu_{\frac{2\pi}{3}+k}+4\nu_{\frac{2\pi}{3}+k}\cos^2\left(k+\frac{\pi}{6}\right)=-\frac{\alpha\sqrt{6}}{\sqrt{N}}\nu
\left[\left(1-2\sin\left(2k+\frac{\pi}{6}\right)\right)\nu_k+
\left(1+2\cos 2k\right)\nu_{\frac{\pi}{3}+k}\right]\\
\end{array}\right.$\\
\hline
B$[a^4,a^3i]$: $\ddot\nu+2\nu=0$\\
$\left\{\begin{array}{l}
\ddot\nu_k+4\nu_k\sin^2k=-\frac{2\alpha\sqrt{2}}{\sqrt{N}}\nu
\left[\left(1-\sqrt{2}\cos\left(2k+\frac{\pi}{4}\right)\right)\nu_{\frac{\pi}{4}+k}+
\left(1-\sqrt{2}\sin\left(2k+\frac{\pi}{4}\right)\right)\nu_{\frac{3\pi}{4}+k}\right]\\
\ddot\nu_{\frac{\pi}{4}+k}+4\nu_{\frac{\pi}{4}+k}\sin^2\left(k+\frac{\pi}{4}\right)=-\frac{2\alpha\sqrt{2}}{\sqrt{N}}\nu
\left[\left(1-\sqrt{2}\cos\left(2k+\frac{\pi}{4}\right)\right)\nu_k+
\left(1+\sqrt{2}\sin\left(2k+\frac{\pi}{4}\right)\right)\nu_{\frac{\pi}{2}+k}\right]\\
\ddot\nu_{\frac{\pi}{2}+k}+4\nu_{\frac{\pi}{2}+k}\cos^2k=-\frac{2\alpha\sqrt{2}}{\sqrt{N}}\nu
\left[\left(1+\sqrt{2}\sin\left(2k+\frac{\pi}{4}\right)\right)\nu_{\frac{\pi}{4}+k}+
\left(1+\sqrt{2}\cos\left(2k+\frac{\pi}{4}\right)\right)\nu_{\frac{3\pi}{4}+k}\right]\\
\ddot\nu_{\frac{3\pi}{4}+k}+4\nu_{\frac{3\pi}{4}+k}\cos^2\left(k+\frac{\pi}{4}\right)=-\frac{2\alpha\sqrt{2}}{\sqrt{N}}\nu
\left[\left(1-\sqrt{2}\sin\left(2k+\frac{\pi}{4}\right)\right)\nu_k+
\left(1+\sqrt{2}\cos\left(2k+\frac{\pi}{4}\right)\right)\nu_{\frac{\pi}{2}+k}\right]\\
\end{array}\right.$\\
\hline
\end{tabular}
\end{table}

\begin{table}
\centering
\caption{\label{table6} Linearized dynamical equations for studying
stability of 1D bushes for the FPU-$\beta$ chain}
\begin{tabular}{|l|}
\hline
B$[a^2,i]$: $\ddot\nu+4\nu=-\frac{16\beta}{N}\nu^3$\\
$\left\{\begin{array}{l}
\ddot\nu_k+4\left(1+\frac{12\beta}{N}\nu^2\right)\nu_k\sin^2k=0
\end{array}\right.$\\
\hline
B$[a^3,a^2i]$: $\ddot\nu+3\nu=-\frac{27\beta}{2N}\nu^3$\\
$\left\{\begin{array}{l}
\ddot\nu_k+4\nu_k\sin^2k=-\frac{18\beta}{N}\nu^2\sin k
\left[2\nu_k\sin k-\nu_{\frac{\pi}{3}+k}\sin\left(k+\frac{\pi}{3}\right)+\nu_{\frac{2\pi}{3}+k}\sin\left(k+\frac{2\pi}{3}\right)\right]\\
\ddot\nu_{\frac{\pi}{3}+k}+4\nu_{\frac{\pi}{3}+k}\sin^2\left(k+\frac{\pi}{3}\right)=-\frac{18\beta}{N}\nu^2\sin\left(k+\frac{\pi}{3}\right)
\left[-\nu_k\sin k+2\nu_{\frac{\pi}{3}+k}\sin\left(k+\frac{\pi}{3}\right)-\nu_{\frac{2\pi}{3}+k}\sin\left(k+\frac{2\pi}{3}\right)\right]\\
\ddot\nu_{\frac{2\pi}{3}+k}+4\nu_{\frac{2\pi}{3}+k}\sin^2\left(k+\frac{2\pi}{3}\right)=-\frac{18\beta}{N}\nu^2\sin\left(k+\frac{2\pi}{3}\right)
\left[\nu_k\sin k-\nu_{\frac{\pi}{3}+k}\sin\left(k+\frac{\pi}{3}\right)+2\nu_{\frac{2\pi}{3}+k}\sin\left(k+\frac{2\pi}{3}\right)\right]\\
\end{array}\right.$\\
\hline
B$[a^4,a^3i]$: $\ddot\nu+2\nu=-\frac{4\beta}{N}\nu^3$\\
$\left\{\begin{array}{l}
\ddot\nu_k+4\left(1+\frac{6\beta}{N}\nu^2\right)\nu_k\sin^2k=0
\end{array}\right.$\\
\hline
B$[a^3,a^2iu]$: $\ddot\nu+3\nu=-\frac{27\beta}{2N}\nu^3$\\
$\left\{\begin{array}{l}
\ddot\nu_k+4\nu_k\sin^2k=-\frac{18\beta}{N}\nu^2\sin k
\left[2\nu_k\sin k+\nu_{\frac{\pi}{3}+k}\sin\left(k+\frac{\pi}{3}\right)-\nu_{\frac{2\pi}{3}+k}\sin\left(k+\frac{2\pi}{3}\right)\right]\\
\ddot\nu_{\frac{\pi}{3}+k}+4\nu_{\frac{\pi}{3}+k}\sin^2\left(k+\frac{\pi}{3}\right)=-\frac{18\beta}{N}\nu^2\sin\left(k+\frac{\pi}{3}\right)
\left[\nu_k\sin k+2\nu_{\frac{\pi}{3}+k}\sin\left(k+\frac{\pi}{3}\right)+\nu_{\frac{2\pi}{3}+k}\sin\left(k+\frac{2\pi}{3}\right)\right]\\
\ddot\nu_{\frac{2\pi}{3}+k}+4\nu_{\frac{2\pi}{3}+k}\sin^2\left(k+\frac{2\pi}{3}\right)=-\frac{18\beta}{N}\nu^2\sin\left(k+\frac{2\pi}{3}\right)
\left[-\nu_k\sin k+\nu_{\frac{\pi}{3}+k}\sin\left(k+\frac{\pi}{3}\right)+2\nu_{\frac{2\pi}{3}+k}\sin\left(k+\frac{2\pi}{3}\right)\right]\\
\end{array}\right.$\\
\hline
B$[a^4,iu]$: $\ddot\nu+2\nu=-\frac{8\beta}{N}\nu^3$\\
$\left\{\begin{array}{l}
\ddot\nu_k+4\left(1+\frac{6\beta}{N}\nu^2\right)\nu_k\sin^2k=-\frac{12\beta}{N}\nu^2\nu_{\frac{\pi}{2}-k}\sin 2k\\
\ddot\nu_{\frac{\pi}{2}-k}+4\left(1+\frac{6\beta}{N}\nu^2\right)\nu_{\frac{\pi}{2}-k}\cos^2k=-\frac{12\beta}{N}\nu^2\nu_k\sin 2k\\
\end{array}\right.$\\
\hline
B$[a^6,a^5i,a^3u]$: $\ddot\nu+\nu=-\frac{3\beta}{2N}\nu^3$\\
$\left\{\begin{array}{l}
\ddot\nu_k+4\nu_k\sin^2k=-\frac{6\beta}{N}\nu^2\sin k
\left[2\nu_k\sin k+\nu_{\frac{\pi}{3}+k}\sin\left(k+\frac{\pi}{3}\right)-\nu_{\frac{2\pi}{3}+k}\sin\left(k+\frac{2\pi}{3}\right)\right]\\
\ddot\nu_{\frac{\pi}{3}+k}+4\nu_{\frac{\pi}{3}+k}\sin^2\left(k+\frac{\pi}{3}\right)=-\frac{6\beta}{N}\nu^2\sin\left(k+\frac{\pi}{3}\right)
\left[\nu_k\sin k+2\nu_{\frac{\pi}{3}+k}\sin\left(k+\frac{\pi}{3}\right)+\nu_{\frac{2\pi}{3}+k}\sin\left(k+\frac{2\pi}{3}\right)\right]\\
\ddot\nu_{\frac{2\pi}{3}+k}+4\nu_{\frac{2\pi}{3}+k}\sin^2\left(k+\frac{2\pi}{3}\right)=-\frac{6\beta}{N}\nu^2\sin\left(k+\frac{2\pi}{3}\right)
\left[-\nu_k\sin k+\nu_{\frac{\pi}{3}+k}\sin\left(k+\frac{\pi}{3}\right)+2\nu_{\frac{2\pi}{3}+k}\sin\left(k+\frac{2\pi}{3}\right)\right]\\
\end{array}\right.$\\
\hline
\end{tabular}
\end{table}

\begin{table}
\centering
\caption{\label{table7} Thresholds of bush stability for the FPU-$\alpha$
and FPU-$\beta$ chains with $N=12$. Here we give the maximal value of the
displacement $x$, as well as the maximal energy for which the corresponding
bushes lose their stability}
\begin{tabular}{|c|c|c|c|c|c|}
\hline
Bush&Displacement pattern&\multicolumn{4}{c}{Threshold of the bush stability}\vline\\
\cline{3-6}
&&\multicolumn{2}{c}{in displacement}\vline&\multicolumn{2}{c}{in energy (per atom)}\vline\\
\cline{3-6}
&&FPU-$\alpha$&FPU-$\beta$&FPU-$\alpha$&FPU-$\beta$\\
\hline
B$[a^2,i]$&$|x,-x|$&$0.302$&$0.112$&$0.183$&$0.025$\\
\hline
B$[a^3,i]$&$|x,0,-x|$&$0.203$&$0.268$&$0.047$&$0.085$\\
\hline
B$[a^4,ai]$&$|0,x,0,-x|$&$0$&$1.161$&$0$&$0.675$\\
\hline
B$[a^4,i,a^2u]$&$|x,x,-x,-x|$&&$>20$&&$>10000$\\
\hline
B$[a^6,ai,a^3u]$&$|x,0,-x,-x,0,x|$&&$0.488$&&$0.079$\\
\hline
B$[a^3,iu]$&$|x,-2x,x|$&&$0.157$&&$0.074$\\
\hline
\end{tabular}
\end{table}

\clearpage

\clearpage

\twocolumn

\clearpage

~

~

~

~

~

~

~

~

~

~

~

~

~

~

\clearpage

\end{document}